\documentclass[preprint,12pt,sort&compress]{elsarticle}

\usepackage{titletoc}
\usepackage{textcomp}
\usepackage{amssymb}
\usepackage{xcolor}
\usepackage{amsmath}
\usepackage{pdflscape}
\usepackage{subcaption}
\usepackage{hyperref}
\usepackage{cleveref}
\usepackage{array}
\usepackage{amsmath,float}

\journal{Solar Energy}
\begin{document}

\begin{frontmatter}



\title{Unveiling architectural and optoelectronic synergies in lead-free perovskite/perovskite/kesterite triple-junction monolithic tandem solar cells}







\author[inst1]{Md. Faiaad Rahman}
\author[inst1,inst2]{Md. Ashaduzzaman Niloy}
\author[inst1]{Ehsanur Rahman\corref{cor1}}
\ead{ehsaneee@eee.buet.ac.bd}
\author[inst1]{Ahmed Zubair\corref{cor1}}
\ead{ahmedzubair@eee.buet.ac.bd}
\cortext[cor1]{Corresponding authors:}

\affiliation[inst1]{organization={Department of Electrical and Electronic Engineering}, 
            addressline={Bangladesh University of Engineering and Technology}, 
            city={Dhaka}, 
            postcode={Dhaka-1205}, 
            country={Bangladesh}}
\affiliation[inst2]{organization={Department of Electrical and Electronic Engineering}, 
            addressline={Green University of Bangladesh}, 
            city={Dhaka}, 
            postcode={Narayanganj-1461}, 
            country={Bangladesh}}

\begin{abstract}
The widespread use of lead-based materials in tandem solar cells raises critical environmental and health concerns due to their inherent toxicity and risk of contamination. To address this challenge, we focused on lead-free tandem architectures based on non-toxic, environmentally benign materials such as tin-based perovskites and kesterites, which are essential for advancing sustainable photovoltaic technologies. In this study, we present the proposition, design, and optimization of two distinct lead-free monolithic tandem solar cell architectures --- an all-perovskite dual-junction device employing potassium tin iodide (KSnI\textsubscript{3}) and formamidinium tin triiodide (FASnI\textsubscript{3}) as absorbers for the top and bottom subcells, respectively, and a triple-junction monolithic tandem structure incorporating KSnI\textsubscript{3}, FASnI\textsubscript{3}, and Ag-doped copper zinc tin selenide (ACZTSe) as absorbers for the top, middle, and bottom subcells, respectively. We simulated the optical and electrical characteristics of these devices using the finite-difference time-domain and finite element methods, explicitly considering radiative, non-radiative, and surface recombination mechanisms. The optimized all-perovskite dual-junction solar cell achieved a power conversion efficiency (PCE) of 27.3\%, with short-circuit current density (J\textsubscript{sc}) of 14.74 mA/cm\textsuperscript{2}, open-circuit voltage (V\textsubscript{oc}) of 2.227 V, and fill factor (FF) of 83.14\%. Conversely, the optimized triple-junction hybrid perovskite–kesterite architecture secured an elevated PCE of 30.69\%, along with J\textsubscript{sc} of 13.184 mA/cm\textsuperscript{2}, V\textsubscript{oc} of 2.766 V, and FF of 84.18\%. These findings reveal the strong potential of lead-free perovskite and kesterite material based absorbers in promoting high-performance hybrid tandem solar cells, highlighting their importance in advancing sustainable and efficient photovoltaic technologies.

\end{abstract}


\end{frontmatter}


\section{Introduction}
The global energy industry is rapidly transitioning towards renewable energy as countries are making tremendous efforts to reduce their dependence on fossil fuels and curb climate change. Among renewable resources, solar energy stands out for its abundance, scalability, and accessibility. Solar photovoltaic (PV) technology has emerged as a key prospect in this shift, with ongoing innovations aimed at improving efficiency and lowering costs. Researchers are advancing solar performance by integrating existing PV technologies with novel materials, making high-efficiency solar power a key strategy for sustainable energy generation.
In general, the evolution of solar PV technology can be grouped into three generations \cite{cite7}. The first generation includes: crystalline silicon (c-Si), further classified into monocrystalline and polycrystalline types. Commercial monocrystalline modules typically exhibit 20–25\% efficiency, while polycrystalline modules provide 18–21\%, with recent research pushing monocrystalline silicon cell efficiency to 27.81\% under laboratory conditions \cite{cite4, cite9}. Second-generation solar cells encompass a diverse class of thin-film technologies consisting of amorphous silicon (a-Si)\cite{Saravanan2022}, cadmium telluride (CdTe)~\cite{Kavanagh2021}, chalcogenide \cite{SAJITHA2024100490}, and most importantly, group III-V materials \cite{Tomasulo2025advances}. A thoroughly studied III-V absorber, GaAs, with a direct bandgap of 1.42 eV, has demonstrated lab efficiencies up to 29.1\% for single-junction devices \cite{nrel_efficiency_2025}, and even higher when used in multi-junction architectures. However, due to high production costs, III–V solar cells are typically reserved for niche applications such as aerospace and concentrator photovoltaics.\\

Third-generation solar cells represent a new wave of emerging photovoltaic technologies designed to overcome the efficiency and cost limitations of first and second-generation devices. These technologies incorporate earth-abundant materials and structural innovations to enhance light absorption and charge separation.  
Notable examples of emerging photovoltaic technologies include plasmonic and metamaterials based solar cells, organic solar cells (OSCs), dye-sensitized solar cells (DSSCs), quantum dot solar cells (QDSCs), and perovskite solar cells (PSCs), as well as multi-junction tandem configurations~\cite{rahman2025hierarchically, dip2023, cite13}.
These emerging photovoltaics possess unique advantages, notably mechanical flexibility, low-temperature solution-based fabrication process, semi-transparent operation, and compatibility with unconventional surfaces such as windows, fabrics, and portable electronic devices~\cite{ugochukwu2025}. Yet, despite such positive aspects, third-generation devices face persistent hurdles like material instability, challenges with large-scale manufacturability, and interfacial degradation. PSCs are particularly vulnerable to environmental degradation and raise concerns about lead toxicity ~\cite{Han2025}, whereas QDSCs struggle with surface traps, heavy-metal constituents, and long-term stability issues~\cite{Najm2023QDSC}. DSSCs are limited by electrolyte leakage and sealing problems ~\cite{Ebenezer2023DSSC}, and OSCs suffer from lower efficiencies, rapid degradation, and morphological instability that impede large-area deployment~\cite{Yi2024OSCs}. Among third-generation cells, perovskite solar cells have emerged as frontrunners, reaching record efficiencies of 26.95\%~\cite{nrel_efficiency_2025}. Perovskites adopt the ABX\textsubscript{3} crystal structure, where A is an organic or inorganic cation, B is a metal cation, and X is a halide. Perovskite materials offer a highly tunable bandgap ranging from 0.96 eV for cesium silver thallium bromide (Cs\textsubscript{2}AgTlBr\textsubscript{6}) to 2.82 eV for cesium lead chloride (CsPbCl\textsubscript{3}), making them attractive for both single and multi-junction applications~\citep{slavney2018small,cite14}.
%
%
Regardless of significant advances, single-junction solar cells remain constrained by the Shockley–Queisser theoretical efficiency limit, primarily due to excess carrier thermalization and sub-bandgap transmission losses. Multi-junction tandems mitigate these limitations by stacking absorbers with complementary bandgaps, and the compositional tunability of perovskites makes them particularly advantageous for achieving optimized spectral splitting and current matching.\\

For multi-junction tandem solar cells, the semiconducting materials employed must combine optimal bandgaps with high absorption coefficients, minimal recombination losses, and good lattice compatibility to suppress interface defects~\cite{cite6}. Dual-junction tandem structures such as III--V/Si~\cite{Essig2017}, perovskite/Si~\cite{Liu2024}, and all--III--V~\cite{Steiner2021} or all--perovskite~\cite{Lin2023} material combinations were studied for their ability to surpass the Shockley-Queisser limit of single-junction cells. More advanced triple-junction architectures, including perovskite/perovskite/Si~\cite{liu2024triple} and III--V/III--V/Si~\cite{Schygulla2025} stacks, can further improve spectral utilization and reduce thermalization losses. Advancing beyond three junctions, devices with four or more junctions, typically composed of III--V materials, have demonstrated record-setting efficiencies. In 2020, Geisz \textit{et al.} reported a six-junction III--V solar cell achieving 47.1\% efficiency under 143 suns concentration through precise bandgap engineering and sophisticated epitaxial growth techniques \cite{Geisz2020}. Meanwhile, perovskite/Si tandem cells have attained over 30.6\% efficiency in recent reports, offering a low-cost and scalable alternative to III--V systems \cite{Green2025}. All-perovskite tandem cells incorporating Sn-Pb-based narrow-bandgap absorbers have achieved 28.2\% efficiency via improved grain passivation and interface quality \cite{Lin2022}. In parallel, chalcogenide-based tandem solar cell architectures such as perovskite/CIGS \cite{Jot2022} and perovskite/CZTS~\cite{ferhati2020106727}, have demonstrated considerable potential as sustainable, earth-abundant photovoltaic solutions. Nonetheless, multi-junction solar cells remain constrained by high fabrication complexity, current and lattice mismatch among subcells, and thermal-induced spectral instability under real-world illumination conditions. Moreover, the need for precise thickness and interface control during epitaxial or layer-by-layer growth imposes scalability challenges and cost barriers for industrial deployment~\cite{Karade2025Opportunities}.
Careful band alignment and interface engineering have pushed tandem perovskite devices to efficiencies near 30\%~\cite{Pei2025}. Yet additional gains can be expected via targeted bandgap tuning, improved interface passivation, efficient optical management, and precise current matching among subcells~\cite{Farias-Basulto2025}. Within the present framework, potassium tin iodide (KSnI\textsubscript{3}) ~\cite{pindolia2023non} and formamidinium tin triiodide (FASnI\textsubscript{3})~\cite{hutchinson2023resolving} have emerged as two highly promising lead-free perovskite materials. Recent theoretical studies reported PCEs for single-junction KSnI3 cells, reaching 23.85\% ~\cite{cite16}. In comparison, FASnI\textsubscript{3}-based devices have demonstrated even higher efficiency of 28.37\%~\citep{pachori2022efficient}. These findings point towards the possibility of coupling KSnI3 and FASnI3 absorbers in tandem configurations to extend device performance even further. On the chalcogenide front, silver-doped copper-zinc-tin-selenide (ACZTSe) stands out as an earth-abundant, non-toxic absorber with tunable bandgap values around 1.04–1.24 eV via the extent of Ag alloying~\cite{cite18}. All-chalcogenide tandem solar cells with ACZTSe as the bottom absorber attained PCEs of 24\% in dual-junction and 36.04\% in triple-junction designs~\cite{saha2018proposition}. Together, KSnI\textsubscript{3} with bandgap ($\sim$1.84 eV), FASnI\textsubscript{3} (1.3$\sim$1.4 eV), and ACZTSe ($\sim$1.0 eV) offer complementary optical characteristics, making them ideal for multi-junction tandem cell integration. Their combined advantages position them as suitable contenders for advancing next-generation, high-efficiency, environmentally viable photovoltaic technologies.\\

In this study, we designed and assessed two novel lead-free tandem solar cells that adopted tin-based perovskite and kesterite materials, both renowned for their strong potential in prompting high-performance, non-toxic, and sustainable photovoltaic technology. Our initial proposed structure included an all-perovskite monolithic dual-junction tandem solar cell, incorporating KSnI\textsubscript{3} and FASnI$_3$ as absorber layers. Building on this architecture, a monolithic two-terminal triple-junction tandem architecture was originated, integrating KSnI\textsubscript{3}, FASnI$_3$, and ACZTSe to enhance spectral utilization and overall device PCE. This research featured a comprehensive analysis of both the dual-junction (KSnI\textsubscript{3}/FASnI$_3$) and triple-junction (KSnI\textsubscript{3}/FASnI$_3$/ACZTSe) configurations. The investigation entailed detailed modeling and optimization of their respective single-junction subcells, followed by the sequential tuning of critical physical parameters, such as absorber layer thickness and doping densities, to maximize device performance. The insights derived from this work lay a foundation for the design and fabrication of high-efficiency next-generation multijunction solar cells with optimized interfacial and electronic properties.

\section{Device architecture and simulation methodology}\label{section 2}
%
We designed and examined four distinct solar cell architectures based on KSnI\textsubscript{3}, FASnI\textsubscript{3}, and ACZTSe, encompassing single-junction (1-J), double-junction (2-J), and triple-junction (3-J) tandem configurations. The investigation expanded progressively from single-junction devices to fully integrated tandem configurations,  with each contact and transport layer cautiously chosen in accordance with literature-guided performance and stability studies.\\

\begin{figure}[!t]
    \centering
    \includegraphics[width=1.0\linewidth]{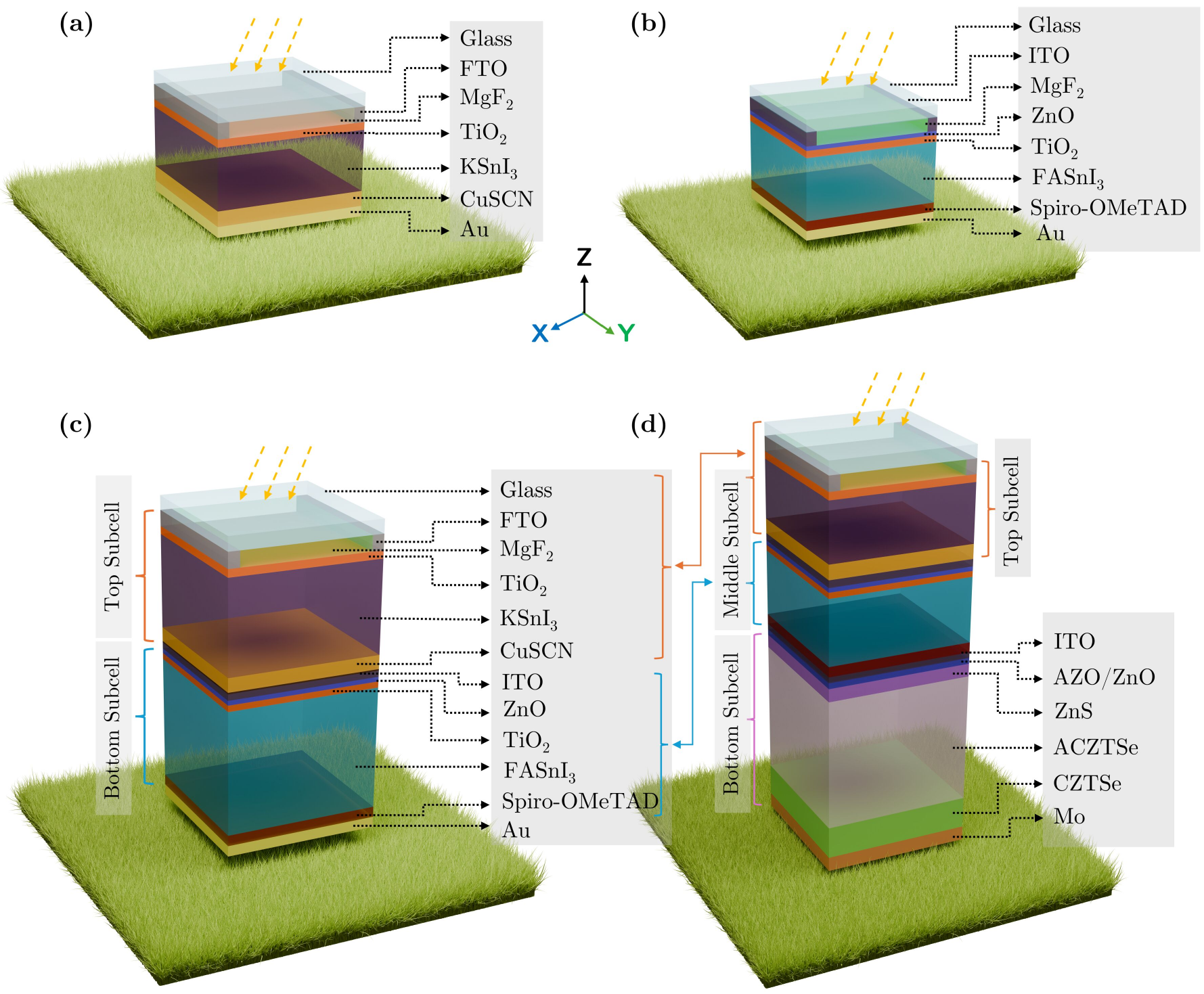}
    \caption{Device schematics of simulated modeled structures featuring: (a) 1-J KSnI\textsubscript{3}, (b) 1-J FASnI\textsubscript{3}, (c) 2-J monolithic KSnI\textsubscript{3}/FASnI\textsubscript{3} tandem, and (d) 3-J monolithic KSnI\textsubscript{3}/FASnI\textsubscript{3}/ACZTSe tandem solar cells.}
    \label{fig:1}
\end{figure}

The first phase focused on designing and simulating a single-junction solar cell with an n/i/p configuration, utilizing KSnI\textsubscript{3} as the primary absorber, as shown in Fig.~\ref{fig:1}(a). With a direct bandgap of $\sim$1.84 eV~\cite{pindolia2023non}, KSnI\textsubscript{3} is well suited as a top absorber in tandem architectures. Fluorine-doped tin oxide (FTO) served as a transparent front contact owing to its high optical transmittance ($>$85\%), wide bandgap ($\sim$3.6 eV), and suitable work function (4.4–-4.7 eV)~\cite{cite26}. Titanium dioxide (TiO\textsubscript{2}) in its anatase phase acted as an electron transport layer (ETL), offering a bandgap of 3.2 eV and favorable band alignment for electron transfer and hole blocking~\cite{cite16}. Copper(I) thiocyanate (CuSCN) functioned as the hole transport layer (HTL) with valence band ($\sim$5.3 eV), wide bandgap ($\sim$3.6 eV), high hole mobility, and strong optical transparency, enabling efficient hole extraction and suppressed recombination~\cite{cite18}. A thin magnesium fluoride (MgF\textsubscript{2}) was introduced as an anti-reflection coating (ARC) to minimize reflection losses. For the back contact layer (BCL), both indium tin oxide (ITO) and gold (Au) were evaluated. 
ITO offers high transmittance and a suitable work function (4.7–5.0~eV), enabling near-ohmic contact and current continuity between stacked subcells~\cite{cite22}.\\

Subsequently, another single-junction cell with an n/n+/p/p+ configuration, featured in Fig.~\ref{fig:1}(b), was modeled employing FASnI\textsubscript{3} as the absorber, with a direct bandgap of $\sim$1.41~eV~\cite{hutchinson2023resolving}. TiO\textsubscript{2} was retained as the ETL for consistency and compatibility, while a ZnO buffer layer was incorporated between TiO\textsubscript{2} and the absorber to improve electron transport. ZnO exhibits good mobility and favorable band alignment, although passivation may be required to reduce reactivity with iodide ions. For HTL, both CuSCN and Spiro-OMeTAD were examined. Spiro-OMeTAD, with a HOMO level of 5.1–5.3 ~eV, aligns closely with the FASnI$_3$ valence band ($\sim$5.2 eV), enabling efficient hole extraction and supporting open-circuit voltages above 0.9 V~\cite{cite18}. Au and ITO were also investigated as BCL to assess their electrical performance and optical transmittance.\\

After optimizing the single-junction devices, a double-junction (2-J) tandem solar cell was constructed by stacking the KSnI\textsubscript{3} cell as the top subcell over the FASnI$_3$ cell as the bottom subcell, as illustrated in Fig.~\ref{fig:1}(c). The same charge transport layers were retained, with TiO\textsubscript{2} serving as the ETL and CuSCN as the HTL. An ITO interlayer was introduced as the intermediate transparent conducting oxide (ITCO). Owing to its wide optical bandgap (3.5–-4.3 eV), high visible transmittance, and strong electrical conductivity, ITO functions effectively as both a transparent electrode and a recombination layer~\cite{cite22}. To further improve band alignment and charge transfer, a ZnO buffer layer was incorporated between the TiO\textsubscript{2} ETL and the ITCO.\\

To extend the architecture further, a triple-junction (3-J) tandem solar cell was developed by introducing ACZTSe as the bottom absorber, as shown in Fig.~\ref{fig:1}(d). ACZTSe, a kesterite-type chalcogenide with a bandgap of ~1.0 eV, exhibits high hole mobility and strong light absorption, with a valence band that aligns well with CZTSe. Zinc sulfide (ZnS) was employed as the ETL due to its wide bandgap (3.6–-3.7 eV), suitable conduction band alignment, and high electron mobility (10–-100 cm$^2$/V.s), ensuring efficient electron extraction with minimal recombination~\cite{cite17}. A CZTSe back surface field (BSF) layer was incorporated below ACZTSe, leveraging its p-type conductivity, high hole mobility (7 cm$^2$/V·s), and slightly deeper valence band to create a favorable gradient for hole transport while blocking electrons~\cite{cite15}. For the rear electrode, molybdenum (Mo) was selected owing to its excellent conductivity, work function alignment, and infrared photon reflectivity, which enhances light harvesting~\cite{cite25}. Between the perovskite middle subcell (FASnI$_3$) and the kesterite bottom subcell (ACZTSe), a ZnO/AZO bilayer was introduced as a buffer and tunneling layer. Aluminum-doped ZnO (AZO) provides low resistivity, high transmittance, and chemical stability, while undoped ZnO improves energy alignment with ZnS and suppresses sulfur/selenium interdiffusion~\cite{cite23, cite24}. Together, this bilayer supports efficient carrier extraction while preserving optical and structural integrity. This progressive simulation strategy, from single-junction cells to 3-J tandem structures, ensured optimized charge extraction, reduced interfacial recombination, and current matching across all subcells, with each layer chosen for compatibility, stability, and scalability.\\

We applied a hybrid optical-electrical simulation framework to design and optimize new proposed tandem solar cell architectures. The finite difference time domain (FDTD) was implemented to examine the optical behavior, whereas the finite element method (FEM) was employed to investigate the electrical performance behavior. Post-processing was conducted to interpret and analyze the results.\\

To analyze the optical responses of the proposed solar cell architectures, the FDTD method solves Maxwell’s wave equations to capture electromagnetic interactions within the device layers. 
FDTD was employed for its broadband capability, high accuracy, and extensive computational flexibility. For optical modeling, the complex refractive indices (n+i$\kappa$) of the constituent materials are utilized as optical input parameters, obtained from reported literature \citep{rodriguez2017self, ball2015optical, konig2014electrically, treharne2011optical, sarkar2019hybridized, tooghi2020high, raoult2019optical, saha2018proposition, ghimire2017optical, pindolia2023non}. For modeling the proposed tandem cells in 2D, a periodic boundary condition was applied along the X-axis, while a perfectly matched layer (PML) was used along the Y-axis to prevent artificial reflection at the simulation boundary. The standard AM1.5G solar spectrum is incident along the Y-axis as an illumination source, covering a photon wavelength $\lambda_{photon}$ ranging from 300 to 1200 nm. However, to exclude contributions from parasitic absorption and intra-band transitions, the photo-generation rate (G) was evaluated solely within the spectral range corresponding to inter-band optical transitions relevant to the solar cell architectures.\\

For electrical stimulation, key performance parameters, such as PCE ($\eta$), short-circuit current density (J\textsubscript{sc}), open-circuit voltage (V\textsubscript{oc}), and fill factor (FF), were evaluated using the FEM method. The electrical simulations were performed by self-consistently solving the Poisson and drift-diffusion equations in two dimensions. Dirichlet boundary conditions were imposed at the metal-semiconductor interfaces, and Neumann conditions were used at the periodic boundaries. The electrical parameters used for theoretical calculations and performing the simulations for absorbers, transport layers, and contacts were extracted from prior studies \cite{tooghi2020high, kuo2014efficiency, saha2017proposition, tabernig2022optically, pindolia2023non, sumona2023optimization, pachori2022efficient, milot2016radiative, hutchinson2023resolving, kumar2020optimized, saha2018proposition, tan2024improved, enkhbat2023high, ivriq2025enhancing, xu2024spiro} and are summarized in Tables~\ref{Table:1} and S1 of the Supplementary Material.\\
\begin{table}[!b]
\centering
\caption{Electrical parameters of absorber materials (KSnI\textsubscript{3}, FASnI\textsubscript{3}, ACZTSe and CZTSe).}
\label{Table:1}
\resizebox{\textwidth}{!}{%
\begin{tabular}{l*{4}{>{\centering\arraybackslash}m{3.2cm}}}
\hline
\textbf{\begin{tabular}[c]{@{}c@{}}Parameters\end{tabular}} &
  \textbf{\begin{tabular}[c]{@{}c@{}}KSnI\textsubscript{3}\\ \citep{pindolia2023non, sumona2023optimization}\end{tabular}} &
  \textbf{\begin{tabular}[c]{@{}c@{}}FASnI\textsubscript{3}\\ \citep{pachori2022efficient, milot2016radiative, hutchinson2023resolving, kumar2020optimized}\end{tabular}} &
  \textbf{\begin{tabular}[c]{@{}c@{}}ACZTSe\\ \citep{saha2018proposition, tan2024improved, enkhbat2023high}\end{tabular}} &
  \textbf{\begin{tabular}[c]{@{}c@{}}CZTSe\\ \citep{saha2018proposition, tan2024improved}\end{tabular}} \\
\hline
Thickness (nm) & 100--800 & 100--800 & 400--1400 & 50--300 \\
Bandgap $E_g$ (eV) & 1.84 & 1.4 & 1.088 & 1.06 \\
DC permittivity $\varepsilon$ & 10.4 & 8.2 & 8.5 & 7 \\
Electron affinity $\chi$ (eV) & 3.44 & 3.9 & 4.05 & 4.05 \\
Mobility $\mu_n/\mu_p$ (cm\textsuperscript{2}/Vs) & 21.28 / 19.46 & 22 / 22 & 75 / 10 & 145 / 35 \\
SRH lifetime $\tau_e/\tau_h$ (ns) & 7.78 / 8.135 & 7.93 / 7.93 & 1.79 / 1.79 & 0.52 / 0.52 \\
Radiative recombination, B\textsubscript{n,p} (cm\textsuperscript{3}s\textsuperscript{-1}) & $1.1\times10^{-10}$ & $4.37\times10^{-10}$ & $3.5\times10^{-10}$ & $1.08\times10^{-10}$ \\
Auger recombination, C\textsubscript{n,p} (cm\textsuperscript{6}s\textsuperscript{-1}) & 1$\times$10\textsuperscript{-30} &  9.3$\times$10\textsuperscript{-30} & 1$\times$10\textsuperscript{-28} & 1$\times$10\textsuperscript{-28} \\
Effective conduction band density, $N_C$ (cm\textsuperscript{-3}) & $1\times10^{18}$ & 1.0$\times$10$^{18}$ & $2.51\times10^{19}$ & $2.2\times10^{18}$ \\
Effective valence band density, $N_V$ (cm\textsuperscript{-3}) & $1\times10^{18}$ & 1.0$\times$10$^{18}$ & $1.79\times10^{19}$ & $1.8\times10^{19}$ \\
Donor doping $N_D$ (cm\textsuperscript{-3}) & $1\times10^{15}$ & -- & -- & -- \\
Acceptor doping $N_A$ (cm\textsuperscript{-3}) & $1\times10^{15}$ & $1\times10^{16}$ & $1\times10^{16}$ & $5\times10^{18}$ \\
Surface Recombination (cm/s) & -- & -- & -- & \begin{tabular}[c]{@{}c@{}}(CZTSe/Mo)\\ $1\times10^{7}$\end{tabular} \\
\hline
\end{tabular}%
}
\end{table}

\begin{figure}[!t]
    \centering
    \includegraphics[width=0.95\linewidth]{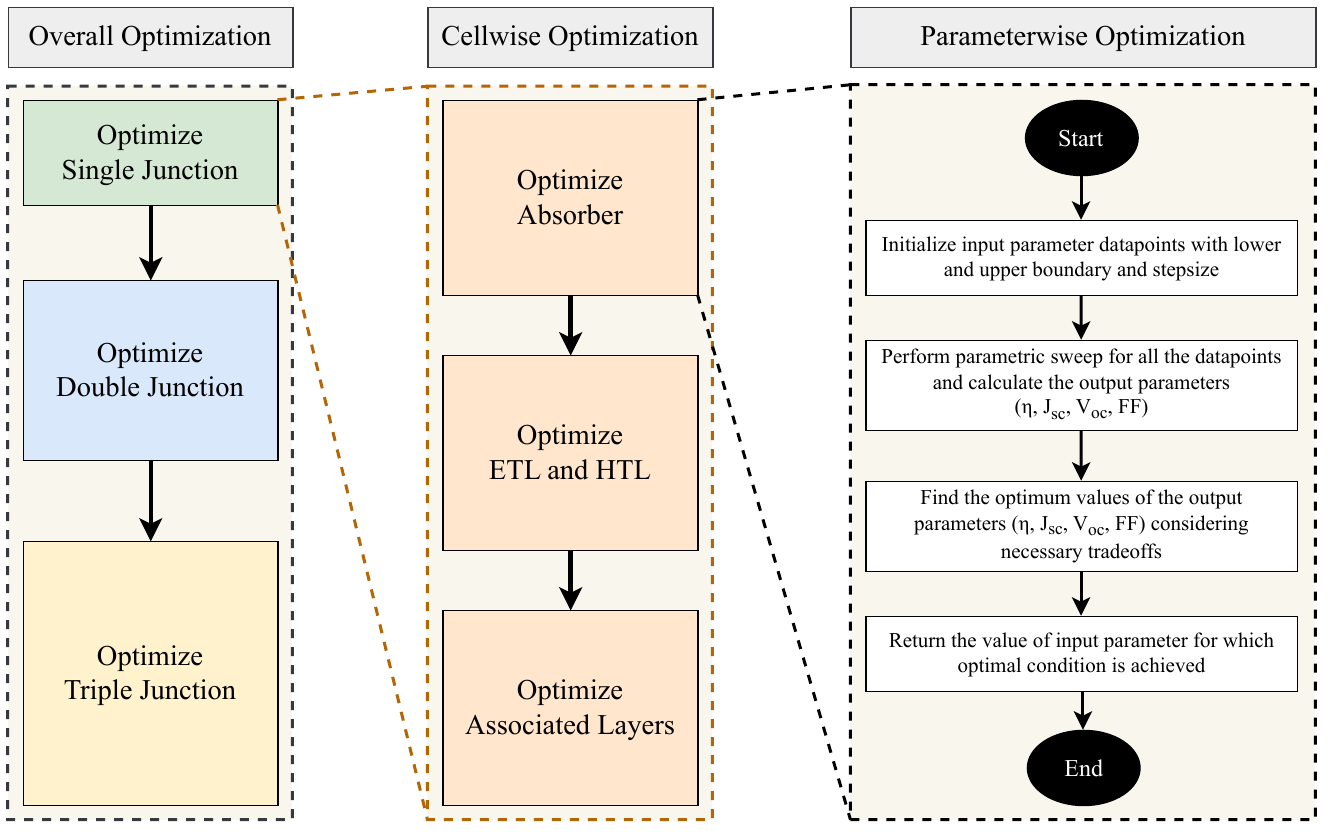}
    \caption{Sequential optimization of the modeled architectures from single junction cells to triple junction tandem cells.}
    \label{fig:2}
\end{figure}

For parametric optimization of the monolithic tandem configurations, current matching was strictly enforced across all subcells to extract key photovoltaic performance parameters such as $\eta$, J\textsubscript{sc}, V\textsubscript{oc}, and FF, ensuring realistic performance evaluation. The definitions and governing equations for these parameters are detailed in the Supplementary Material. A complete workflow, from single-junction modeling to the final triple-junction configuration, is outlined in Fig.~\ref{fig:2}.
The optimization workflow proceeded hierarchically. First, a single-junction cell was modeled, and parametric sweeps were performed until convergence to optimize its parameters. Each layer of the single-junction was then individually refined using the same approach, followed by an overall optimization of the full single-junction considering necessary trade-offs among performance metrics. Once all single-junctions were optimized, dual-junction cells were constructed and optimized with careful consideration of current matching between subcells. Finally, this strategy was extended to triple-junction architectures, applying the same layer and device-level optimization while maintaining spectral complementarity and current matching across all subcells.

\section{Results and discussion}\label{section 3}


\subsection{Structural refinement and light-harvesting in 1-J KSnI\textsubscript{3} cells}
\label{sub-section 3.1} 
To model the MgF\textsubscript{2}/FTO/TiO\textsubscript{2}/KSnI\textsubscript{3}/CuSCN/ITO configuration as an n/i/p single-junction solar cell architecture, as illustrated in Fig.~\ref{fig:1}(a), we considered MgF\textsubscript{2} as ARC, FTO as TCO, TiO\textsubscript{2} as ETL, KSnI\textsubscript{3} as absorber, CuSCN as HTL, and ITO as BCL with initial thickness of 100 nm, 100 nm, 90 nm, 200 nm, 50 nm, and 50 nm, respectively. Initially, the doping densities of ETL, absorber, and HTL were set to 5$\times$10\textsuperscript{18}, 1$\times$10\textsuperscript{15}, and 1$\times$10\textsuperscript{18} cm\textsuperscript{-3}, respectively. 
To establish the baseline device behavior, we performed preliminary thickness sweeps of the KSnI\textsubscript{3} absorber, TiO\textsubscript{2} ETL, and CuSCN HTL to identify performance-sensitive regions and define the parameter space for subsequent detailed optimization. The corresponding results are provided in Figs.~S2 and ~S3 of the Supplementary Material.\\

\begin{figure}[!t]
    \centering
    \includegraphics[width=1.0\linewidth]{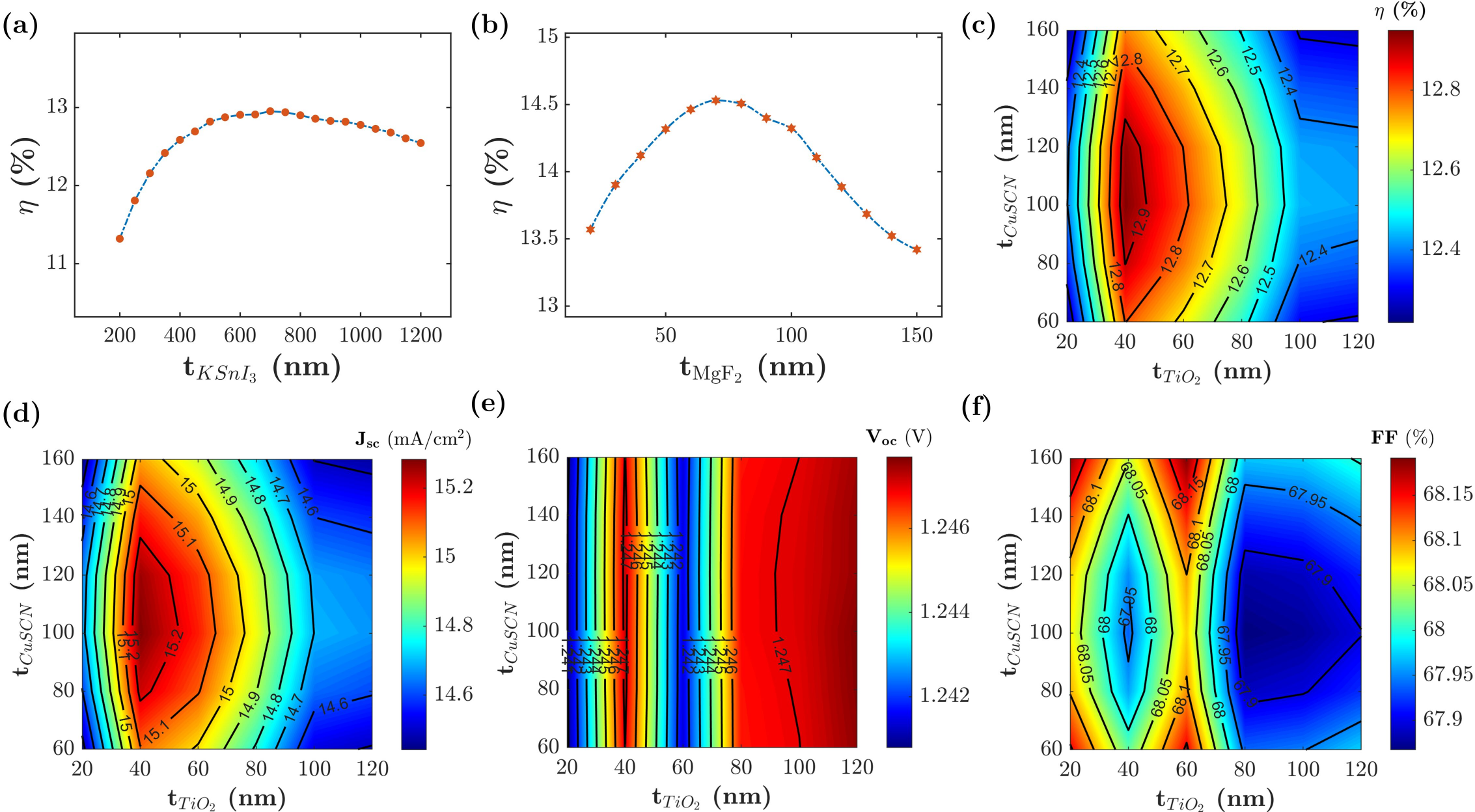}
    \caption{(a) Impact on PCE ($\eta$) of the 1-J KSnI\textsubscript{3} solar cell corresponding to the absorber thickness variation, presented for the final KSnI\textsubscript{3} thickness iterative sweep. (b) Influence of MgF\textsubscript{2} ARC thickness on PCE. Contour plots of key performance metrics (c) $\eta$, (d) J\textsubscript{sc}, (e) V\textsubscript{oc}, and (f) FF as functions of ETL (TiO\textsubscript{2}) and HTL (CuSCN) thicknesses after final parametric sweeping in single junction KSnI\textsubscript{3} cell.}
    \label{fig:3}
\end{figure}

Afterward, following the self-consistent optimization process as featured in Fig.~\ref{fig:2}, final iterative thickness optimization of the KSnI\textsubscript{3} absorber, TiO\textsubscript{2} ETL, and CuSCN HTL were performed to maximize device performance. First, the thickness of KSnI\textsubscript{3} was varied from 200 to 1200 nm while keeping the ETL and HTL thicknesses at 40 and 100 nm, respectively, as shown in Fig.~\ref{fig:3}(a). PCE increased steadily from 11.32\% at 200 nm to a peak of 12.95\% at 700 nm, after which further thickness increases led to marginal decreases in efficiency. This trend reflects the trade-off between enhanced photon absorption and increased carrier recombination in a thicker KSnI\textsubscript{3} absorber. J\textsubscript{sc} increased with thickness due to improved light photon absorption in the 600--700~nm wavelength range, whereas V\textsubscript{oc} and FF decreased slightly, indicating higher recombination and series resistance effects. This trend is observed in Figs.~S4(b-d). Subsequently, a parametric sweep of the TiO\textsubscript{2} ETL (20–160 nm) and CuSCN HTL (60–160 nm) was conducted with the absorber thickness fixed at 700 nm as shown in Figs.~\ref{fig:3}(c-f). 
Analysis revealed that, a 40 nm thickness of TiO\textsubscript{2} and a 100 nm thickness of CuSCN yielded a peak PCE of 12.95\%, with J\textsubscript{sc} of 15.28 mA/cm\textsuperscript{2}, V\textsubscript{oc} of 1.247 V, and FF of 67.94\%. Reducing the ETL thickness weakened the built-in electric field, thereby limiting electron extraction, while excessive thickness introduced additional series resistance and hindered charge transport. Similarly, optimizing the HTL thickness proved essential for maintaining efficient hole transport without compromising carrier collection at the interface. These findings highlight the importance of carefully balancing layer thicknesses to achieve optimal device performance.\\

\begin{figure}[!b]
    \centering
    \includegraphics[width=1.0\linewidth]{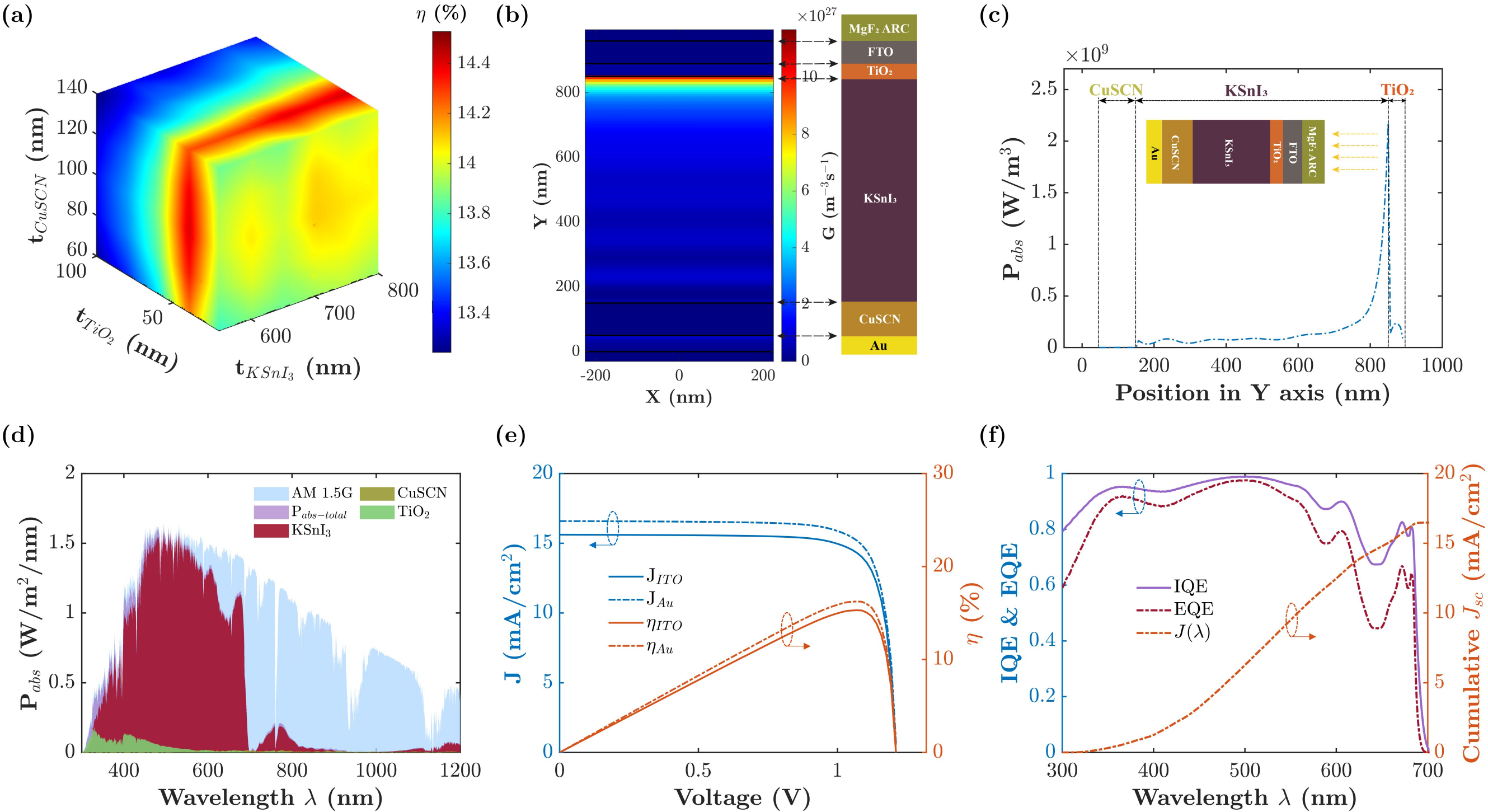}
    \caption{(a) Impact on PCE ($\eta$) through altering KSnI\textsubscript{3}, ETL, and HTL thickness under different configurations. (b) Spatial carrier generation rate profile, G, highlighting the regions of maximum generation. (c) Absorbed power P$_{abs}$ towards the illumination of sunlight in the Y direction. (d) Spectral power absorption of different layers 1-J KSnI\textsubscript{3} cell versus photon wavelength corresponding to AM 1.5G solar spectrum. (e) Comparison between J-V and P-V plot for incorporating Au and ITO as BCL. (f) Internal and external quantum efficiency (IQE, EQE) along with cumulative J\textsubscript{sc} vs photon wavelength of the optimized 1-J KSnI\textsubscript{3} cell.}
    \label{fig:4}
\end{figure}

We then explored doping density variations in the transport layers. The donor density (N\textsubscript{D})  of TiO\textsubscript{2} and acceptor density (N\textsubscript{A}) of CuSCN were varied from 5$\times$10\textsuperscript{16} to 5$\times$10\textsuperscript{19} cm\textsuperscript{-3} with the results visualized in Figs.~S5(a-d). 
The highest PCE of 14.32\% was achieved for N\textsubscript{D} of 5$\times$10\textsuperscript{19} cm\textsuperscript{-3} in TiO\textsubscript{2} and N\textsubscript{A} of 2$\times$10\textsuperscript{19} cm\textsuperscript{-3} in CuSCN, corresponding to J\textsubscript{sc} of 15.29 mA/cm², V\textsubscript{oc} of 1.234 V, and FF of 75.86\%. Higher donor doping in TiO\textsubscript{2} enhanced the built-in electric field, promoting electron extraction and reducing recombination at the absorber/ETL interface. Similarly, increasing acceptor doping in CuSCN improved hole transport; however, acceptor density exceeding 10\textsuperscript{19} cm\textsuperscript{-3} introduced significant defect-assisted recombination, limiting further efficiency gains~\cite{Jin2019}. Limiting N\textsubscript{A} to 5$\times$10\textsuperscript{18} cm\textsuperscript{-3} still yielded a comparable PCE of 14.31\%, indicating the device’s robustness.
Following that, we varied the thickness of MgF\textsubscript{2} ARC from 20-150 nm to reduce the reflection loss. A maximum PCE of 14.52\% was achieved, as shown in Fig.~\ref{fig:3}(b). The optimal ARC thickness of 70~nm resulted in an average normalized reflection (R) of 4.45\%, enhancing photon absorption in the 300–700 nm wavelength range, as highlighted in Fig.~S6(a). Subsequently, a volumetric thicknesses sweep of absorber, ETL, and HTL confirmed an optimum PCE of 14.53\% for absorber, ETL, and HTL thicknesses of 700~nm, 40~nm, and 100~nm, respectively, as shown in Fig.~\ref{fig:4}(a). This analysis demonstrates that, a self-consistent optimization of layer thicknesses and doping densities converges to the global performance maximum, highlighting the robustness and predictive accuracy of the design approach.\\

\begin{table}[!t]
\centering
\small
\renewcommand{\arraystretch}{1.1}
\caption{Summary of the stepwise performance improvement of the single-junction KSnI\textsubscript{3} solar cell.}
\label{Table:2}
\resizebox{\textwidth}{!}{%
\begin{tabular}{@{}lcccc@{}}
\hline
\textbf{Optimization parameters} &
  \textbf{\begin{tabular}[c]{@{}c@{}}$\eta$\\ (\%)\end{tabular}} &
  \textbf{\begin{tabular}[c]{@{}c@{}}J\textsubscript{sc}\\ (mA/cm\textsuperscript{2})\end{tabular}} &
  \textbf{\begin{tabular}[c]{@{}c@{}}V\textsubscript{oc}\\ (V)\end{tabular}} &
  \textbf{\begin{tabular}[c]{@{}c@{}}FF\\ (\%)\end{tabular}} \\ \hline
\begin{tabular}[c]{@{}l@{}} Layers thickness - TiO\textsubscript{2}/KSnI\textsubscript{3}/CuSCN\\ (40 nm/ 750 nm/ 100 nm)  \end{tabular} &
  12.95 & 15.28 & 1.247 & 67.94 \\ \\

\begin{tabular}[c]{@{}l@{}} ETL/HTL doping\\ (ETL: $5\times10^{19}$ cm\textsuperscript{-3}, HTL: $5\times10^{18}$ cm\textsuperscript{-3}) \end{tabular}&
  14.31 & 15.29 & 1.234 & 75.86 \\ \\

\begin{tabular}[c]{@{}l@{}} MgF\textsubscript{2} ARC thickness \\ (70 nm) \end{tabular} &
  14.53 & 15.62 & 1.235 & 75.37 \\ \\


\begin{tabular}[c]{@{}l@{}} UV-ozone treated TiO\textsubscript{2} ETL \\ (Work function: 5.1 eV)  \end{tabular}&
  15.30 & 15.62 & 1.228 & 79.79 \\ \\

\begin{tabular}[c]{@{}l@{}} Substitution of Au BCL with ITO \\ (50 nm)  \end{tabular} &
  16.28 & 16.62 & 1.232 & 79.50 \\
\hline
\end{tabular}%
}
\end{table}
To reduce the interfacial band offset at the KSnI\textsubscript{3}/TiO\textsubscript{2} junction, UV-ozone-treated TiO\textsubscript{2} was employed, shifting the TiO\textsubscript{2} work function ($\phi_{TiO_2}$) from 5.3 eV (oxidized TiO\textsubscript{2} film) to 5.1 eV. Klasen \textit{et al.} reported that UV-ozone treatment significantly increased the surface conductance of TiO\textsubscript{2} films by over two orders of magnitude, mitigated surface oxygen vacancies, and enhanced charge extraction, resulting in a 2\% increase in the average PCE of perovskite solar cells~\cite{klasen2019removal}. Implementing this treatment in the 1-J KSnI\textsubscript{3} device gained an improved PCE of 15.30\%, as presented in Table~S2.
Next, instead of ITO as BCL, 50 nm Au BCL was utilized to further boost the single junction performance by achieving further absorption through back reflection in the wavelength range of 550--690 nm, as shown in Fig.~S6(b). This additional absorption enhanced J\textsubscript{sc} from 15.61 to 16.62 mA/cm\textsuperscript{2}. The final optimum PCE ($\eta$) was found 16.28\% with J\textsubscript{sc} of 16.62 mA/cm\textsuperscript{2}, V\textsubscript{oc} of 1.23V, and FF of 79.50\%.\\ 

Table \ref{Table:2} summarizes the gradually improved results from the initial to the final optimized structure of the 1-J KSnI\textsubscript{3} cell. Spatial generation rate profile, G across the X-Y cross-section, highlighting the maximum carrier generation regions in Fig.~\ref{fig:4}(b), where KSnI\textsubscript{3} spans from 150 to 850 nm (thickness 700 nm). Fig.~\ref{fig:4}(c) highlights absorbed power in different regions of the single junction cell corresponding to the direction of the sunlight in the Y direction. The spectral power absorption, P$_{abs}$ and normalized IQE and EQE, along with cumulative J\textsubscript{sc} as functions of photon wavelength $\lambda$, are presented in Figs.~\ref{fig:4}(d) and (f) respectively. 
For KSnI\textsubscript{3}, photon energies above the optical bandgap of 1.84~eV effectively drive inter-band transitions, generating free electron–hole pairs that contribute to photocurrent. Beyond this threshold, photon energy becomes inadequate to induce inter-band transitions, resulting in weak optical absorption and the rapid non-radiative recombination of short-lived carriers, which therefore remain inactive in photovoltaic operation. Finally, replacing ITO with Au as the BCL significantly enhances device performance, yielding a relative PCE improvement of 6.4\%, as evidenced by Table~\ref{Table:2} and the J–V and $\eta$–V characteristics shown in Fig.~\ref{fig:4}(e).

\subsection{Optoelectronic tuning and performance mapping of 1-J FASnI\textsubscript{3} cells}
For modeling and optimization of the single junction (1-J) FASnI$_3$ solar cell, we employed two distinct hole transport layers, CuSCN and Spiro-OMeTAD, to investigate their respective impacts on device performance along with other layers.
\subsubsection{Incorporating CuSCN as hole transport layer}
To model and perform opto-electronic simulation for 1-J FASnI\textsubscript{3} cell (MgF\textsubscript{2}/ITO/ZnO/FASnI\textsubscript{3}/CuSCN/ITO), the same optimization procedures described previously were followed. We modeled the initial architecture with MgF\textsubscript{2} (100 nm) as ARC, front ITO (100 nm) as TCO, ZnO (10 nm) as buffer layer, and TiO\textsubscript{2} (90 nm) as ETL with donor doping densities of 1.5$\times$10\textsuperscript{17} and 5$\times$10\textsuperscript{18} cm\textsuperscript{-3}, respectively. FASnI\textsubscript{3} absorber (200 nm) was set to an acceptor doping density of 1$\times$10\textsuperscript{16} cm\textsuperscript{-3}, and CuSCN (50 nm) was used as HTL with an acceptor doping of 1$\times$10\textsuperscript{18} cm\textsuperscript{-3}. The bottom ITO (50 nm) acted as BCL. Preliminary thickness sweeps of the FASnI\textsubscript{3} absorber, TiO\textsubscript{2} ETL, and CuSCN HTL were performed to identify performance-sensitive regions and establish the baseline for subsequent optimization, as shown in Figs.~S7 and S8 of the Supplementary Material.\\

\begin{figure}[!b]
    \centering
    \includegraphics[width=1.0\linewidth]{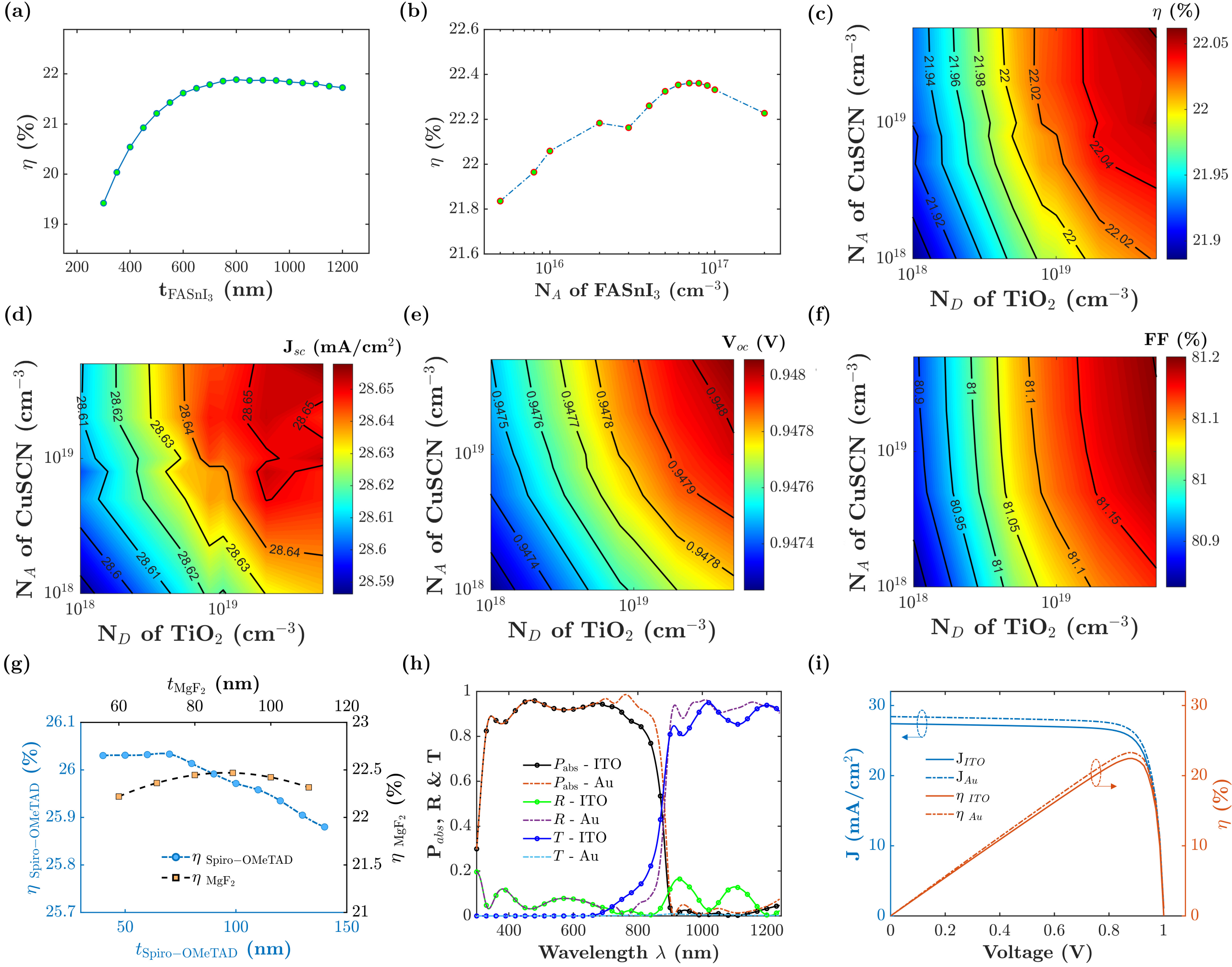}
    \caption{Impact on PCE ($\eta$) through altering (a) FASnI$_3$ absorber thickness and (b) FASnI$_3$ acceptor doping density N$_A$. Contour plot featuring performance matrices (c) $\eta$, (d) J\textsubscript{sc}, (e) V\textsubscript{oc}, and (f) FF corresponding to variation of donor doping density, N$_D$ of TiO\textsubscript{2} ETL and acceptor doping density, N$_A$ of CuSCN HTL for 1-J FASnI$_3$ cell. (g) Impact of MgF\textsubscript{2} ARC and Spiro-OMeTAD thickness on PCE of 1-J FASnI$_3$ cell. (h) Comparison between featuring normalized power absorption, reflection, and transmission corresponding to photon wavelength (i) J-V and $\eta$-V response for 1-J FASnI\textsubscript{3} with ITO and Au configurations.}
    \label{fig:5}
\end{figure}
After identifying the performance-sensitive region, we systematically varied the thicknesses of TiO\textsubscript{2}, FASnI\textsubscript{3}, and CuSCN to achieve self-consistent device performance, as depicted in Figs.~\ref{fig:5}(a) and Figs.~S9(a–d). The absorber thickness was swept in the range of 200–1200 nm, during which the PCE gradually increased and reached a peak value of 21.88\% at a thickness of 800 nm. The improvement with increasing thickness is primarily attributed to enhanced optical absorption and photo-carrier generation within the FASnI\textsubscript{3} layer. Beyond 800 nm, PCE exhibited a slight saturation, reflecting the competing effect of increased recombination losses once the absorber thickness exceeded the carrier diffusion length. Thickness sweeps of the TiO\textsubscript{2} ETL from 20 to 80 nm and CuSCN HTL from 140 to 200 nm yielded nearly invariant PCE, as Fig.~S9 of Supplementary Material, indicating that both transport layers were sufficiently optimized to ensure effective charge extraction without introducing additional series resistance and optical interference. The best PCE of 21.88\% was achieved for a 20 nm TiO\textsubscript{2} and a 160 nm CuSCN layer, confirming the robustness of the device against minor geometrical variations under constant doping conditions.\\

Subsequently, we tuned the doping densities of the ETL and HTL  over a range of 1$\times$10\textsuperscript{18}–5$\times$10\textsuperscript{19} cm\textsuperscript{-3}, as shown in Figs.~\ref{fig:5}(c–f). Enhanced donor and acceptor doping densities improved the built-in potential and reduced contact resistance, resulting in superior carrier selectivity and a higher fill factor. The highest efficiency of 22.05\% was obtained for N\textsubscript{D} of 5$\times$10\textsuperscript{19} cm\textsuperscript{-3} for TiO\textsubscript{2} and N\textsubscript{A} of 5$\times$10\textsuperscript{19} cm\textsuperscript{-3} for CuSCN. However, considering interfacial stability, a moderate acceptor doping of 8$\times$10\textsuperscript{18} cm\textsuperscript{-3} was selected for CuSCN, as it provided nearly identical efficiency with improved long-term reliability. Finally, N\textsubscript{A} of absorber was varied between 10\textsuperscript{15} and 10\textsuperscript{17} cm\textsuperscript{-3}, revealing a clear optimum performance at 7$\times$10\textsuperscript{16} cm\textsuperscript{-3}. In this regime, the enhanced internal electric field strengthened charge separation and reduced interfacial recombination, yielding a peak PCE of 22.36\%.
To minimize optical reflection and maximize photon absorption, MgF\textsubscript{2} ARC thickness was systematically varied from 60 to 110 nm. The optimal PCE of 22.45\% was found at 90 nm, as shown in Fig.~\ref{fig:5}(g), corresponding to the lowest overall reflectance within the device’s active absorption range. Substituting ITO BCL with a 50 nm Au layer further enhanced the PCE to 23.31\%, representing a relative 3.74\% improvement over the ITO-based configuration, as evident from the J–V and $\eta$–V characteristics of Fig.~\ref{fig:5}(i). This enhancement stems from Au’s higher conductivity and favorable energy alignment, which promote efficient carrier extraction and lower interfacial resistance. As shown in Fig.~\ref{fig:5}(h), Au BCL also suppressed reflection and improved near infrared (NIR) absorption in the 700–890 nm wavelength range, increasing J\textsubscript{sc} by 1.01 mA/cm\textsuperscript{2}, as summarized in Table~\ref{Table:3}. These combined optical and electrical enhancements demonstrate that the use of Au as BCL provides a more favorable balance between photon management and charge transport efficiency in the FASnI\textsubscript{3} device.

\subsubsection{Impact of Spiro-OMeTAD for hole extraction}
\begin{figure}[!b]
    \centering
    \includegraphics[width=1.0\linewidth]{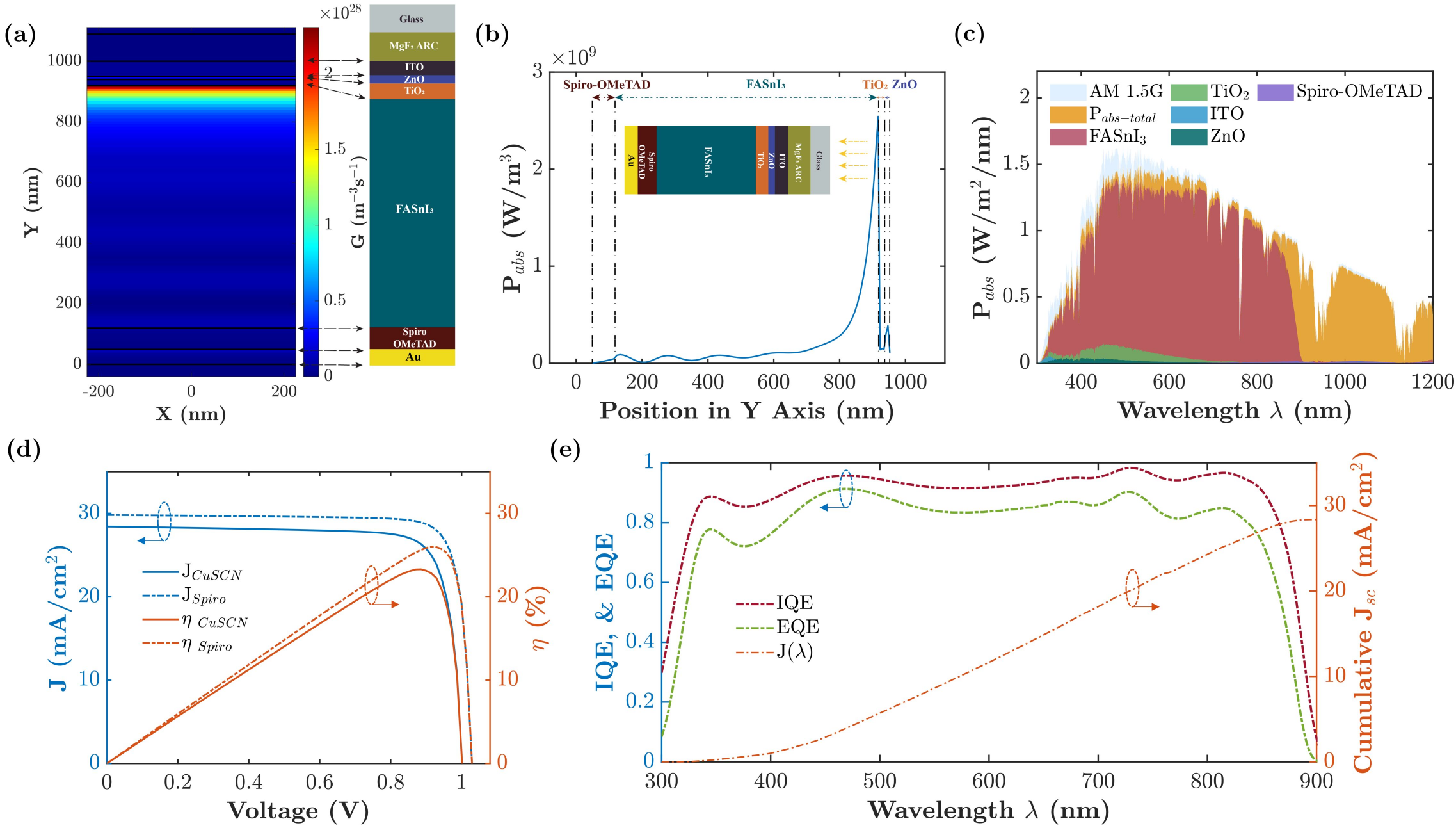}
    \caption{(a) Spatial carrier generation rate profile, G in X-Y cross-section, highlighting the region of maximum generation. (b) Absorbed power, P\textsubscript{abs}, towards the illumination of sunlight in the Y-direction. (c) Spectral power absorption, P\textsubscript{abs} of individual layers in the 1-J FASnI$_3$ solar cell as a function of photon wavelength under the AM1.5G illumination spectrum. (d) Comparison between $J$-$V$ and $\eta$-$V$ curves for incorporating CuSCN and Spiro-OMeTAD as HTL. (e) Response of IQE, EQE, and cumulative J\textsubscript{sc} vs photon wavelength for the optimized 1-J FASnI$_3$ cell.}
    \label{fig:6}
\end{figure}

\begin{table}[!b]
\centering
\caption{Sequential optimization of layer thickness, doping, and interfaces in the 1-J FASnI\textsubscript{3} solar cell.}
\label{Table:3}
\resizebox{\columnwidth}{!}{%
\begin{tabular}{lcccc}
\hline
\textbf{Optimization parameters} &
  \textbf{\begin{tabular}[c]{@{}c@{}}$\eta$\\ (\%)\end{tabular}} &
  \textbf{\begin{tabular}[c]{@{}c@{}}J\textsubscript{sc}\\ (mA/cm\textsuperscript{2})\end{tabular}} &
  \textbf{\begin{tabular}[c]{@{}c@{}}V\textsubscript{oc}\\ (V)\end{tabular}} &
  \textbf{\begin{tabular}[c]{@{}c@{}}FF\\ (\%)\end{tabular}} \\ \hline 

\begin{tabular}[c]{@{}l@{}} Optimized TiO\textsubscript{2}/FASnI\textsubscript{3}/CuSCN thickness \\ (20 nm/800 nm/160 nm)\end{tabular} 
& 21.88 & 28.59 & 0.947 & 80.83 \\ \\



\begin{tabular}[c]{@{}l@{}}Optimized doping in TiO\textsubscript{2}/FASnI\textsubscript{3}/CuSCN \\ 
(N\textsubscript{D,TiO\textsubscript{2}}: 5$\times$10\textsuperscript{19} cm\textsuperscript{-3}, 
N\textsubscript{A,FASnI\textsubscript{3}}: 7$\times$10\textsuperscript{16} cm\textsuperscript{-3},\\ 
N\textsubscript{A,CuSCN}: 8$\times$10\textsuperscript{18} cm\textsuperscript{-3})\end{tabular} 
& 22.36 & 27.28 & 1.001 & 81.89 \\ \\
\begin{tabular}[c]{@{}l@{}}Optimized MgF\textsubscript{2} ARC thickness\\ (90 nm)\end{tabular} 
& 22.47 & 27.40 & 1.001 & 81.92 \\ \\

\begin{tabular}[c]{@{}l@{}}Substituting ITO BCL with Au layer\\ (50 nm)\end{tabular} 
& 23.31 & 28.42 & 1.002 & 81.90 \\ \\

\begin{tabular}[c]{@{}l@{}}Integrated Spiro-OMeTAD as HTL\\ (70 nm)\end{tabular} 
& 26.03 & 29.80 & 1.029 & 84.84 \\ \hline
\end{tabular}%
}
\end{table}

To enhance the effective hole transportation in the FASnI\textsubscript{3}/HTL interface, we incorporated Spiro-OMeTAD instead of CuSCN as HTL in 1-J FASnI\textsubscript{3} cell. The thickness of Spiro-OMeTAD was varied from 40 to 140 nm under a constant acceptor doping density of 8$\times$10\textsuperscript{18} cm\textsuperscript{-3}. The PCE showed a steady increase with Spiro-OMeTAD thickness, reaching a maximum of 26.03\% at 70~nm. Beyond this point, additional thickness led to a decline in performance due to an increase in carrier recombination, as shown in Fig. \ref{fig:5}(g). 
Fig.~S10 of the Supplementary Material presents the band diagrams of the 1-J FASnI\textsubscript{3} cell at short-circuit (V = 0~V) and 
maximum power point (V = V\textsubscript{mpp}) conditions for both CuSCN and Spiro-OMeTAD configurations.  Analyzing Figs. S10(a) and (c), it is evident that the FASnI\textsubscript{3}/Spiro-OMeTAD interface exhibits more favorable energy level alignment compared to the FASnI\textsubscript{3}/CuSCN interface. Specifically, Spiro-OMeTAD introduces a slight negative valence band offset (VBO) $\approx$ -0.06~eV, forming a downhill potential that facilitates hole drift from the absorber to the HTL, while maintaining a sufficient conduction band offset (CBO) $\approx$ 1.56~eV, to block electron back-injection. In contrast, CuSCN provides a nearly flat VBO $\approx$ 0~eV, offering minimal driving force for hole transport despite its higher CBO $\approx$ 2.0~eV, which restricts charge extraction efficiency.\\

Furthermore, the difference in layer thickness amplifies the effect of carrier drift and diffusion. The thinner 70~nm Spiro-OMeTAD layer supports a stronger build-in electric field across the FASnI$_3$/HTL junction, promoting efficient hole drift while minimizing diffusion-induced recombination. Conversely, the thicker 160~nm CuSCN layer imposes higher series resistance and longer diffusion paths, 
causing significant carrier loss and reduced hole mobility. 
The combined influence of favorable band alignment, stronger built-in potential, and reduced transport resistance in Spiro-OMeTAD leads to 
more pronounced band bending and quasi-Fermi level splitting, as evident in Figs.~S10 (c) and (d).\\ 

Table \ref{Table:3} shows the gradual improvement of the performance of the 1-J FASnI$_3$ cell. Incorporation of Spiro-OMeTAD with FASnI$_3$ enhanced the PCE by 2.72\%. Therefore, we considered  FASnI$_3$ with Spiro-OMeTAD configuration as the final optimized 1-J FASnI$_3$ cell. The spatial generation rate profile G, highlighting the region of maximum carrier generation in Fig.~\ref{fig:6}(a). In Fig.\ref{fig:6}(b) features the absorbed power density, P\textsubscript{abs} in different layers of 1-J FASnI\textsubscript{3}. Spectral power absorption in different layers corresponding to AM 1.5G solar spectrum, comparison between $\eta$-V and J-V for CuSCN and Spiro-OMeTAD as HTL, and normalized IQE, EQE with cumulative J\textsubscript{sc} with respect to photon wavelength of optimized 1-J FASnI\textsubscript{3} with Spiro-OMeTAD configuration are illustrated in Figs.~\ref{fig:6}(c-e). Together, these outcomes offer a detailed understanding of the optical and electrical behavior of the optimized 1-J FASnI$_3$ cell.
\subsection{Tandem integration and current matching in 2-J KSnI\textsubscript{3}/FASnI\textsubscript{3} cells}
\begin{figure}[!t]
    \centering
    \includegraphics[width=1.0\linewidth]{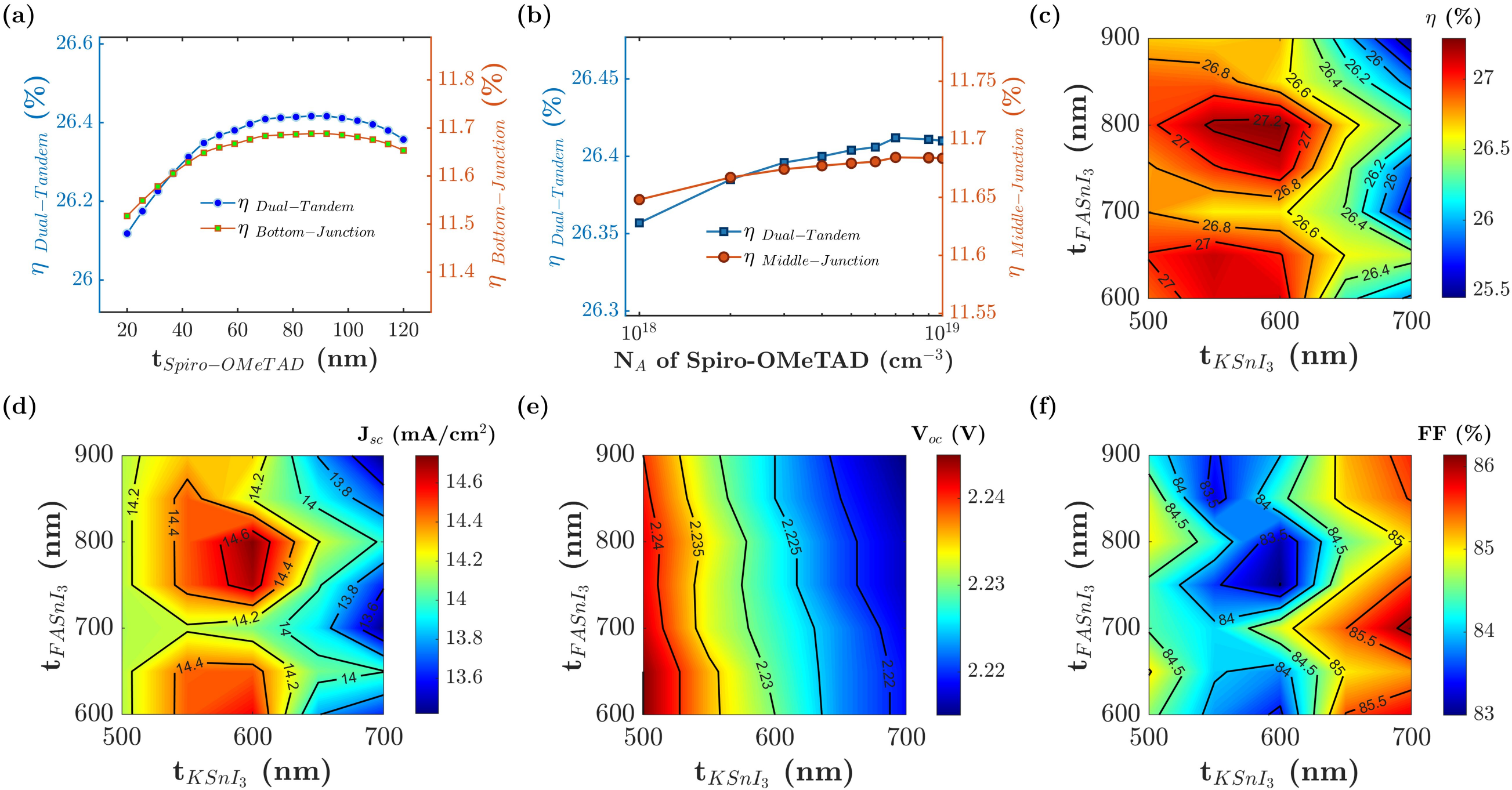}
    \caption{Impact on PCE ($\eta$) of 2-J KSnI\textsubscript{3}/FASnI\textsubscript{3} tandem cell by (a) altering Spiro-OMeTAD HTL (bottom cell) layer's thickness and (b) acceptor doping density. Contour plot of parametric sweep featuring performance metrics -- (c) $\eta$, (d) J\textsubscript{sc}, (e) V\textsubscript{oc}, and (f) FF corresponding to variation of TiO\textsubscript{2} ETL and CuSCN HTL thickness.}
    \label{fig:7}
\end{figure}
To model and optimize the dual-junction monolithic tandem solar cell, we constructed the top subcell as MgF\textsubscript{2}/FTO/TiO\textsubscript{2}/KSnI\textsubscript{3}/CuSCN/ITO, where ITO served as the ITCO and recombination layer, and the bottom subcell as ZnO/TiO\textsubscript{2}/FASnI\textsubscript{3}/Spiro-OMeTAD/Au. Initially, we fixed the top subcell absorber KSnI\textsubscript{3} thickness at 700~nm, along with a 40~nm ETL and 100~nm HTL, while the ITO recombination layer was 50~nm thick. The doping profile of the top subcell followed the optimized parameters from the single-junction KSnI\textsubscript{3} cell. For the bottom subcell in the 2-J tandem configuration, we varied the Spiro-OMeTAD thickness and acceptor doping density, as shown in Figs.~\ref{fig:7}(a) and (b). The highest PCE of 26.415\% was achieved for a Spiro-OMeTAD thickness of 80~nm and an acceptor density N\textsubscript{A} of 7$\times$10\textsuperscript{18}~cm\textsuperscript{-3}, demonstrating the crucial role of hole transport layer optimization in facilitating efficient carrier extraction via drift and diffusion mechanisms.\\
\begin{table}[!b]
\centering
\caption{Summary of performance metrics of the 2-J KSnI\textsubscript{3}/FASnI$_3$ tandem cell as a function of thickness and doping density variations in the bottom subcell.}
\label{Table:4}
\resizebox{\columnwidth}{!}{%
\begin{tabular}{lccccccc}
\hline
\textbf{Solar cell} & \textbf{Parameters} &
  \textbf{\begin{tabular}[c]{@{}c@{}}$\eta$\\ (\%)\end{tabular}} &
  \textbf{\begin{tabular}[c]{@{}c@{}}J\textsubscript{sc}\\ (mA/cm\textsuperscript{2})\end{tabular}} &
  \textbf{\begin{tabular}[c]{@{}c@{}}V\textsubscript{oc}\\ (V)\end{tabular}} &
  \textbf{\begin{tabular}[c]{@{}c@{}}FF\\ (\%)\end{tabular}} &
  \textbf{\begin{tabular}[c]{@{}c@{}}P\textsubscript{mpp}\\ (W/m\textsuperscript{2})\end{tabular}} &
  \textbf{\begin{tabular}[c]{@{}c@{}}V\textsubscript{mpp}\\ (V)\end{tabular}} \\ \hline \\
1-J FASnI\textsubscript{3} cell & \begin{tabular}[c]{@{}c@{}} Spiro-OMeTAD\\ (80 nm and 7$\times$10\textsuperscript{18} cm\textsuperscript{-3}) \end{tabular}&
  26.03 &
  29.81 &
  1.029 &
  84.84 &
  260.33 &
  0.96 \\ \\
2-J tandem cell & \begin{tabular}[c]{@{}c@{}} Spiro-OMeTAD\\ (Thickness: 80 nm)\end{tabular} &
  26.41 &
  13.98 &
  2.217 &
  85.26 &
  264.14 &
  1.994 \\ \\
2-J tandem cell  & \begin{tabular}[c]{@{}c@{}}Spiro-OMeTAD\\(Doping: 7$\times$10\textsuperscript{18} $cm^{-3}$)\end{tabular} &
  26.41 &
  13.98 &
  2.217 &
  85.24 &
  264.14 &
  1.994 \\ \\
2-J tandem cell & \begin{tabular}[c]{@{}c@{}}KSnI\textsubscript{3} and FASnI\textsubscript{3} \\ (Thickness: 600~nm and 800~nm) \end{tabular} & 
27.29 & 14.74 & 2.227 & 83.14& 272.89 & 1.981 \\ \\
\hline
\end{tabular}%
}
\end{table}
To further maximize the performance, we systematically varied the absorber thickness of the top and bottom subcells within 500–700~nm and 600–900~nm ranges, respectively, as featured in Figs.~\ref{fig:7}(c–f). Observing Figs.~\ref{fig:7}(c) and (d), both $\eta$ and J\textsubscript{sc} attain higher values for KSnI\textsubscript{3} thicknesses of 550–600~nm and FASnI$_3$ thicknesses of 700–850~nm, indicating minimal current mismatch and efficient electron-hole transport across the ITCO/recombination interface. This behavior is consistent with reduced bulk and interfacial recombination due to optimal drift-driven carrier extraction in the built-in electric fields of each subcell. Although V\textsubscript{oc} is higher for KSnI\textsubscript{3} thicknesses of 500–550~nm and FASnI$_3$ thicknesses of 600–800~nm, in the two-terminal (2-T) monolithic tandem configuration, the overall V\textsubscript{oc} is predominantly governed by the bottom FASnI$_3$ subcell. Increasing the KSnI\textsubscript{3} thickness beyond this range leads to a gradual decrease in V\textsubscript{oc}, attributable to enhanced bulk recombination and series resistance effects. Meanwhile, FF increases with thicker KSnI\textsubscript{3} and FASnI\textsubscript{3} layers, as seen in Fig.~\ref{fig:7}(f). The optimized 2-J KSnI\textsubscript{3}/FASnI\textsubscript{3} tandem cell achieves a PCE of 27.29\%, with J\textsubscript{sc} of 14.74~mA/cm\textsuperscript{2}, V\textsubscript{oc} of 2.227~V, and FF of 83.14\%.\\

\begin{figure}[!b]
    \centering
    \includegraphics[width=1.0\linewidth]{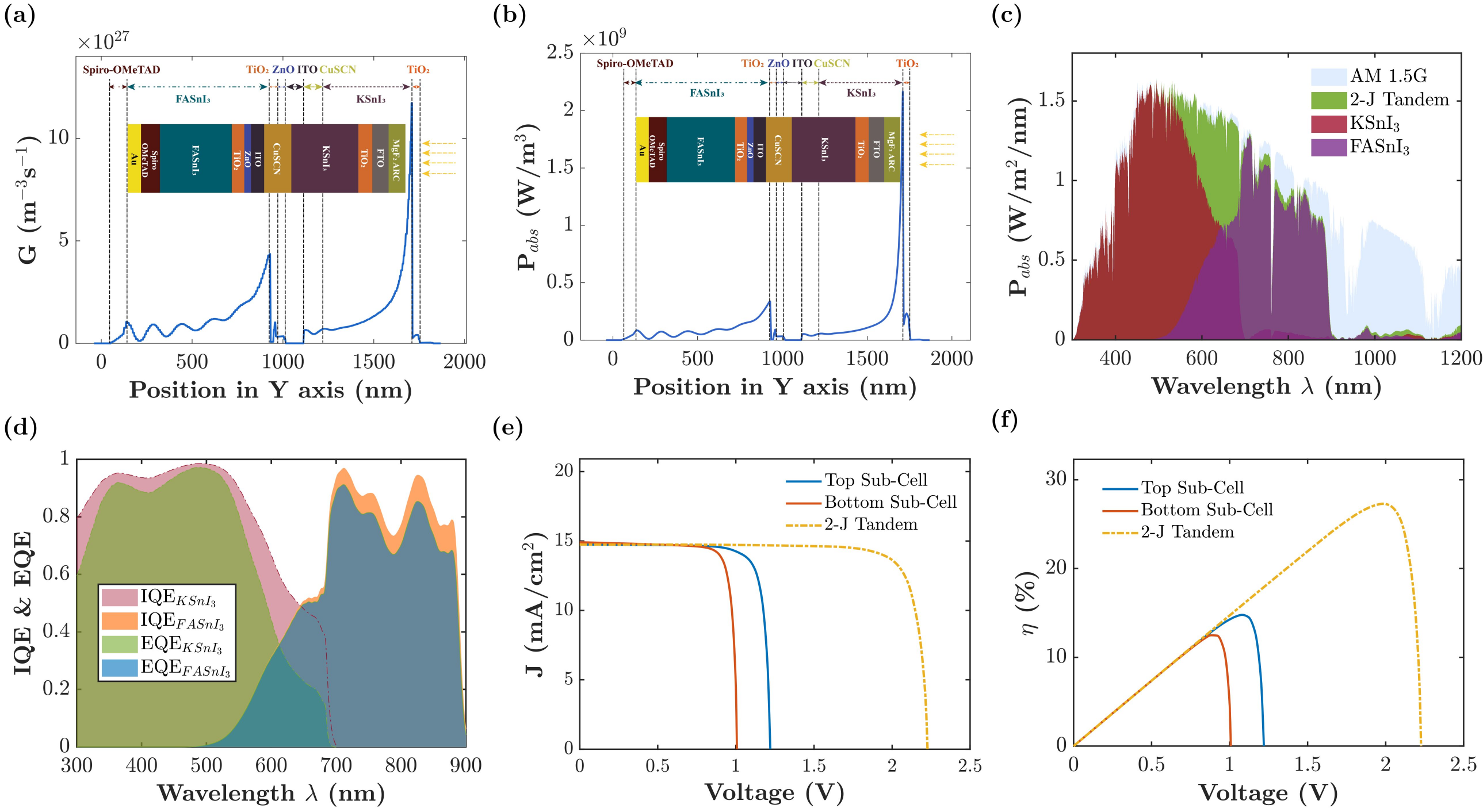}
    \caption{(a) Carrier generation rate, G, towards the illumination of sunlight in the Y direction, highlighting the maximum carrier generation regions. (b) Absorbed power density, P\textsubscript{abs}, of different layers in a 2-J tandem cell towards the illumination of sunlight in the Y direction. (c)  Spectral power absorption of the absorber layers of a 2-J tandem cell as a function of photon wavelength corresponding to the AM 1.5G spectrum. (d) IQE and EQE vs photon wavelength, (e) J-V, and (f) $\eta$-V plots of top, and bottom subcells, along with the optimized 2-T KSnI\textsubscript{3}/FASnI\textsubscript{3} based tandem cell.}
    \label{fig:8}
\end{figure}
Table~\ref{Table:4} presents the progressive enhancement in the performance of the two-terminal (2-T) monolithic tandem solar cell, achieved through systematic optimization of the absorber layer thicknesses in each subcell. Fig.~\ref{fig:8}(a-f) provides a detailed optoelectronic analysis of the optimized tandem cell. Figs.~\ref{fig:8}(a-b) feature the carrier generation rate, highlighting the maximum generation regions, and absorbed power density along the device thickness under the illumination of sunlight in the Y direction. Fig.~\ref{fig:8}(c) compares the spectral absorbed power of the individual subcells and the complete 2-J tandem cell with the AM 1.5G solar spectrum. 
The complementary absorption of KSnI\textsubscript{3} and FASnI\textsubscript{3} enables efficient spectrum splitting from 300 to 900 nm, maximizing photon utilization and minimizing optical losses in the tandem stack.
Fig.~\ref{fig:8}(d) presents the normalized IQE and EQE vs. photon wavelength of both subcells. The broad and overlapping IQE–EQE responses confirm strong carrier generation and collection across the respective absorption regions, with smooth spectral transitions indicating balanced current matching between the top and bottom subcells as depicted in Fig.~\ref{fig:8}(e). These results collectively validate efficient optical coupling and carrier extraction within the optimized 2-J tandem architecture. Additionally, Fig.~\ref{fig:8}(f) includes $\eta$-V characteristics of the top and bottom subcell following the optimized 2-J tandem cell. These analyses collectively elucidate the interplay of optical absorption, drift-diffusion-driven carrier transport, and interfacial recombination, providing a detailed understanding of the superior performance of the optimized monolithic 2-J tandem architecture.
\subsection{Triple-junction architecture: layer-wise optimization and device efficiency}
Before building the triple-junction KSnI\textsubscript{3}/FASnI\textsubscript{3}/ACZTSe tandem cell, we introduced ITO as the ITCO and BCL in the middle FASnI\textsubscript{3} subcell to facilitate the efficient passage of excess low-energy photon injection to the bottom cell. This modification improved optical transmission across the 850–-1240 nm wavelength range, as illustrated in Fig.~S11 of the Supplementary Material. However, an initial drop in performance parameters occurred due to reduced optical back reflection and photon reabsorption, as summarized in Table~S3.\\

\begin{figure}[!t]
    \centering
    \includegraphics[width=1.0\linewidth]{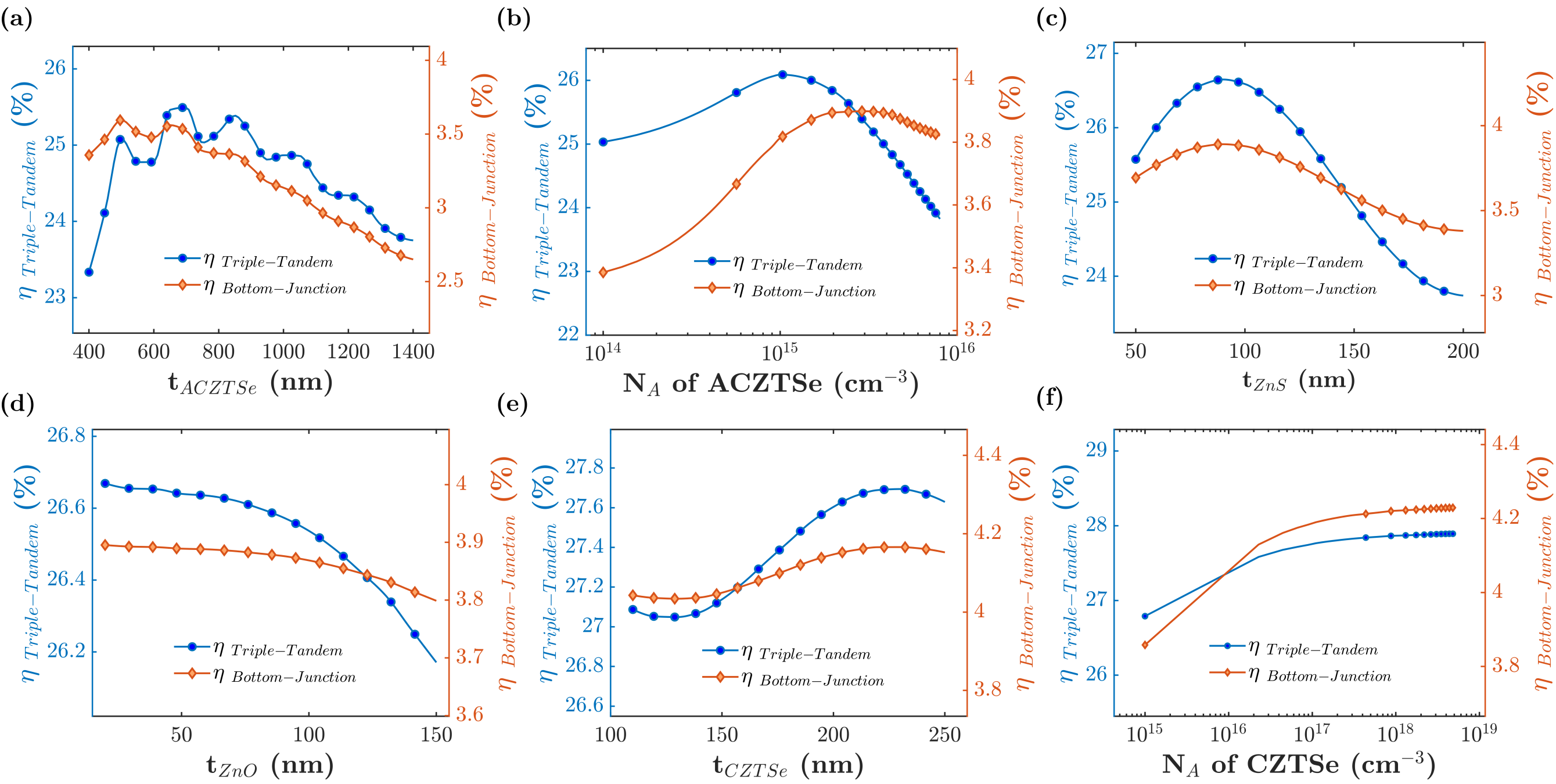}
    \caption{Effect of thickness and doping density variations in the key layers of the ACZTSe bottom subcell on PCE ($\eta$) of a monolithic triple-junction tandem solar cell. The analysis highlights both the overall tandem cell performance and the specific contributions of the bottom subcell. The parameters investigated include: (a) ACZTSe absorber layer thickness, (b) ACZTSe acceptor doping density, (c) ZnS buffer layer thickness, (d) ZnO window layer thickness, (e) CZTSe back surface field (BSF) layer thickness, and (f) CZTSe acceptor doping density.}
    \label{fig:9}
\end{figure}

Next, we modeled the third subcell as follows: AZO/ZnO/ZnS/ACZTSe/\\CZTSe/Mo.
For optimization purposes, the initial thickness and doping density were taken as shown in Table~S4. At first, the ACZTSe absorber thickness was varied from 400 to 1400~nm, under constant acceptor doping. The results showed a peak PCE of 25.77\% for a 750 nm thickness of ACZTSe as Fig.~\ref{fig:9}(a). 
Subsequently, N\textsubscript{A} of ACZTSe absorber layer was varied from 1$\times$10\textsuperscript{14} to 1$\times$10\textsuperscript{15} cm\textsuperscript{-3}. The maximum PCE of 26.08\% was achieved at N\textsubscript{A} of 1$\times$10\textsuperscript{15} cm\textsuperscript{-3}, as shown in Fig.~\ref{fig:9}(b). However, beyond a N\textsubscript{A} value of 9$\times$10\textsuperscript{14} cm\textsuperscript{-3}, a declining trend in
J\textsubscript{sc} was noticed from Table~S5 of Supplementary Material, indicating increased electron–hole recombination at the ACZTSe/CZTSe junction. Therefore, N$_{A}$ of 9$\times$10\textsuperscript{14} cm$^{-3}$ was selected for further optimization. Afterwards, n-doped ZnS and n-doped ZnO layer thicknesses were varied as shown in Figs.~\ref{fig:9}(c) and ~\ref{fig:9}(d).
For a ZnS buffer layer thickness of 90 nm and a ZnO layer thickness of 20 nm, the highest PCE of 26.67\% was achieved. Subsequently, to evaluate the influence of the CZTSe back surface field (BSF) layer, its thickness was varied from 110 to 250 nm, as shown in Fig.~\ref{fig:9}(e). The results indicate a rising trend in PCE, increasing from 27.1\% and peaking at 27.69\% for a BSF thickness of 220 nm. This improvement suggests enhanced lateral absorption in the ACZTSe layer, contributing to better device performance. N$_{A}$ of the ACZTSe layer was varied from  1$\times$10\textsuperscript{16} to 1$\times$10\textsuperscript{17} cm\textsuperscript{-3}. As illustrated in Fig.~\ref{fig:9}(f), the PCE exhibits a steady increase from 1$\times$10\textsuperscript{16} to 1$\times$10\textsuperscript{17} $cm^{-3}$ and continues to rise with higher doping densities. However, Li \textit{et al.} reported that increasing the N$_{A}$ of CZTSe from 10\textsuperscript{15} to 10\textsuperscript{18} cm\textsuperscript{-3} significantly reduced the trap-assisted carrier lifetime from 5 ns to 13 ps, indicating higher recombination losses ~\cite{li2020effects}. Therefore, N\textsubscript{A} of 6$\times$10\textsuperscript{16} cm\textsuperscript{-3} was selected as a trade-off, yielding a PCE of 27.71\%, with J\textsubscript{sc} of 12.11 mA/cm\textsuperscript{2}, V\textsubscript{oc} of 2.726 V, and FF of 83.98\% for the 3-J tandem cell.

\begin{figure}[!t]
    \centering
    \includegraphics[width=0.9\linewidth]{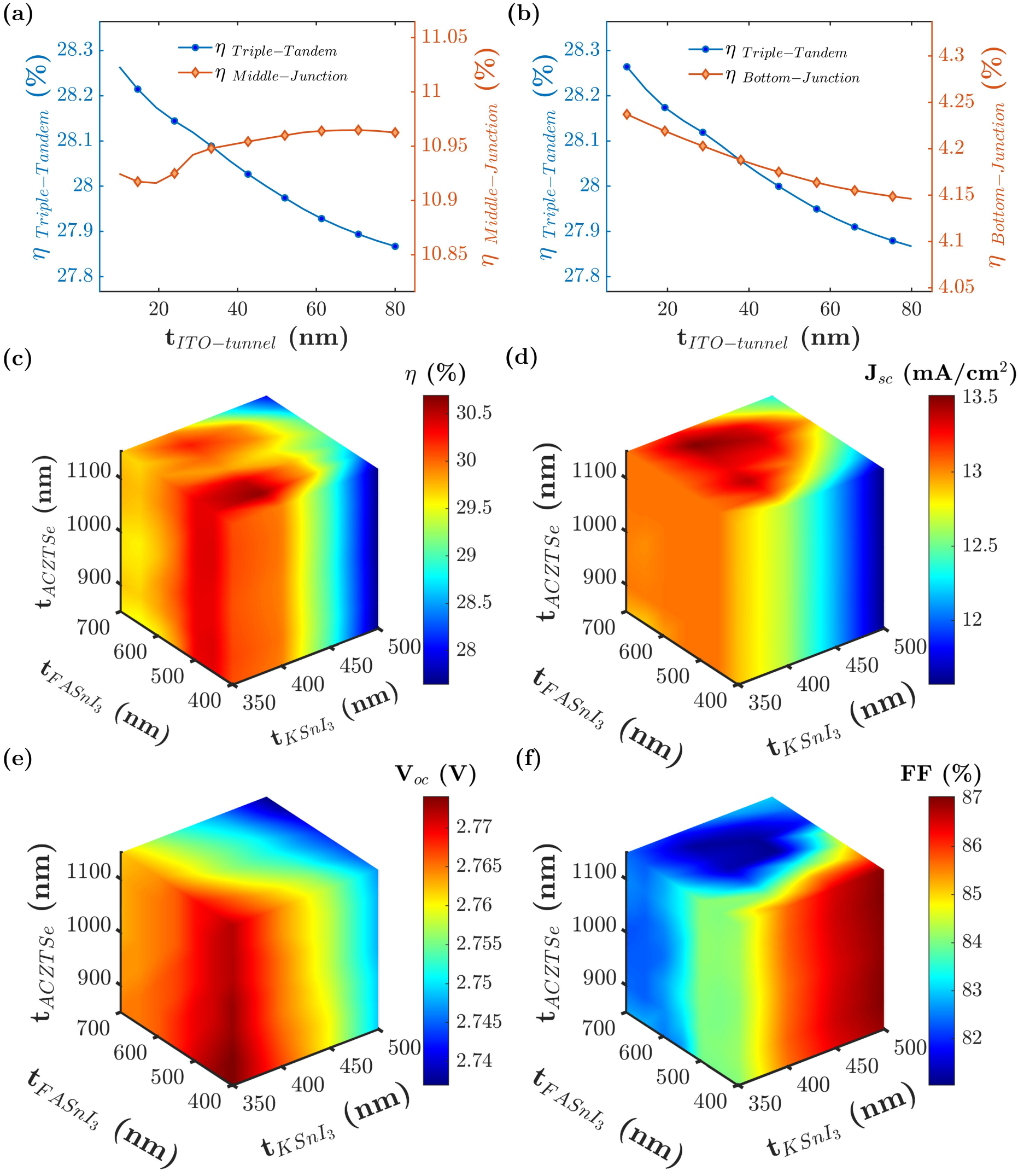}
    \caption{Influence of ITO tunnel junction thickness on PCE ($\eta$) of the device featuring (a) overall tandem PCE and middle junction, and (b) overall tandem PCE and bottom junction. Volumetric contour plot of (c) $\eta$, (d) J\textsubscript{sc}, (e) V\textsubscript{oc}, and (f) FF featuring the performance of triple junction tandem cell corresponding to various thickness ranges of KSnI\textsubscript{3}, FASnI$_3$, and ACZTSe.}
    \label{fig:10}
\end{figure}

Afterward, the thickness of the ITO tunnel layer between the middle and bottom subcell was varied as shown in Figs.~\ref{fig:10}(a) and (b). As the ITO thickness increases, enhanced reflection of the overall tandem cell and carrier recombination are observed. This phenomenon led to a decrease in the PCE of both the middle and bottom subcells, along with the overall 3-J tandem cell. Based on this analysis, a 20 nm ITO thickness was selected as an optimal trade-off, balancing device stability with efficient charge transport.\\

To determine the overall optimal performance of the 3-J tandem cell, the thicknesses of the primary absorber layers in each subcell were systematically varied, while maintaining constant thicknesses for the transport and recombination layers. Specifically, KSnI\textsubscript{3} thickness was varied from 350 to 550 nm, FASnI\textsubscript{3} from 400 to 700 nm, and ACZTSe from 800 to 1150 nm. Figs. \ref{fig:10}(c-f) present the performance metrics corresponding to the volumetric parametric thickness sweep for the KSnI\textsubscript{3}, FASnI\textsubscript{3}, and ACZTSe.
As shown in Fig.~\ref{fig:10}(c), the PCE of the 3-J tandem cell was highly sensitive to the thickness of the top subcell absorber KSnI\textsubscript{3} (350–450 nm) and the middle subcell absorber FASnI\textsubscript{3} (400–550 nm), particularly when the bottom subcell absorber ACZTSe had a thickness in the range of 850–-1100 nm. This behavior can be attributed to the monolithic 2-T configuration of the tandem structure, which requires a uniform J\textsubscript{sc} across all subcells. Any mismatch in the photocurrent among the subcells leads to increased internal carrier recombination, thereby reducing the overall device efficiency. For KSnI\textsubscript{3} thickness ranging from
\begin{table}[!b]
\centering
\caption{Summary of performance metrics of the 3-J tandem cell as a function of thickness and doping density variations in the bottom subcell.}
\label{tab:com_3-J_tandem_bottom_cell}
\resizebox{\columnwidth}{!}{%
\begin{tabular}{lcccccc}
\hline
\textbf{Optimization Parameters} &
  \textbf{\begin{tabular}[c]{@{}c@{}}$\eta$\\ (\%)\end{tabular}} &
  \textbf{\begin{tabular}[c]{@{}c@{}}J\textsubscript{sc}\\ (mA/cm\textsuperscript{2})\end{tabular}} &
  \textbf{\begin{tabular}[c]{@{}c@{}}V\textsubscript{oc}\\ (V)\end{tabular}} &
  \textbf{\begin{tabular}[c]{@{}c@{}}FF\\ (\%)\end{tabular}} &
  \textbf{\begin{tabular}[c]{@{}c@{}}P\textsubscript{mpp} \\ (W/m\textsuperscript{2})\end{tabular}} &
  \textbf{\begin{tabular}[c]{@{}c@{}}V\textsubscript{mpp} \\ (V)\end{tabular}} \\ \hline \\
\begin{tabular}[c]{@{}l@{}}ACZTSe thickness\\ (750 nm)\end{tabular} &
  25.77 &
  11.74 &
  2.695 &
  81.46 &
  257.68 &
  2.343 \\ \\
\begin{tabular}[c]{@{}l@{}}ACZTSe doping\\ (9.00$\times$10\textsuperscript{14} cm\textsuperscript{-3})\end{tabular} &
  26.06 &
  11.46 &
  2.714 &
  83.77 &
  260.60 &
  2.414 \\ \\
\begin{tabular}[c]{@{}l@{}}ZnS thickness\\ (90 nm)\end{tabular} &
  26.64 &
  11.78 &
  2.716 &
  83.30 &
  266.45 &
  2.402 \\ \\
\begin{tabular}[c]{@{}l@{}}ZnO thickness\\ (20 nm)\end{tabular} &
  26.67 &
  11.79 &
  2.716 &
  83.29 &
  266.69 &
  2.402 \\ \\
\begin{tabular}[c]{@{}l@{}}CZTSe Thickness\\ (220 nm)\end{tabular} &
  27.69 &
  12.10 &
  2.725 &
  83.95 &
  276.89 &
  2.410 \\ \\
\begin{tabular}[c]{@{}l@{}}CZTSe doping\\ (6.00$\times$10\textsuperscript{16} cm\textsuperscript{-3})\end{tabular} &
  27.71 &
  12.11 &
  2.726 &
  83.98 &
  277.11 &
  2.411 \\ \\
\begin{tabular}[c]{@{}l@{}}Tandem absorbers thickness\\ KSnI\textsubscript{3}/FASnI\textsubscript{3}/ACZTSe\\ (400 nm/450 nm/900 nm)\end{tabular} &
  30.69 &
  13.18 &
  2.766 &
  84.18 &
  30.692 &
  2.432 \\ \\ \hline
\end{tabular}%
}
\end{table}
400 to 450 nm and FASnI\textsubscript{3} thickness ranging from 550 to 650 nm, elevated J\textsubscript{sc} values are observed, peaking at 13.51 mA/cm\textsuperscript{2} for thicknesses of 400 nm of KSnI\textsubscript{3}, 650 nm of FASnI\textsubscript{3}, and 1100 nm of ACZTSe, as depicted in Fig.~\ref{fig:10}(d). However, V\textsubscript{oc} was observed to be higher for thinner regions of KSnI\textsubscript{3} and FASnI\textsubscript{3} as shown in Fig.~\ref{fig:10}(e). 
This trend arises from increased carrier recombination associated with a larger absorber thickness, which in turn elevates the dark saturation current density (J$_{0}$) and, consequently, reduces V\textsubscript{oc}.
The FF shows a consistent increase with ACZTSe thickness, likely due to enhanced carrier collection and reduced recombination. For KSnI\textsubscript{3}, FF improved as thickness increased, providing better charge transport and current matching. However, FASnI\textsubscript{3} demonstrated an optimal FF at an intermediate thickness, beyond which resistive losses became dominant. This consistent behaviour underscores the crucial role of KSnI\textsubscript{3} in maintaining a high FF as illustrated in Fig.~\ref{fig:10}(f). An optimal PCE of 30.69\% was achieved with J\textsubscript{sc} of 13.184 mA/cm$^2$, V\textsubscript{oc} of 2.766 V, and FF of 84.18\%, corresponding to absorber thicknesses of 400 nm for KSnI\textsubscript{3}, 450 nm for FASnI\textsubscript{3}, and 900 nm for ACZTSe. A 2D contour plot, as shown in Figs.~S12(a-d), along with the data in Table~S6, supports these findings. A summary of the progressive enhancement in performance metrics of the ACZTSe subcell within the 3-J tandem configuration, as a result of volumetric absorber thickness optimization across all subcells, is presented in Table~\ref{tab:com_3-J_tandem_bottom_cell}.\\

\begin{figure}[!htb]
    \centering
    \includegraphics[width=1.0\linewidth]{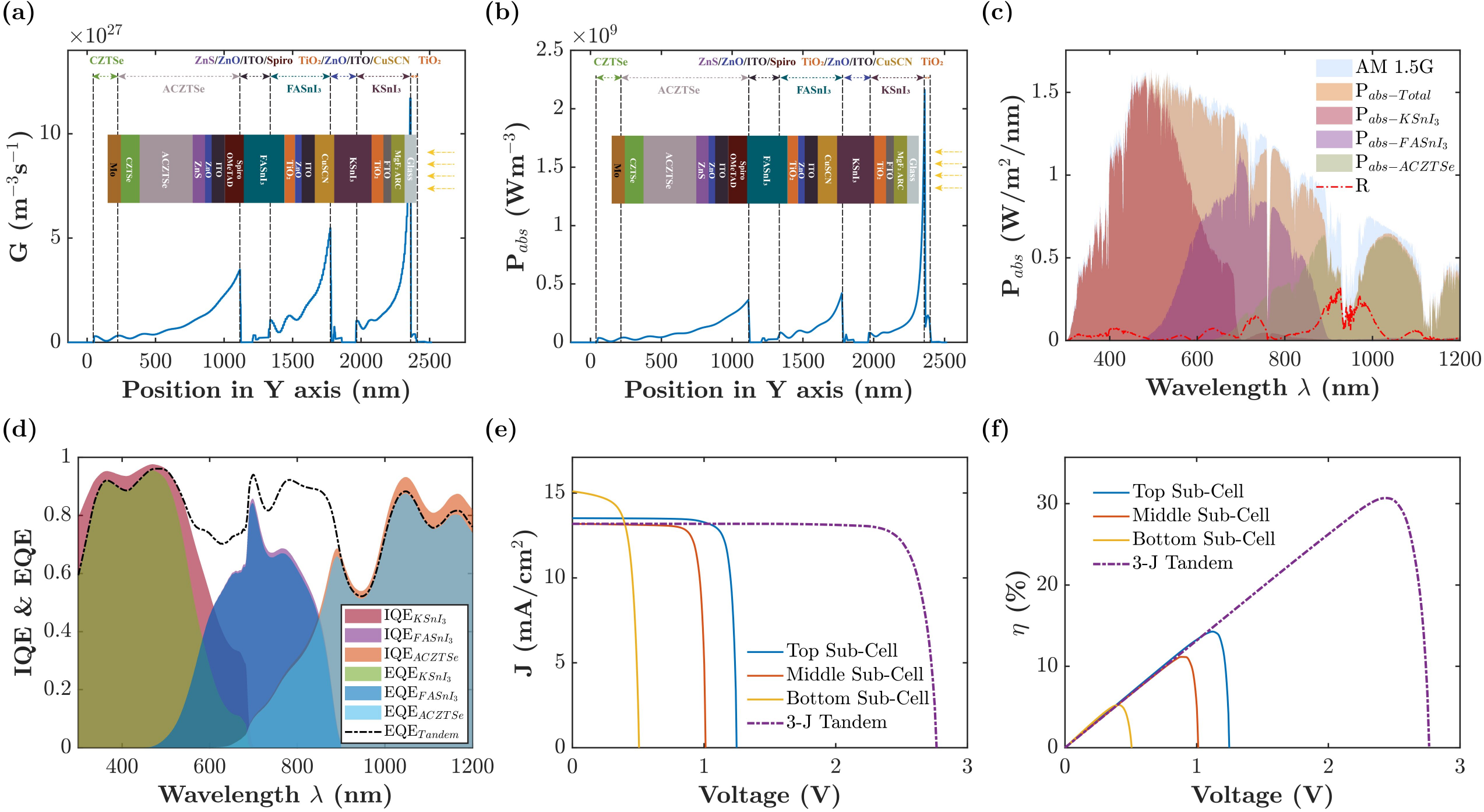}
    \caption{(a) Carrier generation rate, G as a function of depth (Y-axis), under solar illumination, highlighting the maximum generation regions. (b) Absorbed power density, $P_{abs}$, along the Y-direction, for different layers in 3-J tandem cell. (c) Normalized optical power absorption, (d) spectral optical power absorption across the primary absorber layers of the 3-J tandem cell as a function of photon wavelength corresponding to AM 1.5G solar spectrum, (e) IQE, and EQE versus photon wavelength $\lambda$, (e) J–V, and (f) $\eta$–V curve of the optimized 2-T KSnI\textsubscript{3}/FASnI\textsubscript{3}/ACZTSe-based tandem cell and its different subcells.}
    \label{fig:11}
\end{figure}
Fig.~\ref{fig:11} presents a comprehensive opto-electronic analysis of the optimized 2-T KSnI\textsubscript{3}/FASnI$_3$/ACZTSe-based tandem solar cell. Fig.~\ref{fig:11}(a) showcases the generation rate profile, G, highlighting the carrier generation regions as a function of depth (Y-axis) under solar illumination. Fig.~\ref{fig:11}(b) highlights the absorbed power density, P$_{abs}$, in different layers of the tandem cell, along the Y-direction. Fig.~\ref{fig:11}(c) compares the wavelength-resolved spectral absorbed power, P$_{abs}$, of the optimized triple-junction tandem with the AM 1.5G spectrum. The distinct absorption regions of each subcell demonstrate effective spectral splitting from 300–1200 nm, while the low reflectance verifies strong optical confinement and minimal photon loss. The high-energy photons are predominantly absorbed in the KSnI\textsubscript{3} top cell, intermediate photons in FASnI\textsubscript{3}, and near-infrared photons in the ACZTSe bottom cell. Fig.~\ref{fig:11}(d) presents the corresponding EQE and IQE spectra, showing high carrier collection efficiency within each subcell’s absorption range. The smooth spectral transitions and broad EQE response confirm efficient current matching and complementary photon harvesting across the solar spectrum.
Finally, Figs.~\ref{fig:11}(e–f) represent the J–V and corresponding $\eta$–V characteristics of the optimized 3-J tandem solar cell. The individual subcells exhibit distinct V\textsubscript{oc} determined by their bandgaps, while the combined 3-J tandem curve demonstrates series-connected behavior with the total current limited by the lowest photocurrent. The close current matching among subcells confirms efficient carrier extraction, minimal interfacial recombination, and balanced photo-response across the device stack.\\

\begin{table}[!b]
\centering
\caption{Summary of the performance matrices from single junction cells to triple junction tandem cells.}
\label{Table:6}
\resizebox{\columnwidth}{!}{%
\begin{tabular}{lcccccc}
\hline
\textbf{Solar cell types} &
  \textbf{\begin{tabular}[c]{@{}c@{}}$\eta$\\ (\%)\end{tabular}} &
  \textbf{\begin{tabular}[c]{@{}c@{}}J\textsubscript{sc}\\ (mA/cm\textsuperscript{2})\end{tabular}} &
  \textbf{\begin{tabular}[c]{@{}c@{}}V\textsubscript{oc}\\ (V)\end{tabular}} &
  \textbf{\begin{tabular}[c]{@{}c@{}}FF\\ (\%)\end{tabular}} &
  \textbf{\begin{tabular}[c]{@{}c@{}}P\textsubscript{mpp}\\ (mW/cm\textsuperscript{2})\end{tabular}} &
  \textbf{\begin{tabular}[c]{@{}c@{}}V\textsubscript{mpp}\\ (V)\end{tabular}} \\ \hline \\
\begin{tabular}[c]{@{}l@{}} 1-J KSnI\textsubscript{3} Cell\end{tabular}                       & 16.28 & 16.63 & 1.232 & 79.50 & 16.28 & 1.08 \\ \\
\begin{tabular}[c]{@{}l@{}} 1-J FASnI\textsubscript{3} Cell\end{tabular}                      & 26.03 & 29.81 & 1.030 & 84.84 & 26.03 & 0.92 \\ \\
\begin{tabular}[c]{@{}l@{}}2-J tandem cell\\ (KSnI\textsubscript{3}/FASnI\textsubscript{3})\end{tabular}        & 27.29 & 14.74 & 2.227 & 83.14 & 27.29 & 1.98 \\ \\
\begin{tabular}[c]{@{}l@{}}3-J tandem cell\\ (KSnI\textsubscript{3}/FASnI\textsubscript{3}/ACZTSe)\end{tabular} & 30.69 & 13.18 & 2.766 & 84.18 & 30.69 & 2.43 \\ \\ \hline
\end{tabular}%
}
\end{table}

\begin{figure}[!t]
    \centering
    \includegraphics[width=1.0\linewidth]{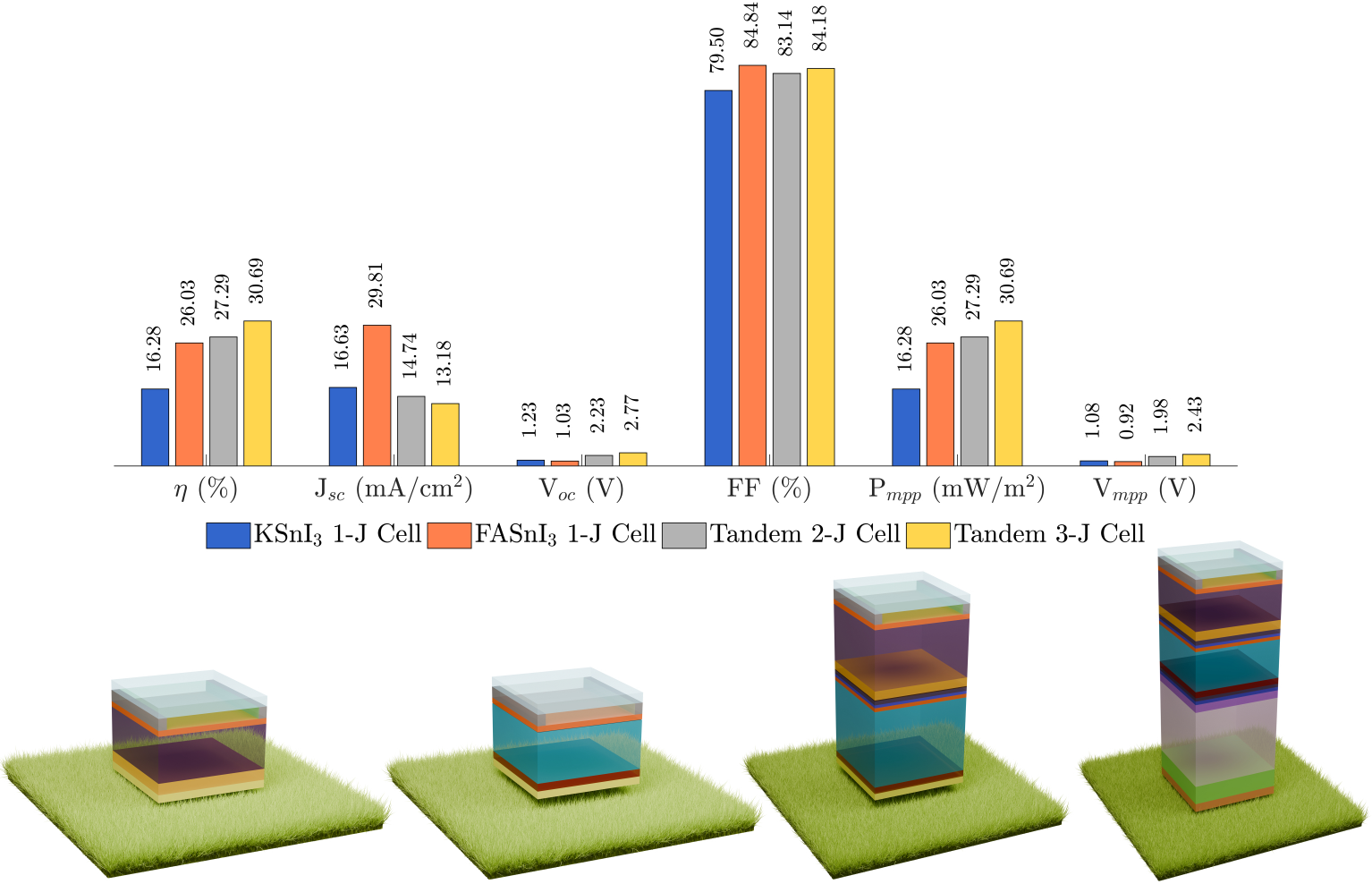}
    \caption{Bar plot illustrating the optimized performance metrics of both single-junction and different multi-junction tandem solar cells shown in the bottom row.}
    \label{fig:12}
\end{figure}

Table~\ref{Table:6} and a bar plot as shown in Fig.~\ref{fig:12} highlight the progressive enhancement in device performance as the architecture evolved from single to tandem structures, demonstrating the effectiveness of absorber stacking and junction engineering. After optimization, 1-J KSnI\textsubscript{3} and 1-J FASnI$_3$ solar cells secured 16.28\% and 26.03\% respectively, whereas the 2-J and 3-J tandem cell secured 27.29\% and 30.69\% accordingly.
\begin{table}[!b]
\centering
\caption{Photon flux ($\phi$) and power consumption (P\textsubscript{consump}) in 2-J and 3-J tandem solar cells.}
\label{tab:merged-power-consump}
\small
\begin{tabular}{lcc}
\hline
\multicolumn{3}{c}{\textbf{2T KSnI\textsubscript{3}/FASnI\textsubscript{3} Tandem Cell}} \\ \hline
\textbf{Subcells} &
  \textbf{$\phi$ (cm\textsuperscript{-2}s\textsuperscript{-1})} &
  \textbf{P\textsubscript{consump} (Wm\textsuperscript{-2})} \\ \hline
Top KSnI\textsubscript{3} subcell & 1.136$\times$10\textsuperscript{21} & 387.36 \\
Bottom FASnI$_3$ Subcell & 1.1665$\times$10\textsuperscript{21} & 270.63 \\ \hline
\multicolumn{3}{c}{\textbf{2T KSnI\textsubscript{3}/FASnI\textsubscript{3}/ACZTSe Tandem Cell}} \\ \hline
Top KSnI\textsubscript{3} subcell & 1.19$\times$10\textsuperscript{21} & 349.90 \\
Middle FASnI$_3$ subcell & 1.03$\times$10\textsuperscript{21} & 238.31 \\
Bottom ACZTSe subcell & 1.28$\times$10\textsuperscript{21} & 204.17 \\ \hline
\end{tabular}
\end{table}
Table \ref{tab:merged-power-consump} showcases the photon flux, $\phi$, and absorbed optical power per unit cross-sectional area for the subcells of 2-J and 3-J tandem cells. In the case of the 2-J tandem cell, as the top subcell is responsible for absorbing higher-energy photons, it necessitates a larger portion of the incident optical power to ensure proper current matching with the bottom subcell. Moreover, the photon fluxes across the two subcells are closely aligned, with the top subcell ultimately determining the current limitation of the overall tandem device as seen in Fig.~\ref{fig:8}(e). For the 3-J tandem cell, the top subcell primarily absorbs near-UV and visible photons, the middle subcell absorbs visible and near-infrared (NIR) photons, and the bottom subcell predominantly absorbs NIR photons from the incident spectrum to facilitate current matching with the subsequent subcells as featured in Fig.~\ref{fig:11}(c). From the absorbed photon fluxes among the three subcells, the middle subcell emerged as the current-limiting junction in the 3-J tandem cell, as evidenced in Table~\ref{tab:merged-power-consump}.

\subsection{Performance benchmarking with literature}
Table~\ref{tab:com-work-with-publish-work} presents a comparative analysis of key performance metrics from this work alongside previously reported perovskite-based tandem solar cells. Liu \textit{et al.} demonstrated an all-perovskite tandem solar cell achieving 28.1\% efficiency by combining a wide-bandgap absorber, formamidinium–cesium lead iodide–bromide, (FA$_{0.8}$Cs$_{0.2}$Pb(I$_{0.62}$Br$_{0.38}$)$_3$), with a narrow-bandgap absorber, formamidinium–methylammonium lead–tin iodide, (FA$_{0.7}$MA$_{0.3}$Pb$_{0.5}$\\Sn$_{0.5}$I$_{3}$), and employing metal-oxide nanocrystal hole transport layers to minimize optical losses \cite{liu2023efficient}. Despite the impressive efficiency and stability, the device’s reliance on lead and tin poses concerns due to toxicity and the instability of tin-based perovskites. Sun \textit{et al.} reported scalable solution processing of hybrid fullerene electron transport layers in an all-perovskite tandem module, reaching 23.3\% efficiency. Their approach improves energy-level alignment and reduces interfacial recombination but faces challenges related to the complexity of fullerene blends and slightly lower performance \cite{sun2024scalable}. Lin \textit{et al.} introduced a bilayer heterojunction design with mixed-dimensional Pb–Sn perovskites to suppress surface recombination and facilitate charge extraction, achieving a certified efficiency of 28.0\% \cite{Lin2023}. However, the mixed-dimensional phases can introduce charge transport losses, limiting fill factor and complicating fabrication. Pan \textit{et al.} developed a surface reconstruction method using diammonium halide salts to passivate Sn–Pb mixed perovskites, minimizing nonradiative recombination and achieving a certified 28.49\% efficiency for two-junction tandems \cite{Pan2024}. Their work advances film quality and stability but still contends with tin oxidation challenges inherent to Sn–Pb perovskites. Wang \textit{et al.} tackled halide heterogeneity in Br-rich perovskites to reduce recombination and realized a 25.1\% efficient triple-junction all-perovskite tandem, though challenges remain in balancing halide distribution and improving device stability \cite{wang2024halide}. Xu \textit{et al.} combined co-additives to stabilize wide-bandgap perovskites integrated in perovskite–silicon triple-junction tandems, achieving 26.4\% efficiency, but the inclusion of silicon introduces additional processing complexity and cost \cite{xu2024monolithic}. In contrast, our work proposes novel double and triple-junction tandem architectures integrating lead-free, tin-based perovskites
\begin{table}[!t]
\centering
\caption{Comparative analysis of previously reported performance matrices of different perovskite (PVSK) material-based tandem solar cells with this work.}
\label{tab:com-work-with-publish-work}
\resizebox{\columnwidth}{!}{%
\begin{tabular}{cccccccc}
\hline
\textbf{Solar Cell Type} & 
\textbf{Structure} & 
\begin{tabular}[c]{@{}c@{}}$\eta$ \\ (\%)\end{tabular} & 
\begin{tabular}[c]{@{}c@{}}J$_{\text{sc}}$ \\ (mA/cm\textsuperscript{2})\end{tabular} & 
\begin{tabular}[c]{@{}c@{}}V$_{\text{oc}}$ \\ (V)\end{tabular} & 
\begin{tabular}[c]{@{}c@{}}FF \\ (\%)\end{tabular} & 
\textbf{Ref} \\ 
\hline\\
PVSK/PVSK & 
\begin{tabular}[c]{@{}c@{}c@{}}Glass/ITO/SAM-NiO$_x$/FA$_{0.8}$Cs$_{0.2}$Pb(I$_{0.62}$Br$_{0.38}$)$_3$/\\C$_{60}$/ALD-SnO$_2$/IC-CH/FA$_{0.7}$MA$_{0.3}$Pb$_{0.5}$Sn$_{0.5}$I$_3$/\\C$_{60}$/ALD-SnO$_2$/Cu\end{tabular} & 
28.10 & 
16.70 & 
2.11 & 
79.50 & 
\cite{liu2023efficient} \\ \\ 
PVSK/PVSK & 
\begin{tabular}[c]{@{}c@{}c@{}}Glass/ITO/SAMs-NiO$_x$/Cs$_{0.35}$FA$_{0.65}$PbI$_{1.8}$Br$_{1.2}$/\\HF(C$_{60}$:PCBM:ICBA)/ALD-SnO$_2$/Au/PEDOT:PSS/\\FA$_{0.7}$MA$_{0.3}$Pb$_{0.5}$Sn$_{0.5}$I$_3$/HF/ALD-SnO$_2$/Cu\end{tabular} & 
27.40 & 
15.90 & 
2.09 & 
82.40 & 
\cite{sun2024scalable} \\ \\
 PVSK/PVSK & \begin{tabular}[c]{@{}c@{}c@{}}
ITO/PEDOT:PSS/FA$_{0.7}$Cs$_{0.3}$Pb(I$_{0.85}$Br$_{0.14}$)$_3$/\\FL-FA$_{0.8}$Cs$_{0.2}$Pb(I$_{0.62}$Br$_{0.38}$)$_3$/C$_{60}$/BCP/Cu\end{tabular}& 28.5 & 16.5 & 0.873 & 82.6 &\cite{Lin2023}\\ \\
PVSK/PVSK & \begin{tabular}[c]{@{}c@{}c@{}} Glass/ITO/Me-4PACz/Al$_2$O$_3$/FA$_{0.8}$Cs$_{0.2}$Pb(I$_{0.6}$Br$_{0.4}$)$_3$/\\PDAI$_2$/C$_{60}$/ALD-SnO$_2$/Au/PEDOT:PSS/\\Al$_2$O$_3$/FA$_{0.6}$MA$_{0.3}$Cs$_{0.1}$Pb$_{0.5}$Sn$_{0.5}$I$_3$/\\BDA-EDAI$_2$/C$_{60}$/BCP/Ag \end{tabular}& 28.49 & 16.02 & 2.12 & 83.88 &\cite{Pan2024}\\ \\
\hline\\ 
PVSK/PVSK/PVSK & 
\begin{tabular}[c]{@{}c@{}}Glass/IOH/NiO$_x$/Me-4PACz/Cs$_{0.15}$FA$_{0.85}$Pb(I$_{0.4}$Br$_{0.6}$)$_3$/\\PCBM/PEI/SnO$_x$/ITO/NiO$_x$/Me-PACz/\\Cs$_{0.05}$FA$_{0.9}$MA$_{0.05}$Pb(I$_{0.85}$Br$_{0.15}$)$_3$/C$_{60}$/SnO$_x$/Au/\\PEDOT:PSS/Cs$_{0.05}$FA$_{0.7}$MA$_{0.25}$Pb$_{0.5}$Sn$_{0.5}$I$_3$/\\C$_{60}$/SnO$_x$/Ag\end{tabular} & 
25.10 & 
9.70 & 
3.33 & 
78.00 & 
\cite{wang2024halide} \\ \\

PVSK/PVSK/Si & 
\begin{tabular}[c]{@{}c@{}}SHJ/IZO/MeO$_2$PACz/Rb$_{0.05}$Cs$_{0.1}$FA$_{0.85}$PbI$_3$/C$_{60}$/\\SnO$_2$/IZO/NiO$_x$/2PACz/Cs$_{0.1}$FA$_{0.9}$PbI$_{0.9}$Br$_{2.1}$/\\C$_{60}$/SnO$_2$/IZO/MgF\textsubscript{2}/Ag\end{tabular} & 
26.40 & 
11.90 & 
3.04 & 
72.90 & 
\cite{xu2024monolithic} \\ \\

PVSK/PVSK/Si & 
\begin{tabular}[c]{@{}c@{}c@{}}Si/ITO/Me-4PACz/FA$_{0.9}$Cs$_{0.1}$PbI$_3$/C$_{60}$/SnO$_2$/ITO\\/NiO$_x$/Me-4PACz/ FA$_{0.6}$MA$_{0.15}$Cs$_{0.25}$Pb(I$_{0.45}$Br$_{0.5}$OCN$_{0.05}$)$_3$\\/C$_{60}$/SnO$_2$/ITO/LiF\end{tabular} & 
27.60 & 
11.58 & 
3.13 & 
76.30 & 
\cite{liu2024triple} \\ \\
\hline\\
PVSK/PVSK & 
\begin{tabular}[c]{@{}c@{}c@{}}Glass/FTO/TiO\textsubscript{2}/KSnI\textsubscript{3}/CuSCN/ITO/ZnO/\\ TiO\textsubscript{2}/FASnI$_3$/Spiro-OMeTAD/Au\end{tabular} & 
27.29 & 
14.74 & 
2.227 & 
83.14 & 
This work \\ \\

\begin{tabular}[c]{@{}c@{}}PVSK/PVSK/\\ Kesterite\end{tabular} & 
\begin{tabular}[c]{@{}c@{}c@{}}Glass/FTO/TiO\textsubscript{2}/KSnI\textsubscript{3}/CuSCN/ITO\\/ZnO/TiO\textsubscript{2}/FASnI$_3$/Spiro-OMeTAD/ITO/\\AZO/ZnO/ZnS/ACZTSe/CZTSe/Mo\end{tabular} & 
30.69 & 
13.184 & 
2.766 & 
84.18 & 
This work \\ \\
\hline
\end{tabular}%
}
\end{table}
(KSnI\textsubscript{3} and FASnI$_3$) and an earth-abundant kesterite absorber (ACZTSe). This design achieved competitive PCEs of 27.29\% for 2-J tandem and 30.69\% for 3-J tandem cells, respectively, while simultaneously addressing critical issues such as toxicity, material scarcity, and infrared spectral coverage. 
The absorption edge of the ACZTSe layer boosted infrared photon absorption, enhancing current-matching and device efficiency. Moreover, band alignment and intermediate transportation layers were optimized to provide efficient charge extraction as depicted from high fill factors exceeding several reported perovskite tandems. Even though the oxidation of tin is still an issue, it can be alleviated by using encapsulation and interface engineering methods. By avoiding toxic lead and scarce dopants like rubidium or bromide, our tandem cells offer an environmentally friendly, scalable, and high-performance platform that balances efficiency, stability, and sustainability more effectively than many existing perovskite tandem designs.

\section{Feasibility of the proposed 3-J tandem cell architecture}\label{section 4}
The fabrication of the proposed triple junction tandem cell architecture requires precise and sequential deposition techniques optimized according to the thermal and chemical sensitivities of each constituent layer. The process commences with the deposition of a thin MgF\textsubscript{2} ARC via thermal or electron beam evaporation, designed to reduce optical losses at the air/glass interface \cite{Li2024}. Subsequently, highly conductive FTO films are deposited using radio frequency (RF) magnetron sputtering, which offers strong adhesion, uniformity, and scalability \cite{Ahmadipour2018}. Anatase-phase TiO\textsubscript{2} thin films are then prepared on the FTO surface through spin-coating followed by annealing. Post-annealing, UV–ozone treatment is employed to enhance surface properties for subsequent layer integration \cite{klasen2019removal}. The deposition of KSnI\textsubscript{3} absorber layer, though still limited experimentally, can be achieved via spin-coating a precursor solution of potassium iodide (KI) and tin(II) iodide (SnI$_2$) in a solvent mixture of N,N-dimethylformamide (DMF) and dimethyl sulfoxide (DMSO). This is followed by thermal annealing to promote crystallization, following methodologies adapted from tin-based perovskite literature \cite{Wang2024}. Afterwards, CuSCN HTL is applied through spin-coating from a dipropyl sulfide solution and mildly annealed to form a compact and hydrophobic interface \cite{Arora2017}. ITO interlayer is then deposited by low-temperature sputtering to preserve the integrity of the underlying layers \cite{Bush2017}. ZnO is introduced as ETL layer either by atomic layer deposition (ALD) or low-temperature sol–gel techniques, followed by a second compact TiO\textsubscript{2} layer. The FASnI$_3$ absorber is deposited using an antisolvent-assisted spin-coating method, incorporating SnF$_2$ as an additive to improve phase stability and reduce hysteresis \citep{Wu2022}. Subsequently, Spiro-OMeTAD is spin-coated from a chlorobenzene solution containing Li-TFSI and 4-tert-butylpyridine (tBP) as dopants to enhance hole mobility \cite{Kasparavicius2021}. A second transparent ITO layer (~20 nm) is sputtered, followed by an Al-doped ZnO (AZO) layer to reduce sheet resistance and enhance optical transmission. ZnO and ZnS buffer layers are deposited via ALD or chemical bath deposition (CBD), ensuring favorable band alignment and reduced interfacial recombination \citep{Hernndez-Caldern2020}. ACZTSe/CZTSe absorber stack is subsequently formed through co-sputtering of Cu, Zn, Sn, Se, and Ag, followed by high-temperature selenization to achieve the desired kesterite phases and interface quality \cite{Shi2024}. Molybdenum (Mo) back contact layer is deposited via DC magnetron sputtering to serve as a chemically stable, low-resistance electrode. To ensure both environmental and operational stability, the device is encapsulated with a UV-curable epoxy and sealed with a glass barrier. 

\section{Conclusion}\label{section 5}
The design and optimization of high-efficiency tandem solar cells calls for a holistic approach, that encompasses several interconnected aspects, such as the selection and incorporation of complementary absorber materials, precise tunnel junction engineering, and accurate matching of photocurrent among the subcells. In this paper, we have addressed these challenges by introducing two new lead-free tandem solar cell architectures, an all-perovskite dual-junction solar cell (KSnI\textsubscript{3}/FASnI\textsubscript{3}) and a hybrid perovskite-kesterite triple-junction solar cell (KSnI\textsubscript{3}/FASnI\textsubscript{3}/ACZTSe). The chosen absorbers effectively cover from near-UV to NIR spectral region, reducing bandgap disparities and enhancing overall photon utilization. By systematically optimizing the absorber thickness, doping density, and tunnel-junction band alignment, the proposed architectures achieved effective current matching and enhanced charge transport throughout the device. The dual-junction cell delivered a PCE of 27.29\% with J\textsubscript{sc} of 14.74 mA/cm\textsuperscript{2}, V\textsubscript{oc} of 2.227 V, and FF of 83.14\%. Subsequently, after integration and careful optimization of the triple junction subcell, the optimized triple-junction configuration reached a champion PCE of 30.69\% with J\textsubscript{sc} of 13.184 mA/cm\textsuperscript{2}, V\textsubscript{oc} of 2.766 V, and FF of 84.18\%, corresponding to a relative PCE gain of 12.5\% over the dual-junction architecture. These results and insights demonstrated the ability of hybrid tandem architectures to exceed the efficiency limits of conventional designs while employing sustainable, earth-abundant, and non-toxic materials.


\section{CRediT authorship contribution statement}
\textbf{Md. Faiaad Rahman}: conceptualization, methodology, visualization, simulation, investigation, writing – original draft, and writing – review \& editing. \textbf{Md. Ashaduzzaman Niloy}: conceptualization, visualization, investigation, writing – original draft, and writing – review \& editing. \textbf{Ehsanur Rahman}: methodology, visualization, supervision, writing – original draft, and writing – review \& editing, \textbf{Ahmed Zubair}: conceptualization, methodology, visualization, resources, supervision, writing – original draft, and writing – review \& editing.
\section{Data availability statement}
The data that support the findings of this study are available from the
corresponding authors upon reasonable request.
\section{Declaration of competing interest}
The authors declare that they have no known competing financial interests or personal relationships that could have appeared to influence the work reported in this paper.
\section{Acknowledgments}
The authors express their sincere gratitude to the Department of Electrical and Electronic Engineering at Bangladesh University of Engineering and Technology (BUET) for providing access to the Ansys Lumerical software and the necessary technical support.\\

\bibliographystyle{elsarticle-num}
\bibliography{references}

\newpage
\setcounter{section}{0}
\setcounter{page}{1}
\setcounter{figure}{0}
\setcounter{equation}{0}
\setcounter{table}{0}
\renewcommand{\thesection}{S\arabic{section}}
\renewcommand{\thepage}{S\arabic{page}}
\renewcommand{\thetable}{S\arabic{table}}
\renewcommand{\thefigure}{S\arabic{figure}}

\setcounter{affn}{0}
\resetTitleCounters

\makeatletter
\let\@title\@empty
\makeatother
\title{Supplementary Material: Unveiling architectural and optoelectronic synergies in lead-free perovskite/perovskite/kesterite triple-junction monolithic tandem solar cells}

\makeatletter
\renewenvironment{abstract}{\global\setbox\absbox=\vbox\bgroup
  \hsize=\textwidth\def\baselinestretch{1}%
  \noindent\unskip\textbf{Contents}
 \par\medskip\noindent\unskip}
 {\egroup}
\def\ps@pprintTitle{%
     \let\@oddhead\@empty
     \let\@evenhead\@empty
     \def\@oddfoot{\footnotesize\itshape
        Supplementary Material for \ifx\@journal\@empty Elsevier
       \else\@journal\fi\hfill\today}%
     \let\@evenfoot\@oddfoot}
\makeatother
\vspace{-40pt}
\startlist{toc}
\begin{abstract}
\vspace{-57pt}
\printlist{toc}{}{\section*{}}
\end{abstract}
\maketitle
\vspace{-10pt}
\section*{}
\parindent0pt
\vspace{-20pt}
\section{Simulation methodology}\label{S01:methodlogy}
First, absorption and optical generation needs to be calculated by solving Maxwell's curl electromagnetic wave equation using Finite Difference Time Domain (FDTD) analysis for optical electric field distribution in different layers:
\begin{equation}
    \frac{\partial \vec{D}}{\partial t}=\nabla \times \vec{H}
\end{equation}
\begin{equation}
    \vec{D}(\omega)=\epsilon_o \epsilon_r(\vec{r}, \omega)\vec{E_{op}}
\end{equation}
\begin{equation}
    \frac{\partial \vec{H}}{\partial t}=-\frac{1}{\mu_o}\nabla \times \vec{E_{op}}
\end{equation}
Where, $\vec{H}$, $\vec{E_{op}}$ and $\vec{D}(\omega)$ are the magnetic, electric and displacement fields respectively. $\epsilon_r(\vec{r}, \omega)$ is the complex relative dielectric constant, where, $\omega$ is the angular frequency. Each material has been modeled according to its respective refractive index ($n$) and extinction coefficient ($k$). The absorbed power ($P_{abs}$) is then calculated by the following equation:
\begin{equation}
    P_{abs}=-\frac{1}{2} \omega |\vec{E_{op}}(\vec{r}, \omega)|^2 \mathfrak{J} \{\epsilon(\vec{r}, \omega)\}
\end{equation}
Afterwards, the generation rate is calculated as:
\begin{equation}
    G(\vec{r})=\int g(\vec{r}, \omega) d\omega
\end{equation}
\begin{equation}
    g(\vec{r}, \omega)=\frac{P_{abs}}{\hbar \omega}=-\frac{\pi}{h}|\vec{E_{op}}(\vec{r}, \omega)|^2 \mathfrak{J} \{\epsilon(\vec{r}, \omega)\}
\end{equation}
Where, $h$ is the Planck's constant. In X-axis, periodic boundary condition and in Y-axis, perfectly matched layer (PML) boundary condition is maintained. To perform optical simulation, the AM 1.5G solar spectrum has been set as the input irradiation source.\\
Performance metrics like efficiency ($\eta$), open circuit voltage (V$_{oc}$), short circuit current (J$_{sc}$), and fill factor FF can be found from the J-V characteristics of each cell independently by solving Poisson's equation, drift-diffusion equations, and continuity equations as mentioned below:
\begin{equation}
    -\nabla.(\epsilon_{dc} \nabla V)=q\rho
    \label{Eq:8}
\end{equation}
\begin{equation}
    \vec{J}_n=q \mu_n n \vec{E}+qD_n\nabla n
    \label{Eq:9}
\end{equation}
\begin{equation}
    \vec{J}_p=q \mu_n p \vec{E}-qD_p\nabla p
    \label{Eq:10}
\end{equation}
\begin{equation}
    \frac{\partial n}{\partial t}=\frac{1}{q} \nabla.\vec{J}_n-R_n
    \label{Eq:11}
\end{equation}
\begin{equation}
    \frac{\partial p}{\partial t}=-\frac{1}{q} \nabla.\vec{J}_p-R_p
    \label{Eq:12}
\end{equation}
Where, $\epsilon_{dc}$ is the dielectric dc permittivity, $V$ is the electrostatic potential (electric field, $\vec{E}=-\nabla V$), $\rho$ is the net charge density ($\rho=p-n+C$, $C$=ionized impurity density), $\vec{J}_{n(p)}$ is the electron (hole) current density, $q$ is the positive electron charge, $\mu_{n(p)}$ is the mobility of electron (hole), $D_{n(p)}$ is the diffusivity of electron (hole) ($D_{n(p)}=\mu_{n(p)} \frac{k_B T}{q}$), $n$ and $p$ are electron and hole densities respectively, $R_{n(p)}$ is the net recombination rate, $k_B$ is the Boltzmann constant and $T$ is the temperature. The computed generation rate retrieved from the optical simulation is then employed as an input in the continuity equations, and \cref{Eq:8,Eq:9,Eq:10,Eq:11,Eq:12} are solved by implementing Dirichlet boundary conditions at contacts, and Neumann boundary conditions at the insulating boundaries and interfaces, respectively.

To quantify the charge collection efficiency, the external quantum efficiency (EQE) is evaluated as the ratio between the photocurrent density and the incident photon flux at each wavelength. Under illumination, EQE at a given wavelength $\lambda_{0}$ is calculated using:
\begin{equation}
EQE(\lambda_0) = \frac{J(\lambda_0)}{q \cdot \Phi_{\text{inc}}(\lambda_0)}
\end{equation}
whereas, J($\lambda_0$) is the photocurrent density (mA/cm\textsuperscript{2}), $\phi$($\lambda_0$) is the incident photon flux density (photons/cm\textsuperscript{2}·s), q is the elementary charge. Again, the photon flux $\phi_{\lambda}$ for a given wavelength, $\lambda_0$ (nm) can be expressed as:
\begin{equation}
\Phi_{\text{inc}}(\lambda) = \frac{P_{\text{inc}}(\lambda)}{hc/\lambda} = \frac{P_{\text{inc}}(\lambda) \cdot \lambda}{h c}
\end{equation}
Combining the above, the EQE becomes:
\begin{equation}
EQE(\lambda) = \frac{hc}{q \lambda} \cdot \frac{J(\lambda)}{P_{\text{inc}}(\lambda)}
\end{equation}
The obtained EQE is a unitless quantity. 
Unlike EQE, IQE considers only those photons that are absorbed in the photoactive layer, $P_{abs}^{\ absorber}$, excluding optical losses due to reflection, transmission, and parasitic absorption in non-active layers (e.g., electrodes, transport layers).
IQE is defined at a specific wavelength $\lambda_0$ as:
\begin{equation}
    IQE(\lambda_0) = \frac{EQE(\lambda_0)}{A(\lambda_0)}
\end{equation}
\begin{equation}
    A(\lambda_0) = \frac{P_{abs}^{\ absorber}(\lambda_0)}{P_{inc}(\lambda_0)}
\end{equation}
Here, A($\lambda_0$) is the absorbance of the photoactive layer at given wavelength $\lambda_0$. These overall frameworks allow assessment of device performance under spectrally resolved illumination and are essential for optimizing the design of high-efficiency single junction solar cells.\\

In the case of calculating the performance parameters of a tandem solar cell, the short-circuit current density is determined by the subcell that delivers the lowest current denisty, due to the monolithic architecture:

\begin{equation}
    J_{\text{sc}}^{\text{Tandem}} = \min \left( J_{\text{sc}}^{1}, J_{\text{sc}}^{2}, \ldots, J_{\text{sc}}^{m} \right)
\end{equation}

The open-circuit voltage, V\textsubscript{oc} of the tandem cell is the sum of the open-circuit voltages of all subcells if the current density J is the same in all the subcells.

\begin{equation}
    V_{\text{oc}}^{\text{Tandem}} = V_{\text{oc}}^{1} + V_{\text{oc}}^{2} + \ldots + V_{\text{oc}}^{m}
\end{equation}

Alternatively, at a matched current $J$, the total operating voltage of the tandem cell is given by:

\begin{equation}
    V_{\text{Tandem}}(J) = V_{1}(J) + V_{2}(J) + \ldots + V_{m}(J)
\end{equation}

Here, $m$ denotes the number of subcells in the tandem configuration, and $V_{i}(J)$ is the voltage of subcell $i$ when operating at the common current $J$.

\clearpage
\section{Optimization procedure and electrical parameters}

\label{appendix}
\begin{figure}[!ht]
    \centering
    \includegraphics[width=0.78\linewidth]{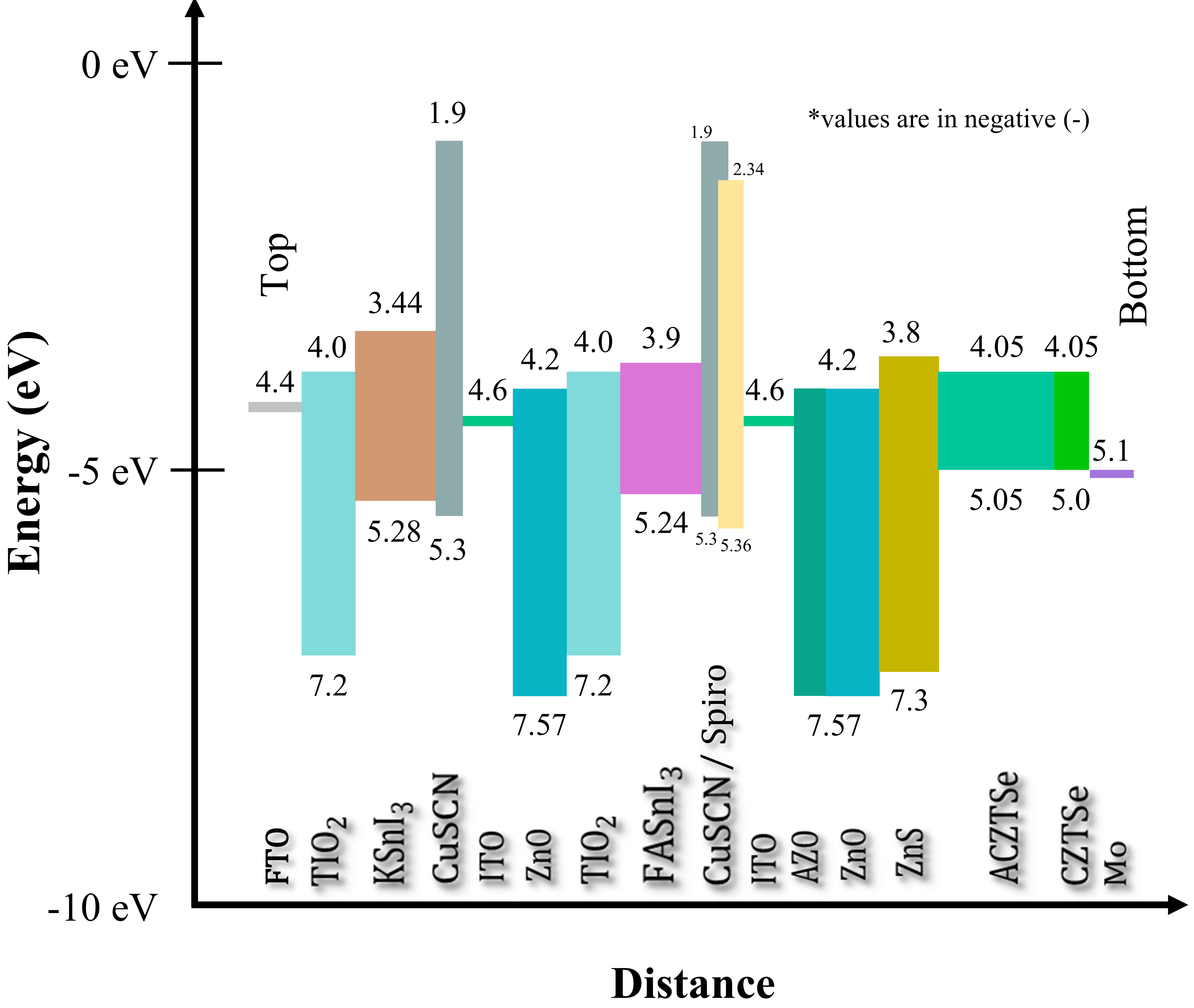}
    \caption{Energy (eV) vs space diagram of proposed 2-T monolithic triple junction tandem solar cell architecture.}
    \label{Fig:S01}
\end{figure}
\hspace{0.2cm}
For the modeling and simulation of the solar cell architectures, we considered the work functions of FTO and ITO as 4.4~eV and 4.6~eV, respectively.
\begin{table}[!ht]
\centering
\caption{Electrical parameters of TiO\textsubscript{2}, ZnS, ZnO, AZO, CuSCN and Spiro-OMeTAD.}
\label{Table:S1}
\resizebox{\textwidth}{!}{
\begin{tabular}{l*{6}{>{\centering\arraybackslash}m{2.8cm}}}
\hline
\textbf{Parameters} &
  \textbf{\begin{tabular}[c]{@{}c@{}}TiO\textsubscript{2}\\ \cite{tooghi2020high}\end{tabular}} &
  \textbf{\begin{tabular}[c]{@{}c@{}}ZnS\\ \citep{kuo2014efficiency, saha2017proposition}\end{tabular}} &
  \textbf{\begin{tabular}[c]{@{}c@{}}ZnO\\ \citep{kuo2014efficiency, saha2017proposition, tabernig2022optically}\end{tabular}} &
  \textbf{\begin{tabular}[c]{@{}c@{}}AZO\\ \citep{kuo2014efficiency, saha2017proposition}\end{tabular}} &
  \textbf{\begin{tabular}[c]{@{}c@{}}CuSCN\\ \citep{tooghi2020high}\end{tabular}} &
  \textbf{\begin{tabular}[c]{@{}c@{}}Spiro-OMeTAD\\ \citep{ivriq2025enhancing, xu2024spiro}\end{tabular}} \\
\hline
Thickness (nm) & 20--150 & 20--100 & 20--100 & 10 & 20--150 & 20--150 \\
Bandgap $E_g$ (eV) & 3.2 & 3.58 & 3.37 & 3.37 & 3.4 & 3.02 \\
DC permittivity $\varepsilon$ & 9 & 9 & 9 & 9 & 10 & 3.2 \\
Electron affinity $\chi$ (eV) & 4.0 & 3.8 & 4.0 & 4.0 & 1.9 & 2.34 \\
Mobility $\mu_n/\mu_p$ (cm\textsuperscript{2}/Vs) & 20 / 10 & 230 / 40 & 150 / 50 & 50 / 5 & $1\times10^{-4}$ / $1\times10^{-2}$ & $2\times10^{-4}$ / $2\times10^{-4}$ \\
SRH lifetime $\tau_e/\tau_h$ (ns) & 5 / 2 & 0.5 / 0.5 & 1 / 1 & -- & 5 / 5 & 5 / 5 \\
Radiative recomb. (cm\textsuperscript{3}s\textsuperscript{-1}) & -- & $1.5\times10^{-10}$ & -- & -- & -- & -- \\
Effective conduction band density, $N_C$ (cm\textsuperscript{-3}) & $1\times10^{19}$ & 2.7$\times$10$^{18}$ & 2.2$\times$10$^{18}$ & $2.2\times10^{18}$ & 2.51$\times$10$^{19}$ & 1$\times$10$^{19}$ \\
Effective valence band density, $N_V$ (cm\textsuperscript{-3}) & $1\times10^{19}$ & 1.7$\times$10$^{19}$ & 1.8$\times$10$^{19}$ & $1.9\times10^{19}$ & $1.79\times10^{19}$ & 1$\times$10$^{19}$ \\
Donor doping $N_D$ (cm\textsuperscript{-3}) & $5\times10^{18}$ & $5\times10^{16}$ & $1.5\times10^{17}$ & $8\times10^{18}$ & -- & -- \\
Acceptor doping $N_A$ (cm\textsuperscript{-3}) & -- & 1.79$\times$10$^{19}$ & -- & $1\times10^{15}$ & $5\times10^{18}$ & $2\times10^{18}$ \\
Surface Recombination (cm/s) &
  \begin{tabular}[c]{@{}c@{}}(FTO/TiO$_{2}$)\\ \&\\ (MgF$_{2}$/TiO$_{2}$)\\ $1\times10^{7}$\end{tabular} &
  -- & -- &
  \begin{tabular}[c]{@{}c@{}}(AZO/ITO)\\ $1\times10^{7}$\end{tabular} &
  -- &
  -- \\
\hline
\end{tabular}%
}
\end{table}

\clearpage
\section{Optimization results: from single junction to triple junction}
\subsection{Single Junction KSnI\textsubscript{3} Cell}
\subsubsection{Thickness optimization of layers}
\begin{figure}[!b]
    \centering
    \includegraphics[width=0.85\linewidth]{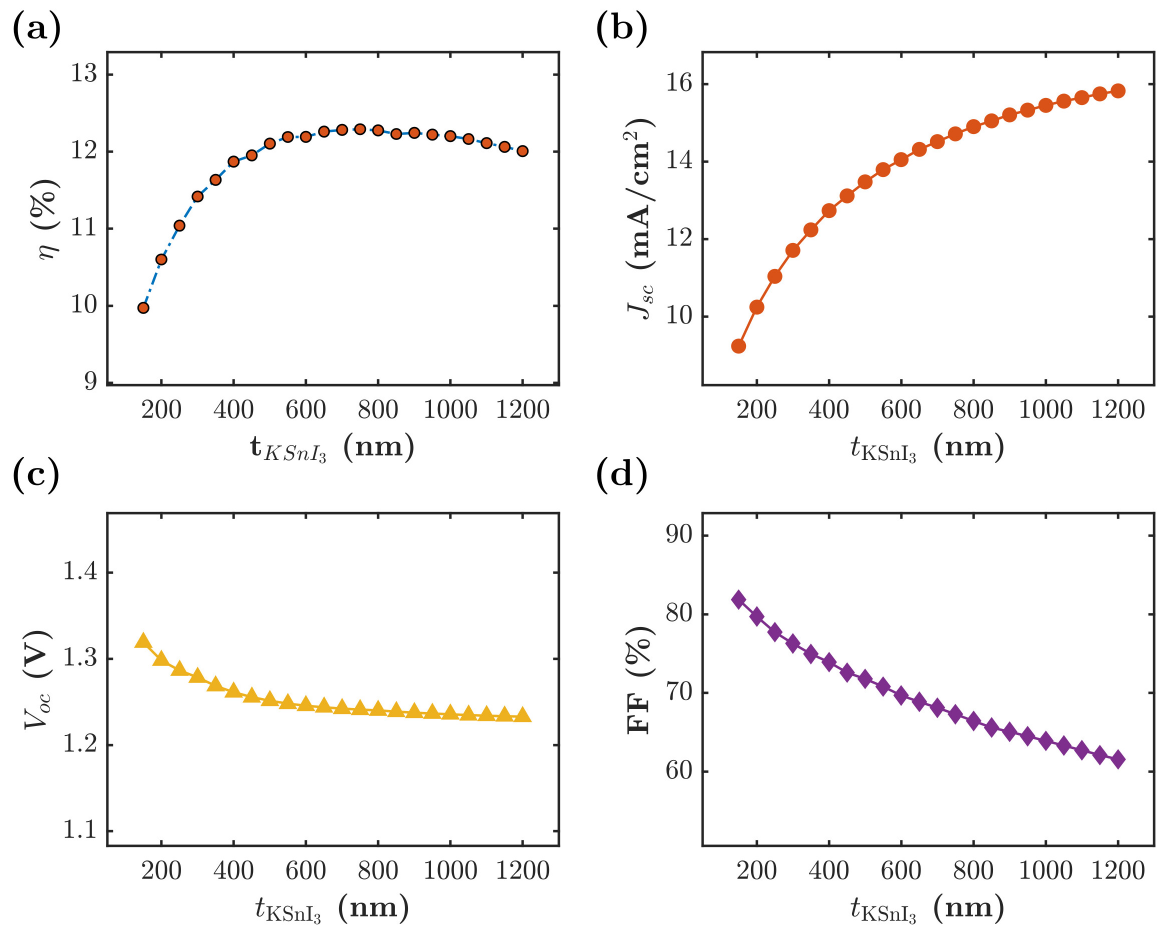}
    \caption{Initial iterative analysis showing the impact on performance metrics—(a) $\eta$, (b) $J_{\mathrm{sc}}$, (c) $V_{\mathrm{oc}}$, and (d) FF—of the single-junction KSnI$_3$ cell corresponding to the thickness variation of the KSnI$_3$ absorber. The peak PCE ($\eta$) was obtained at a thickness of 750~nm.}
    \label{Fig:S2}
\end{figure}

To establish the baseline device behavior, we first performed a preliminary thickness sweep of the KSnI\textsubscript{3} absorber from 200 to 1200 nm, keeping TiO\textsubscript{2} ETL, and CuSCN HTL layer thickness fixed. As shown in Fig.~\ref{Fig:S2}(b), increasing the KSnI\textsubscript{3} absorber thickness from 150 to 1200~nm led to a steady rise in J\textsubscript{sc} due to enhanced optical absorption and
improved photon harvesting within the bulk region. However, as the absorber thickness increases, carriers generated deeper inside the absorber experience longer transport paths and higher recombination probability, especially beyond 750~nm for our case. This resulted in a gradual decrease in V\textsubscript{oc} and FF as shown in Figs.~\ref{Fig:S2}(c) and (d), attributed to increased bulk recombination and reduced internal electric field strength that hindered efficient carrier extraction. Consequently, the PCE ($\eta$) exhibited a saturation trend, reaching its peak value of 12.29\% at 750~nm, where the trade-off between optical absorption and carrier transport was best balanced.\\

\begin{figure}[!b]
    \centering
    \includegraphics[width=0.88\linewidth]{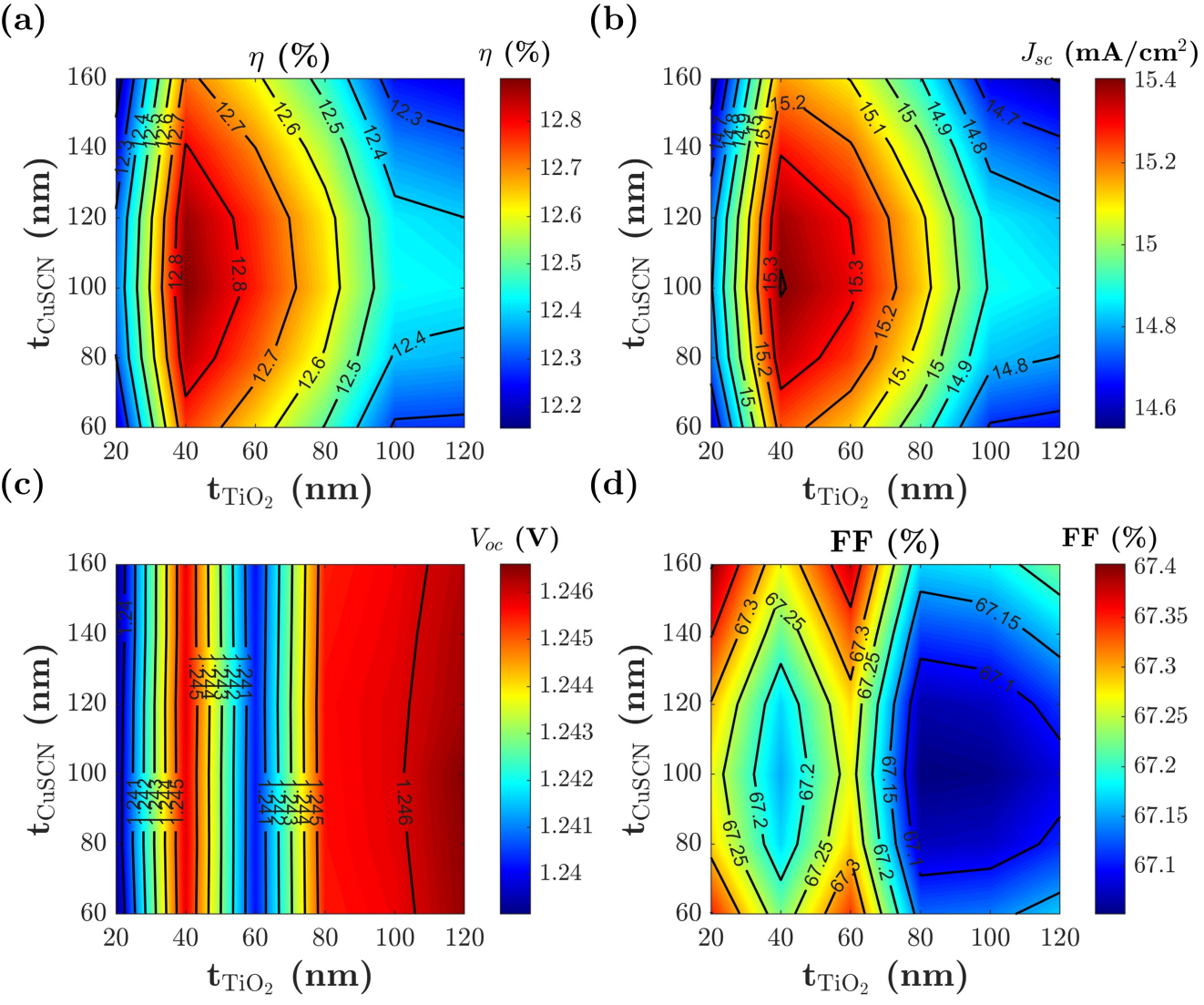}
    \caption{Initial parametric sweep showing the impact on performance metrics—(a) $\eta$, (b) J\textsubscript{sc}, (c) V\textsubscript{oc}, and (d) FF of 1-J KSnI$_3$ cell corresponding to the thickness variation of TiO$_2$ ETL and CuSCN HTL layers. The peak $\eta$ was secured 12.89\% for 40 nm thickness of TiO$_2$ ETL and 100 nm CuSCN HTL.}
    \label{Fig:S3}
\end{figure}

Keeping the KSnI\textsubscript{3} absorber fixed at 750~nm, we analyzed the influence of ETL (TiO\textsubscript{2}) and HTL (CuSCN) thicknesses on device performance as shown in Fig.~\ref{Fig:S3}(a-d). From Figs.~\ref{Fig:S3}(a) and (b), increasing the TiO\textsubscript{2} thickness from 40 to 60~nm improved electron extraction by providing more uniform coverage and better contact at the ETL/absorber interface, reducing interface recombination and enhancing J\textsubscript{sc} and PCE. Beyond 60~nm, further thickening slightly increased the carrier transport path and series resistance, leading to marginal reductions in performance. For the HTL, thinner CuSCN layers led to insufficient hole extraction and higher recombination at the absorber/HTL interface, whereas thicker layers increased the transport distance for holes, slightly reducing V\textsubscript{oc} and FF due to slower charge collection, as seen from Fig.~\ref{Fig:S3}(c) and (d). 
\begin{figure}[!b]
    \centering
    \includegraphics[width=0.88\linewidth]{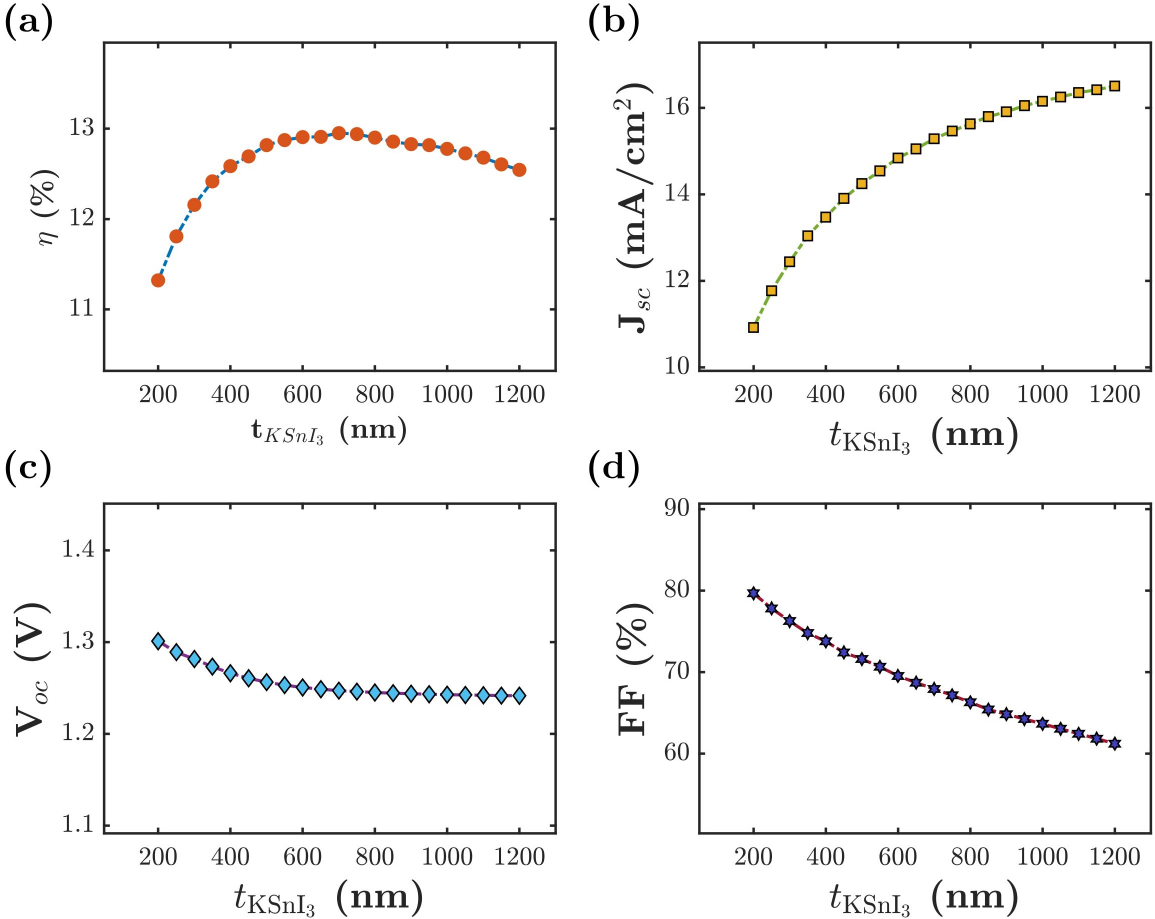}
    \caption{Final iterative analysis showing the impact on performance metrics—(a) PCE ($\eta$), (b) $J_{\mathrm{sc}}$, (c) $V_{\mathrm{oc}}$, and (d) FF—of the single-junction KSnI$_3$ cell corresponding to the thickness variation of the KSnI$_3$ absorber. The peak PCE is obtained at a thickness of 700~nm.}
    \label{Fig:S4}
\end{figure}
The highest PCE ($\eta$) of 12.89\% was obtained for TiO\textsubscript{2} of 40~nm and CuSCN of 100~nm, reflecting a balance between effective carrier extraction, minimized interface recombination, and favorable electric field distribution across the junction.\\

The final iterative analysis of KSnI\textsubscript{3} thickness, as shown in Fig.~\ref{Fig:S4}, confirms the trends from our preliminary study. Increasing the absorber up to 700~nm steadily improves J\textsubscript{sc} and PCE, while further thickening offers only marginal J\textsubscript{sc} gains that are offset by additional recombination losses and resistive effects, resulting in a modest decrease in PCE. The maximum PCE of 12.95\% is achieved at 700 nm, aligning with our preliminary estimate.\\

\subsubsection{Doping optimization of ETL and HTL}
\begin{figure}[!ht]
    \centering
    \includegraphics[width=0.95\linewidth]{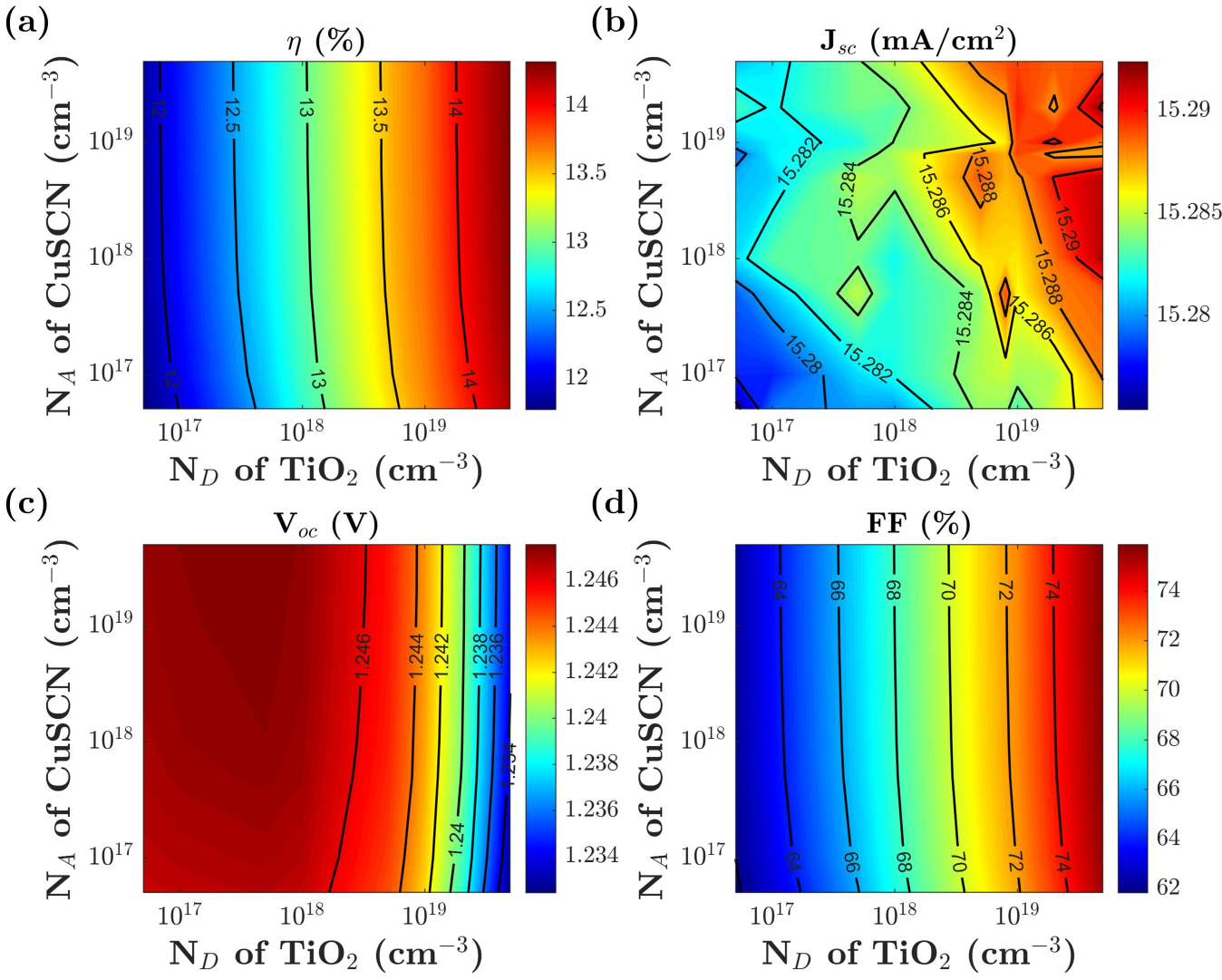}
    \caption{Contour plot of parametric sweep of the performance metrics -- (a) $\eta$, (b) J$_{sc}$, (c) V$_{oc}$, and (d) FF of 1-J KSnI$_3$ cell corresponding to variation of donor density, N$_{D}$ of TiO$_2$ ETL and acceptor density, N$_{A}$ of CuSCN HTL.}
    \label{Fig:S5}
\end{figure}
We varied the donor doping density, N$_{D}$ in the TiO$_2$ ETL from $5\times10^{16}$ to $5\times10^{19}\ \text{cm}^{-3}$, while we adjusted the acceptor doping density, N$_{A}$ in the CuSCN HTL over the same range, from $5\times10^{16}$ to $5\times$10\textsuperscript{19} cm\textsuperscript{-3}.
The results indicate that increasing the acceptor doping in CuSCN and donor doping in TiO\textsubscript{2} improves the overall device performance. Higher doping enhances the built-in electric field, promoting more efficient carrier drift and reducing recombination losses, which is reflected in the gradual increase of PCE, J\textsubscript{sc}, V\textsubscript{oc}, and FF. The maximum PCE of 14.32\% is achieved for TiO\textsubscript{2} ETL and CuSCN HTL doping densities of 5$\times$10\textsuperscript{19} cm\textsuperscript{-3}, demonstrating the critical role of optimized doping in facilitating charge extraction and improving optical-to-electrical conversion in the KSnI\textsubscript{3} solar cell. 
\subsubsection{Impact of MgF\textsubscript{2} ARC thickness}
\begin{figure}[!b]
    \centering
    \includegraphics[width=1.0\linewidth]{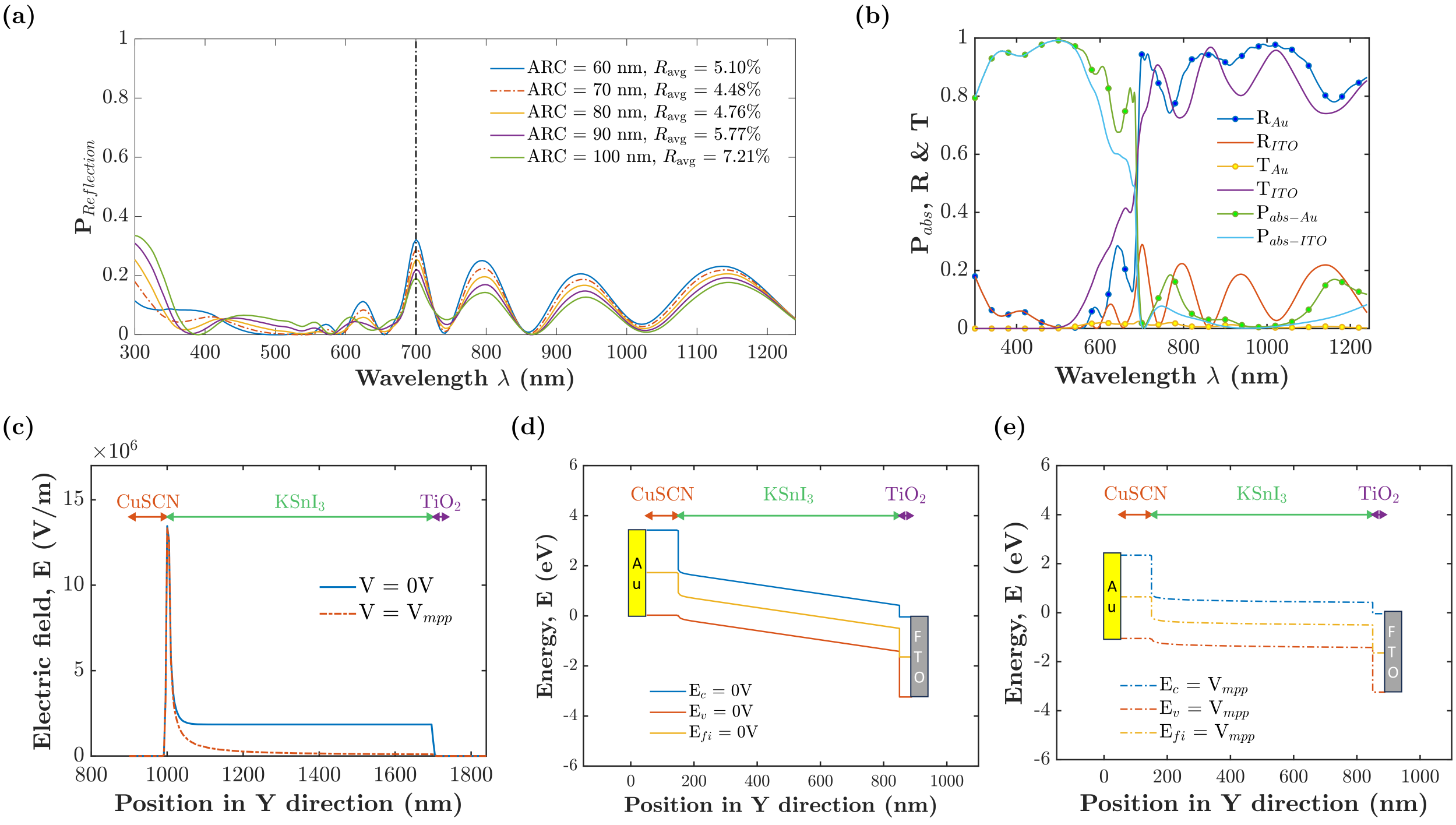}
    \caption{(a) Impact on overall reflection from 1-J KSnI\textsubscript{3} cell via  altering MgF\textsubscript{2} ARC thickness.(b) Normalized power absorption (P$_{abs}$), reflection (R), and transmission (T) corresponding to photon wavelength ($\lambda$). (c) Spatial E-field profile, and (d) band-diagram of ITO/TiO$_2$/KSnI$_3$/CuSCN/Au corresponding to device thickness at thermal equilibrium at short circuit condition (V = 0V) and maximum power point (V = V$_{mpp}$).}
    \label{Fig:S6}
\end{figure}

To reduce the optical reflection, we varied the thickness of MgF\textsubscript{2} ARC from 20 to 150~nm. A minimum average reflection of 4.48\% has been attained for photon wavelength of 300-700 nm range as depicted from Fig.~\ref{Fig:S6}(a). This reduced reflection boost the carrier-generation rate, resulting increase in PCE from 14.31\% to 14.51\%. \\

As shown in Figs.~\ref{Fig:S6}(c–e), the spatial distribution of the electric field and the band diagram collectively illustrate the charge transport mechanism across the FTO/TiO\textsubscript{2}/KSnI\textsubscript{3}/CuSCN/Au stack. The electric field profile exhibits a pronounced peak at the KSnI\textsubscript{3}/CuSCN interface, originating from the built-in potential, V$_{bi}$ between the p-type CuSCN and the intrinsic KSnI\textsubscript{3} absorber. This high-field region corresponds to the strong band bending seen in the thermal-equilibrium band diagram, which drives drift-dominated carrier transport, promoting efficient hole extraction toward CuSCN. Across the KSnI\textsubscript{3} layer, the field gradually diminishes, consistent with the band flattening that leads to a diffusion-dominated regime governed by carrier concentration gradients. Near the TiO\textsubscript{2}/KSnI\textsubscript{3} interface, the bands align smoothly with a weak and flat field, indicating minimal energy barriers for electron extraction. Under operating bias at V = V\textsubscript{mpp}, partial band flattening and a reduced field magnitude confirm the transition toward balanced drift–diffusion transport, ensuring efficient charge collection and limited interfacial recombination.
\subsubsection{Incorporation of UV-ozone treated ETL and Au BCL}
At the FTO/TiO\textsubscript{2}/KSnI\textsubscript{3} interface, the conduction band of TiO\textsubscript{2} lies above that of KSnI\textsubscript{3}, having a conduction band offset (CBO) of $\approx$ 0.56~eV, enabling efficient electron extraction while blocking holes. To further optimize this interface, UV-ozone treatment was applied to TiO\textsubscript{2}, lowering its work function from 5.3~eV to 5.1~eV. This reduced interfacial recombination by mitigating surface defects and improving electron transport, resulting in enhanced charge collection, achieving a PCE of 15.30\% as shown in Table~\ref{Table:S2}.\\

Incorporating Au as the back contact layer (BCL) enhances photon absorption in the 550–690~nm range via back-reflection, increasing carrier generation and collection. This leads to higher J\textsubscript{sc} and an overall ~8.6\% relative improvement in PCE, reaching 16.28\% for the optimized 1-J KSnI\textsubscript{3} cell.
\begin{table}[!ht]
\centering
\small
\renewcommand{\arraystretch}{1.1}
\caption{Impact of UV-ozone treated TiO$_2$ ETL on single-junction KSnI$_3$ solar cell performance.}
\label{Table:S2}
\resizebox{\textwidth}{!}{%
\begin{tabular}{@{}ccccccc@{}}
\hline
\textbf{\begin{tabular}[c]{@{}c@{}}Back Contact Layer\\(BCL)\end{tabular}} &
\textbf{TiO\textsubscript{2} ETL} &
\textbf{\begin{tabular}[c]{@{}c@{}}$\phi_{TiO_{2}}$\\(eV)\end{tabular}} &
\textbf{\begin{tabular}[c]{@{}c@{}}$\eta$\\(\%)\end{tabular}} &
\textbf{\begin{tabular}[c]{@{}c@{}}J\textsubscript{sc}\\(mA/cm$^2$)\end{tabular}} &
\textbf{\begin{tabular}[c]{@{}c@{}}V\textsubscript{oc}\\(V)\end{tabular}} &
\textbf{\begin{tabular}[c]{@{}c@{}}FF\\(\%)\end{tabular}} \\
\hline
ITO & Oxidized & 5.3 & 14.53 & 15.62 & 1.235 & 75.37 \\
ITO & UV-ozone treated & 5.1 & 15.30 & 15.62 & 1.228 & 79.79 \\
Au & UV-ozone treated & 5.1 & 16.28 & 16.62 & 1.232 & 79.50 \\
\hline
\end{tabular}%
}
\end{table}

\clearpage
\subsection{Single Junction FASnI\textsubscript{3} Cell}
\subsubsection{Thickness optimization of layers}
\begin{figure}[!ht]
    \centering
    \includegraphics[width=0.9\linewidth]{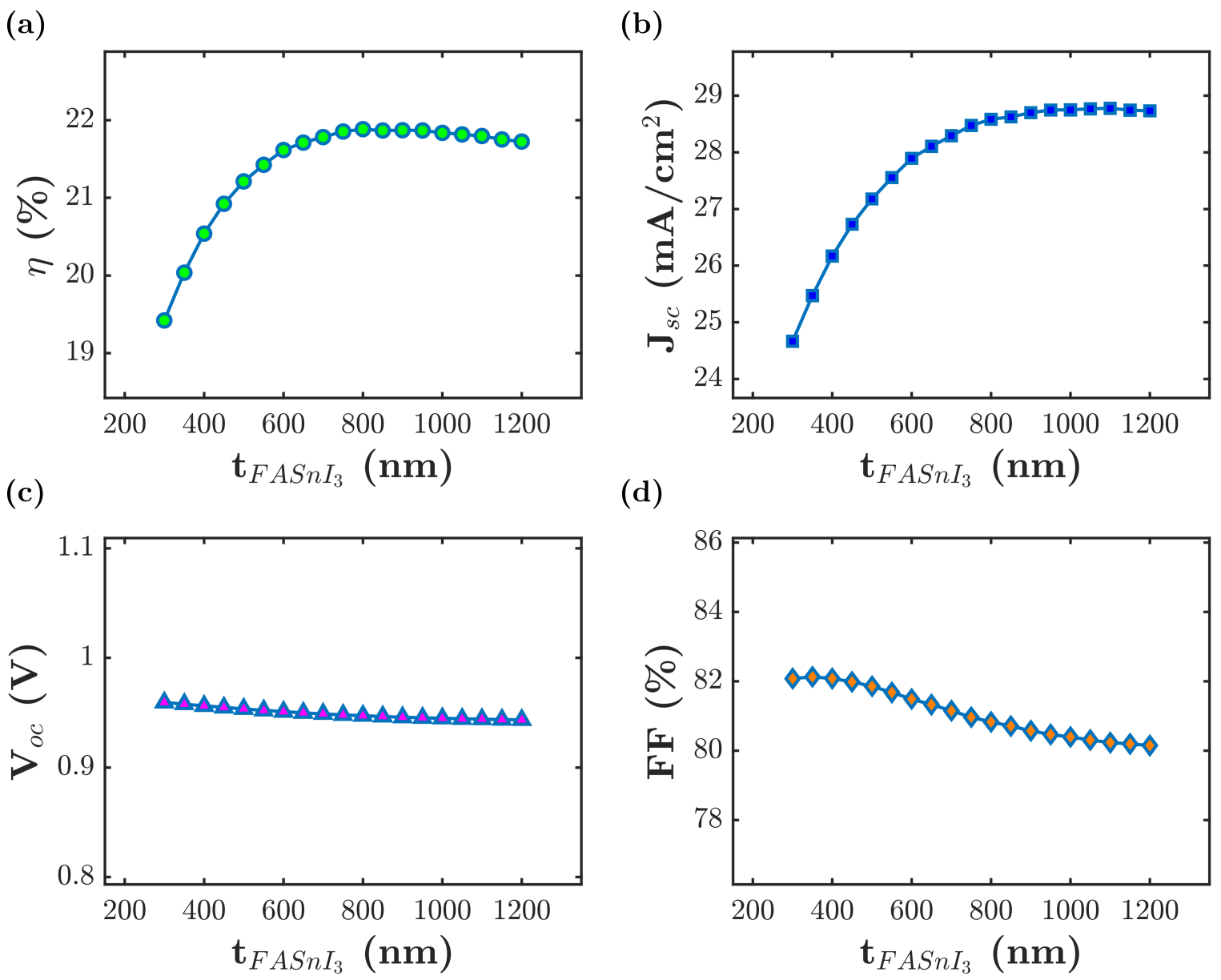}
    \caption{Initial iterative analysis showing the impact on performance metrics—(a) $\eta$, (b) J\textsubscript{sc}, (c) V\textsubscript{oc}, and (d) FF of 1-J FASnI$_3$ cell corresponding to the initial thickness variation of FASnI$_3$ absorber.}
    \label{Fig:S7}
\end{figure}

To identify performance-sensitive regions and optimize the 1-J FASnI\textsubscript{3} cell (MgF\textsubscript{2}/ITO/ZnO/FASnI\textsubscript{3}/CuSCN/ITO), we initialized an iterative thickness sweep of the FASnI\textsubscript{3} absorber from 300 to 1200~nm as shown in Fig.~\ref{Fig:S7}. Increasing the absorber thickness enhances photon absorption across the visible–NIR spectrum, raising J\textsubscript{sc} from 24.52 to 28.53 mA/cm\textsuperscript{2}, which saturates around 900~nm, indicating efficient collection of photogenerated carriers at the selective contacts. As the absorber thickens, longer carrier diffusion paths increase bulk recombination, leading to a gradual reduction in V\textsubscript{oc} and FF as noticed in Fig.~\ref{Fig:S7}(c–d). Consequently, the PCE rises with thickness, reaching a maximum of 21.69\% at 900~nm, beyond which the trade-off between enhanced absorption and recombination causes a slight decline.

\begin{figure}[!ht]
    \centering
    \includegraphics[width=0.9\linewidth]{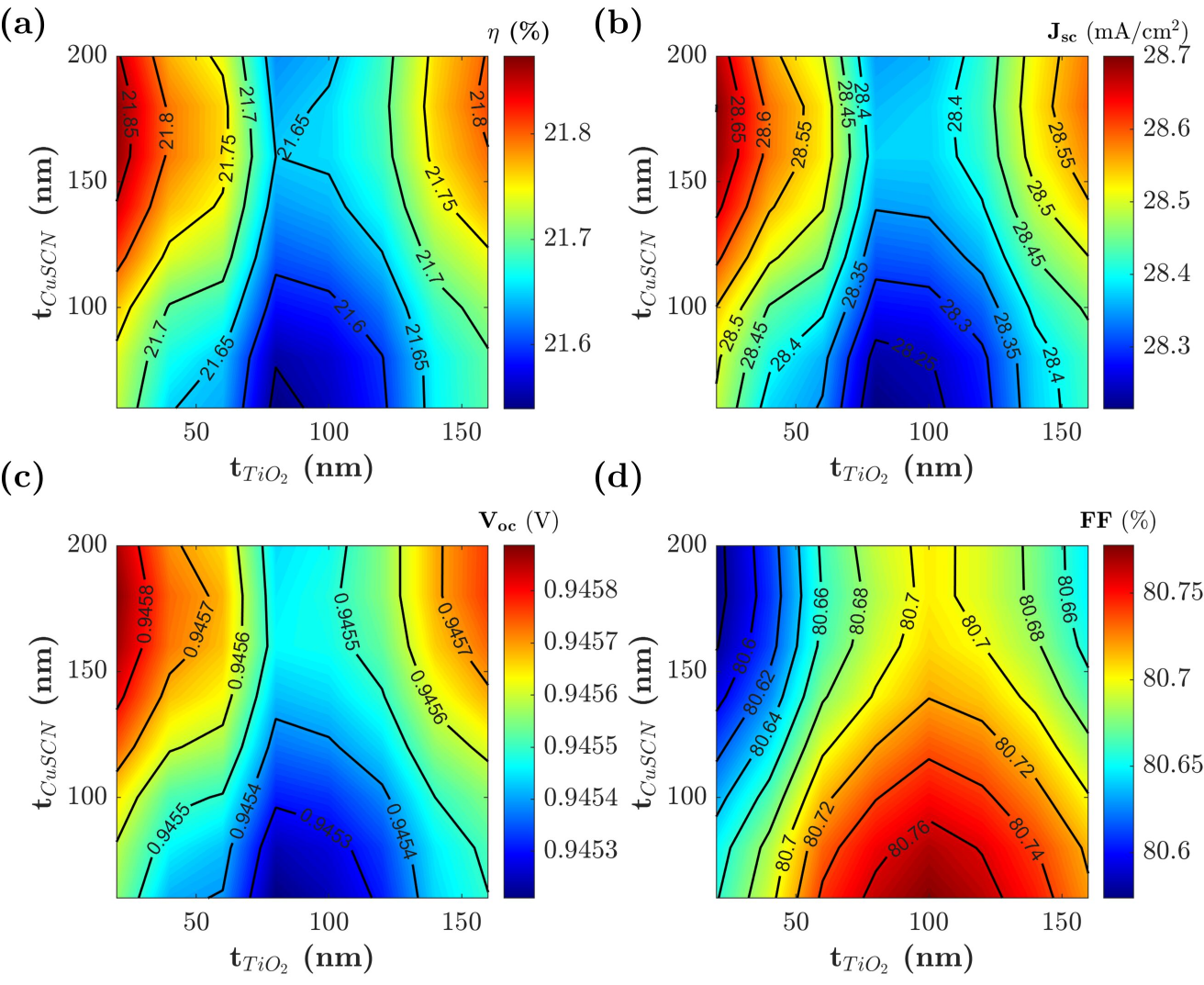}
    \caption{Initial parametric sweep showing the impact on performance metrics—(a) $\eta$, (b) J\textsubscript{sc}, (c) V\textsubscript{oc}, and (d) FF of 1-J FASnI$_3$ cell—by varying thickness of TiO$_2$ ETL and CuSCN HTL layers for ITO/ZnO/TiO\textsubscript{2}/FASnI\textsubscript{3}/CuSCN/ITO architecture.}
    \label{Fig:S8}
\end{figure}

After optimizing the FASnI\textsubscript{3} absorber at 900~nm, we performed iterative thickness sweeps of the TiO\textsubscript{2} ETL from 20 to 160~nm and CuSCN HTL from 60 to 200~nm as featured in Fig.~\ref{Fig:S8}. The analysis shows that a thin ETL of 20~nm is sufficient for efficient electron extraction, minimizing series resistance and preserving the built-in field, whereas a thicker HTL in 160--180~nm range improves hole collection and reduces interfacial recombination at the back contact. J\textsubscript{sc} remains nearly saturated, and both V\textsubscript{oc} and FF exhibit minor variations, indicating that the absorber governs photo-generation while the transport layers optimize charge extraction. The peak PCE of 21.87\% is achieved for this configuration, confirming that thinner ETL and thicker HTL layers provide optimal carrier collection without compromising overall device efficiency.\\

 After obtaining optimum FASnI\textsubscript{3} absorber thickness of 800 nm, the final parametric sweep of transport layers demonstrates that the device performance exhibits a clear plateau across variations in TiO\textsubscript{2} ETL and CuSCN HTL thickness. The maximum performance was achieved for 20 nm and 160 nm thickness of TiO\textsubscript{2} ETL and CuSCN HTL, confirming that the simulation has converged to a self-consistent optimum. The performance matrices shows minimal changes beyond these ETL and HTL values, indicating that the device has reached an optimized state where optical absorption, carrier generation, and drift–diffusion transport are balanced.
\begin{figure}[!ht]
    \centering
    \includegraphics[width=0.9\linewidth]{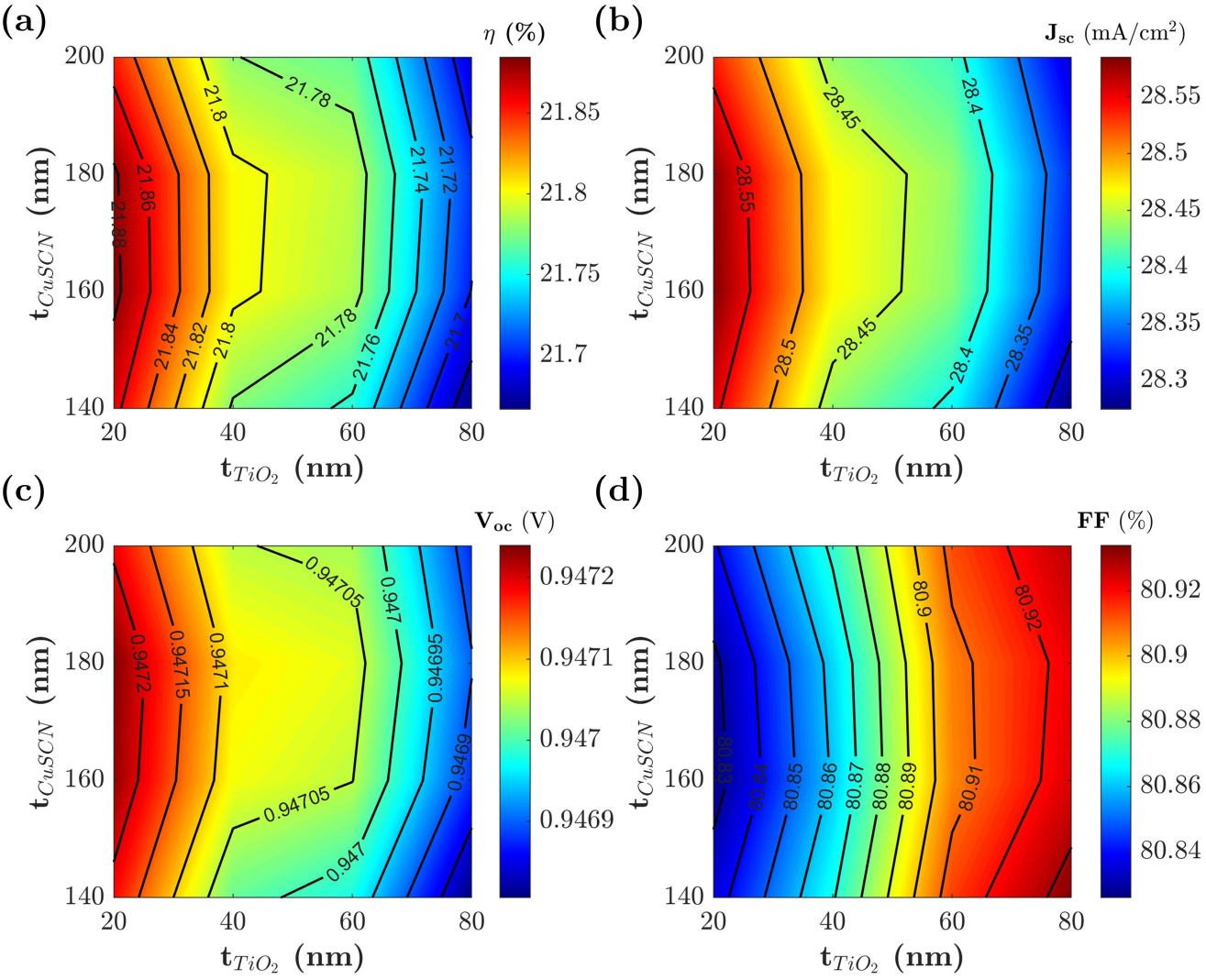}
    \caption{Final parametric sweep showing the impact on performance metrics—(a) $\eta$, (b) J\textsubscript{sc}, (c) V\textsubscript{oc}, and (d) FF of 1-J FASnI$_3$ cell—by varying thickness of TiO$_2$ ETL and CuSCN HTL layers for ITO/ZnO/TiO\textsubscript{2}/FASnI\textsubscript{3}/CuSCN/ITO architecture.}
    \label{Fig:S9}
\end{figure}

\pagebreak
\subsubsection{Incorporation of Spiro-OMeTAD HTL}
\begin{figure}[!b]
     \centering
     \includegraphics[width=0.95\linewidth]{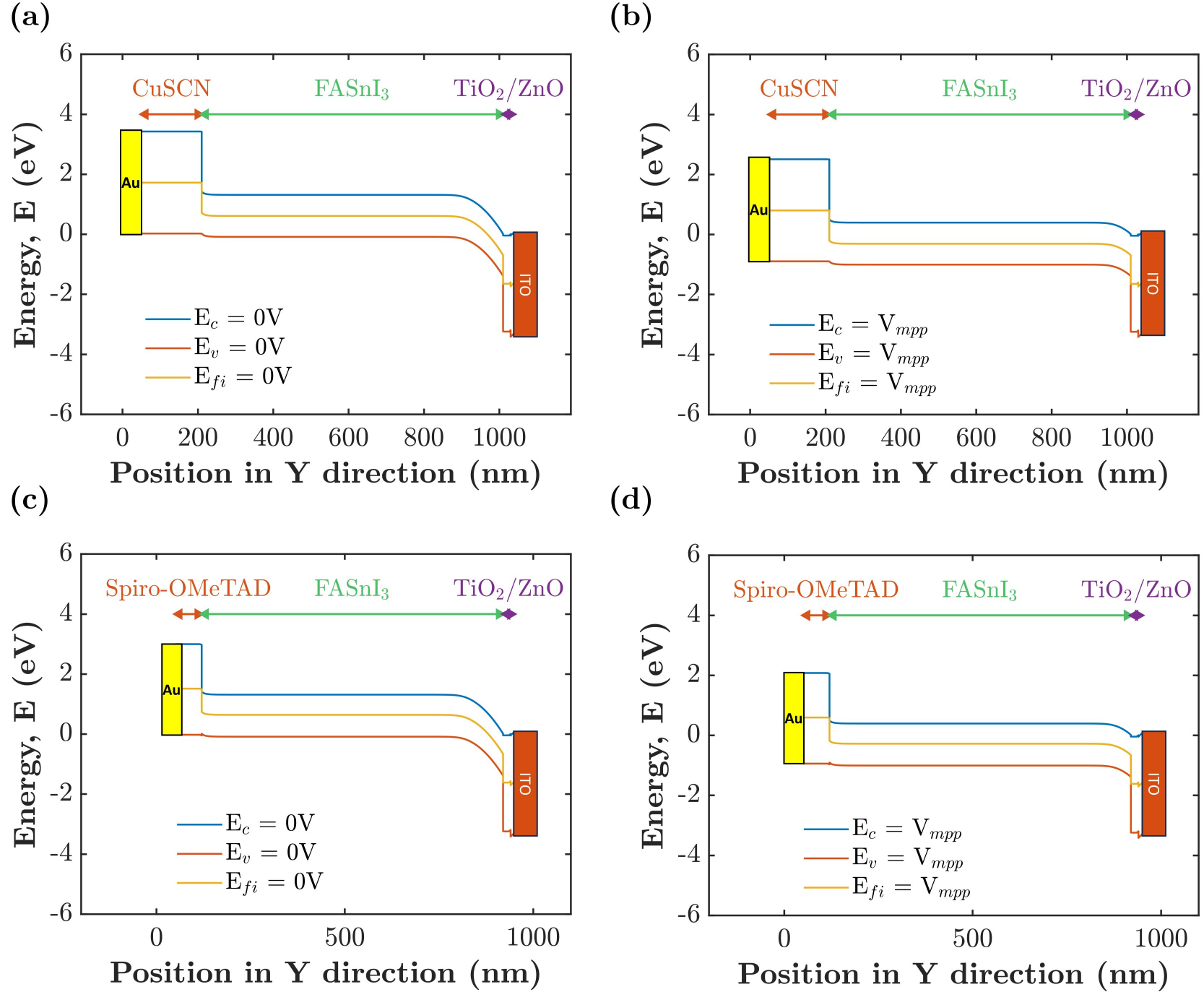}
     \caption{Band-diagram of ITO/ZnO/TiO$_2$/FASnI$_3$/CuSCN/Au at thermal equilibrium at (a) short circuit condistion ($V = 0V$) and (b) maximum power point condition ($V = V_{mpp}$). Band-diagram of ITO/ZnO/TiO$_2$/FASnI$_3$/Spiro-OMeTAD/Au at thermal equilibrium at (c) short-circuit condition (V = 0V) and (d) maximum power point (V = V$_{mpp}$).}
     \label{Fig:S10}
\end{figure}

Fig.~\ref{Fig:S10} presents the band diagrams of the 1-J FASnI\textsubscript{3} cell (ITO/ZnO/FASnI\textsubscript{3}\\/HTL/Au) using CuSCN and Spiro-OMeTAD as HTLs at short-circuit condition (V = 0 V) and maximum power point condition (V = V\textsubscript{mpp}). The FASnI\textsubscript{3}/Spiro-OMeTAD interface exhibits a slightly negative VBO $\approx$~--0.06~eV, providing a downhill potential that promotes hole drift toward the Au contact while maintaining a sufficient CBO $\approx$~1.56~eV, to suppress electron back-injection. In contrast, the FASnI$_3$/CuSCN interface shows a nearly flat VBO of $\approx$~0~eV, resulting in a weaker hole driving potential despite its larger CBO of $\approx$~2.0~eV.
The thinner Spiro-OMeTAD layer of 70~nm, induces a stronger built-in electric field, enhancing hole drift and reducing recombination, whereas the thicker CuSCN layer of 160~nm, increases series resistance and diffusion length, leading to higher carrier loss. Consequently, the Spiro-OMeTAD-based cell exhibits stronger band bending and quasi-Fermi level splitting, confirming more efficient field-assisted charge extraction and superior photovoltaic performance.
\subsection{Dual Junction KSnI\textsubscript{3}/FASnI\textsubscript{3} Tandem Cell}
\begin{figure}[!ht]
    \centering
    \includegraphics[width=0.8\linewidth]{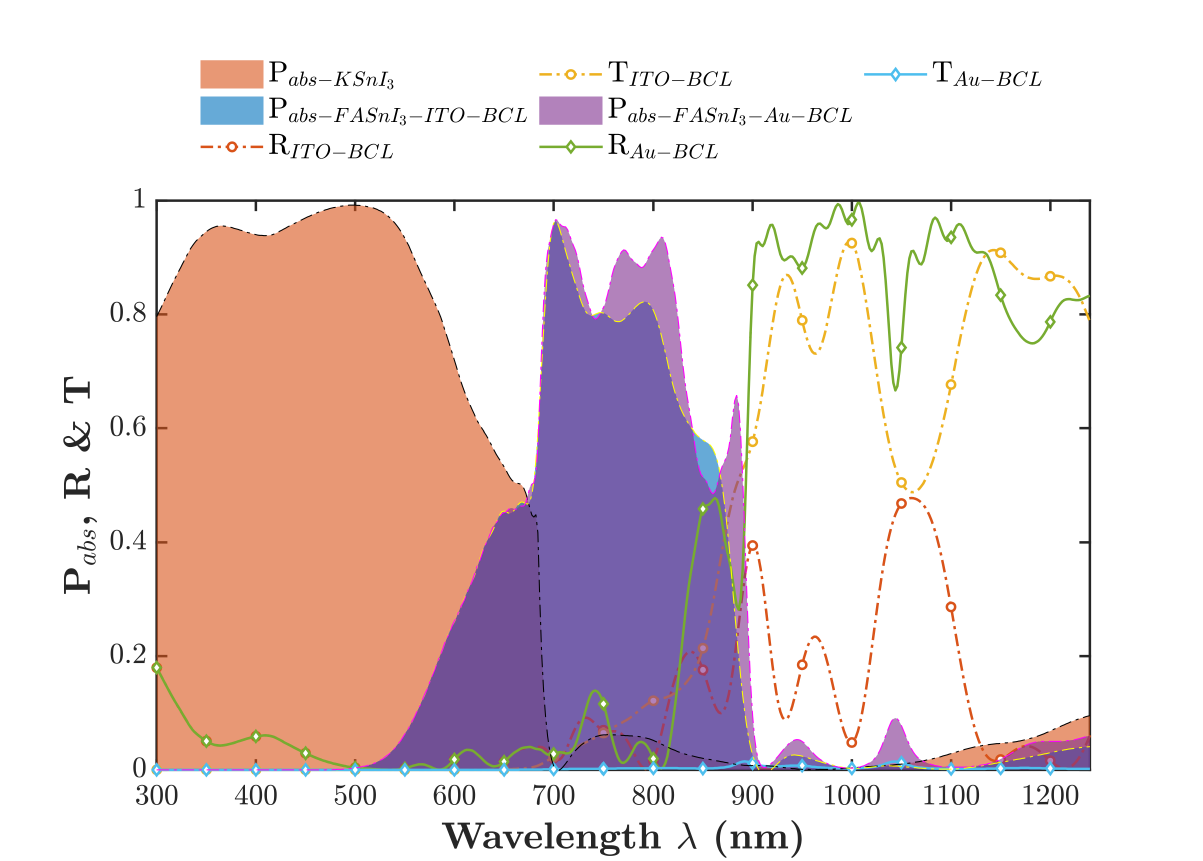}
    \caption{Normalized power absorption (P\textsubscript{abs}), reflection (R), and transmission (T) as a function of photon wavelength ($\lambda$), illustrating the effect of Au and ITO back contacts on bottom absorber (FASnI$_3$) absorption, total device reflection, and transmission through the bottom cell in the 2-J KSnI$_3$/FASnI$_3$ tandem configuration.}
    \label{Fig:S11}
\end{figure}

\begin{table}[!ht]
\centering
\caption{Impact of incorporating ITO instead of Au as BCL in 2-J bottom subcell’s BCL.}
\label{Table:S3}
\resizebox{\columnwidth}{!}{%
\begin{tabular}{lcccccc}
\hline
\textbf{Bottom subcell’s BCL} &
  \textbf{\begin{tabular}[c]{@{}c@{}}$\eta$\\ (\%)\end{tabular}} &
  \textbf{\begin{tabular}[c]{@{}c@{}}J\textsubscript{sc}\\ (mA/cm\textsuperscript{2})\end{tabular}} &
  \textbf{\begin{tabular}[c]{@{}c@{}}V\textsubscript{oc}\\ (V)\end{tabular}} &
  \textbf{\begin{tabular}[c]{@{}c@{}}FF\\ (\%)\end{tabular}} &
  \textbf{\begin{tabular}[c]{@{}c@{}}P\textsubscript{mpp}\\ (W/m\textsuperscript{2})\end{tabular}} &
  \textbf{\begin{tabular}[c]{@{}c@{}}V\textsubscript{mpp}\\ (V)\end{tabular}} \\ \hline
Au BCL &
  27.29 &
  14.74 &
  2.227 &
  83.14 &
  272.89 &
  1.981 \\
ITO BCL &
  24.89 &
  12.97 &
  2.224 &
  86.31 &
  248.89 &
  2.012 \\ \hline
\end{tabular}%
}
\end{table}
\pagebreak
\subsection{Triple Junction KSnI\textsubscript{3}/FASnI\textsubscript{3}/ACZTSe Tandem Cell}
To construct the initial triple-junction (3-J) tandem architecture, the bottom subcell was designed as AZO/ZnO/ZnS/ACZTSe/CZTSe/Mo, while the optimized 1-J KSnI$_3$ and 1-J FASnI$_3$ layers were employed as the top and middle subcell, respectively. The initial layer thicknesses and doping concentrations used for each subcell are summarized in Table~\ref{Table:S4}.

\begin{table}[!ht]
    \centering
    \small
    \caption{Initial thickness and doping density of KSnI\textsubscript{3}, FASnI\textsubscript{3}, and ACZTSe-based triple junction tandem solar cell.}
    \label{Table:S4}
    \resizebox{\textwidth}{!}{%
    \begin{tabular}{lccc}
        \hline
        \textbf{Material} & 
        \textbf{\begin{tabular}[c]{@{}c@{}}Thickness\\(nm)\end{tabular}} & 
        \textbf{\begin{tabular}[c]{@{}c@{}}Doping density\\(cm$^{-3}$)\end{tabular}} & 
        \textbf{Doping type} \\
        \hline
        FTO & 100 & $-$ & $-$\\
        MgF\textsubscript{2} & 90 & $-$ & $-$\\
        TiO\textsubscript{2} & 40 & $5\times10^{19}$ & n\\
        KSnI\textsubscript{3} & 600 & $1\times10^{15}$ & i\\
        CuSCN & 100 & $5\times10^{18}$ & p\\
        ITO & 50 & $-$ & $-$\\
        ZnO & 10 & $8\times10^{18}$ & n\\
        TiO\textsubscript{2} & 20 & $5\times10^{19}$ & n\\
        FASnI\textsubscript{3} & 800 & $7\times10^{16}$ & p\\
        Spiro-OMeTAD & 80 & $7\times10^{18}$ & p\\
        ITO & 20 & $-$ & $-$\\
        AZO & 10 & $8\times10^{18}$ & n\\
        ZnO & 60 & $1.5\times10^{17}$ & n\\
        ZnS & 100 & $5\times10^{16}$ & n\\
        ACZTSe & 400 & $5\times10^{14}$ & p\\
        CZTSe & 200 & $1\times10^{16}$ & p\\
        Mo & 100 & $-$ & $-$\\\hline
    \end{tabular}%
    }
\end{table}

\subsubsection{Impact of ACZTSe doping on 3-J tandem cell.}
Table~\ref{Table:S5} summarizes the effect of ACZTSe doping concentration on the photovoltaic performance of the 3-J tandem cell. As the acceptor density increases from $5\times10^{14}$ to $1\times10^{15}$~cm$^{-3}$, PCE improves slightly from 25.72\% to 26.08\%, primarily due to enhanced built-in potential and improved carrier extraction, leading to higher V\textsubscript{oc} and FF. Beyond this optimum range, further increase in doping reduces J\textsubscript{sc} and overall efficiency owing to increased carrier scattering and recombination losses in the highly doped absorber. The optimal doping concentration for ACZTSe is identified at $9\times10^{14}$~cm\textsuperscript{-3}, where the balance between electric field strength and carrier transport yields maximum PCE.
\begin{table}[!ht]
\centering
\caption{Impact of acceptor doping, N$_{A}$ of ACZTSe on 3-J tandem cell.}
\label{Table:S5}
\resizebox{\columnwidth}{!}{%
\begin{tabular}{ccccccc}
\hline
\textbf{\begin{tabular}[c]{@{}c@{}}ACZTSe acceptor density\\ N$_{A}$ (cm\textsuperscript{-3})\end{tabular}} & 
\textbf{\begin{tabular}[c]{@{}c@{}}$\eta$\\ (\%)\end{tabular}} & 
\textbf{\begin{tabular}[c]{@{}c@{}}J\textsubscript{sc}\\ (mA/cm\textsuperscript{2})\end{tabular}} & 
\textbf{\begin{tabular}[c]{@{}c@{}}V\textsubscript{oc}\\ (V)\end{tabular}} & 
\textbf{\begin{tabular}[c]{@{}c@{}}FF\\ (\%)\end{tabular}} & 
\textbf{\begin{tabular}[c]{@{}c@{}}P\textsubscript{mpp}\\ (W/m\textsuperscript{2})\end{tabular}} & 
\textbf{\begin{tabular}[c]{@{}c@{}}V\textsubscript{mpp}\\ (V)\end{tabular}} \\ \hline
$5.00 \times 10^{14}$ & 25.72 & 11.44 & 2.705 & 83.17 & 257.25  & 2.392 \\
$6.00 \times 10^{14}$ & 25.84 & 11.45 & 2.707 & 83.39 & 258.45 & 2.394 \\
$7.00 \times 10^{14}$ & 25.93 & 11.46 & 2.710 & 83.56 & 259.39 & 2.396 \\
$8.00 \times 10^{14}$ & 26.01 & 11.46 & 2.712 & 83.68 & 260.11 & 2.412 \\
$9.00 \times 10^{14}$ & 26.06 & 11.46 & 2.714 & 83.77 & 260.60 & 2.414 \\
$1.00 \times 10^{15}$ & 26.08 & 11.45 & 2.717 & 83.82 & 260.89 & 2.416 \\
$2.00 \times 10^{15}$ & 25.82 & 11.27 & 2.733 & 83.89 & 258.28 & 2.431 \\
$3.00 \times 10^{15}$ & 25.34 & 10.99 & 2.745 & 83.99 & 253.47 & 2.455 \\
$4.00 \times 10^{15}$ & 24.93 & 10.76 & 2.752 & 84.24 & 249.39 & 2.461 \\
$5.00 \times 10^{15}$ & 24.60 & 10.56 & 2.757 & 84.46 & 246.00 & 2.466 \\ \hline
\end{tabular}%
}
\end{table}

\begin{figure}[!b]
    \centering
    \includegraphics[width=0.9\linewidth]{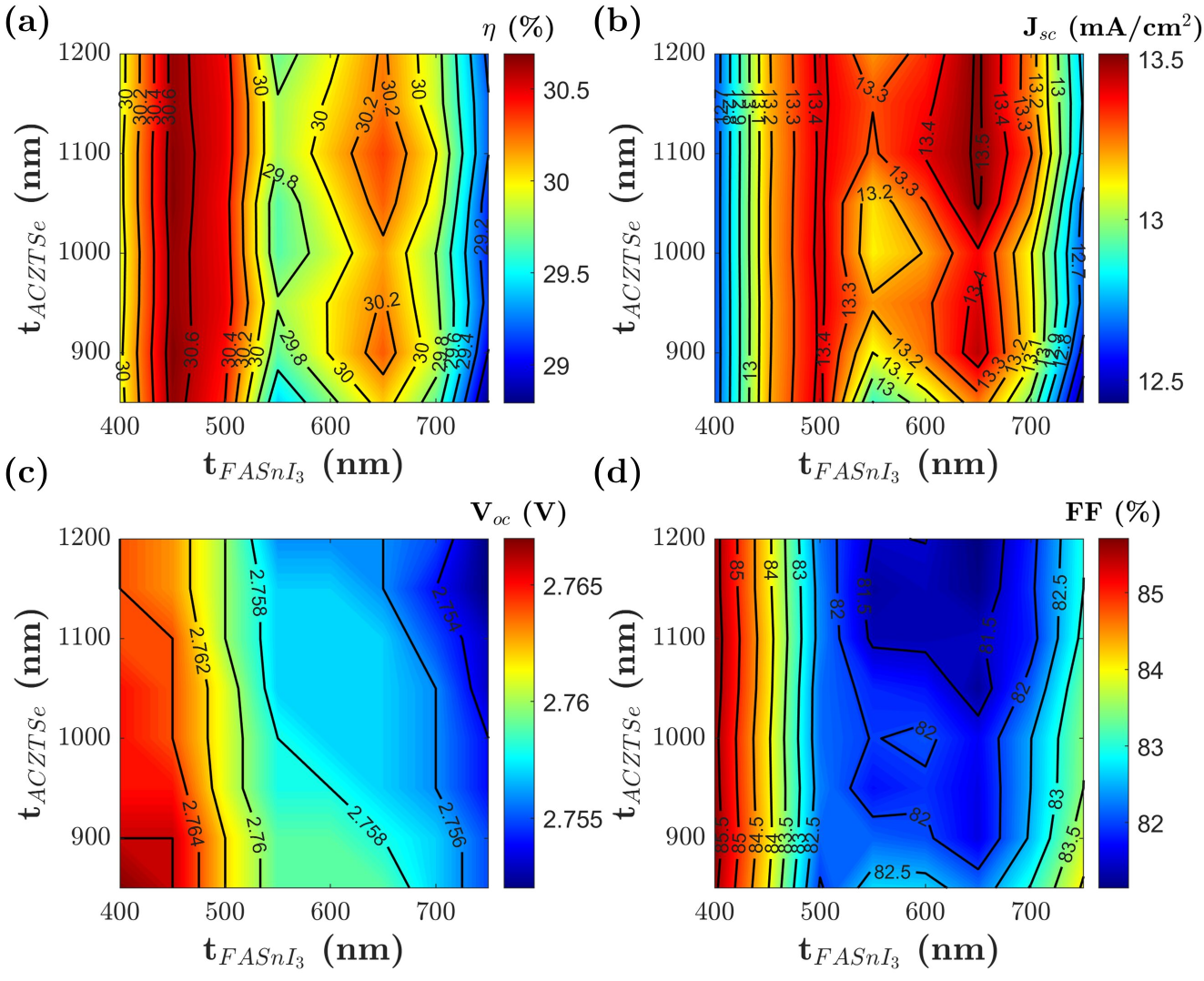}
    \caption{Contour plot of parametric sweep showing performance metrics--(a) $\eta$ (b) J$_{sc}$ (c) V$_{oc}$ (d) FF corresponding to thickness variation of FASnI$_3$ and ACZTSe thickness keeping top cell's KSnI$_3$ absorber thickness 400 nm highlighiting the optimized performance of the 3-J tandem cell with a champion power conversion efficiency ($\eta$) of 30.69\%.}
    \label{fig:S12}
\end{figure}
\subsubsection{Volumetric thickness optimization of absorbers}
To identify the optimum configuration of the 3-J tandem device, we carried out an extensive volumetric optimization analysis, where the absorber thicknesses of KSnI\textsubscript{3}, FASnI\textsubscript{3}, and ACZTSe were systematically tuned to achieve optimal current matching and maximize device efficiency. Specifically, the KSnI\textsubscript{3} thickness was varied from 350 to 500~nm, and for each selected KSnI\textsubscript{3} thickness, a detailed two-dimensional sweep of FASnI\textsubscript{3} from 400 to 700~nm and ACZTSe from 850 to 1150~nm, was carried out. This comprehensive approach allowed us to capture the performance-sensitive regions and identify the most balanced combinations of subcell thicknesses. The summarized results in Table~\ref{Table:S6} present the best-performing configurations corresponding to each KSnI\textsubscript{3} region, highlighting the depth of our optimization analysis and the considerable effort devoted to achieving a finely tuned, high-efficiency 3-J tandem design. Fig.~\ref{fig:S12} features the contour plots of the parametric sweep, showing the variation of performance metrices with FASnI\textsubscript{3} and ACZTSe thicknesses while keeping the top cell’s KSnI\textsubscript{3} thickness fixed at 400~nm. The plot highlights the most efficient region of the parameter space, where a champion PCE($\eta$) of 30.69\% was achieved.
\begin{table}[!t]
\centering
\caption{Tandem 3D thickness sweep (KSnI\textsubscript{3}, FASnI\textsubscript{3} \& ACZTSe)}
\label{Table:S6}
\resizebox{\columnwidth}{!}{%
\begin{tabular}{ccccccccc}
\hline
\textbf{\begin{tabular}[c]{@{}c@{}}KSnI\textsubscript{3}\\ (nm)\end{tabular}} &
  \textbf{\begin{tabular}[c]{@{}c@{}}FASnI\textsubscript{3}\\ (nm)\end{tabular}} &
  \textbf{\begin{tabular}[c]{@{}c@{}}ACZTSe \\(nm)\end{tabular}} &
  \textbf{\begin{tabular}[c]{@{}c@{}}$\eta$\\ (\%)\end{tabular}} &
  \textbf{\begin{tabular}[c]{@{}c@{}}J\textsubscript{sc}\\ (mA/cm\textsuperscript{2})\end{tabular}} &
  \textbf{\begin{tabular}[c]{@{}c@{}}V\textsubscript{oc}\\ (V)\end{tabular}} &
  \textbf{\begin{tabular}[c]{@{}c@{}}FF\\ (\%)\end{tabular}} &
  \textbf{\begin{tabular}[c]{@{}c@{}}P\textsubscript{mpp}\\ (mW/cm\textsuperscript{2})\end{tabular}} &
  \textbf{\begin{tabular}[c]{@{}c@{}}V\textsubscript{mpp}\\ (V)\end{tabular}} \\ \hline
350 & 450 & 900  & 30.40 & 13.05 & 2.772 & 84.07 & 304.04 & 2.438 \\
400 & 450 & 900  & 30.69 & 13.18 & 2.766 & 84.18 & 306.92 & 2.432 \\
450 & 600 & 1100 & 29.99 & 13.45 & 2.751 & 81.04 & 299.94 & 2.405 \\
500 & 550 & 1100 & 29.17  & 12.64 & 2.743 & 84.16 & 291.70  & 2.412 \\
550 & 500 & 850  & 28.44  & 12.14 & 2.738 & 85.55 & 284.40  & 2.422 \\ \hline
\end{tabular}%
}
\end{table}
\clearpage


\end{document}